\documentclass[]{pasj01}
\draft

\usepackage{natbib}

\begin{document} 
\Received{}%{yyyy/mm/dd}
\Accepted{}%{yyyy/mm/dd}
%\Published{yyyy/mm/dd}

\title{Clustering of galaxies around AGN in the HSC Wide survey
\thanks{Based on data collected at Subaru Telescope, which is operated 
by the National Astronomical Observatory of Japan.}}

%%% begin:list of authors
% Do NOT capitalize all letters in "textsc".

\author{Yuji \textsc{Shirasaki}\altaffilmark{1,2}}%
\author{Masayuki \textsc{Akiyama}\altaffilmark{3}}%
\author{Tohru \textsc{Nagao}\altaffilmark{4}}%
\author{Yoshiki \textsc{Toba}\altaffilmark{5}}%
\author{Wanqiu \textsc{He}\altaffilmark{3}}%

\author{Masatoshi \textsc{Ohishi}\altaffilmark{1,2}}%
\author{Yoshihiko \textsc{Mizumoto}\altaffilmark{1,2}}%

\author{Satoshi \textsc{Miyazaki}\altaffilmark{1,2}}%
\author{Atsushi J. \textsc{Nishizawa}\altaffilmark{6}}%
\author{Tomonori \textsc{Usuda}\altaffilmark{1,2}}%

\altaffiltext{1}{National Astronomical Observatory of Japan,
2-21-1 Osawa, Mitaka, Tokyo 181-8588, Japan}
\email{yuji.shirasaki@nao.ac.jp}

\altaffiltext{2}{Department of Astronomical Science, School 
of Physical Sciences, SOKENDAI (The Graduate University for 
Advanced Studies), 
2-21-1 Osawa, Mitaka, Tokyo 181-8588, Japan}

\altaffiltext{3}{Astronomical Institute, Tohoku University,
6-3 Aramaki, Aoba-ku Sendai, Miyagi 980-8578, Japan}

\altaffiltext{4}{Research Center for Space and Cosmic Evolution,
Ehime University, 
10-13 Dogo-Himata, Matsuyama, Ehime 790-8577, Japan}

\altaffiltext{5}{Academia Sinica Institute of Astronomy and
Astrophysics, P.O. Box 23-141, Taipei 10617, Taiwan}

\altaffiltext{6}{Institute for Advanced Research, Nagoya University,
Furocho, Chikusa-ku, Nagoya 464-8602, Japan}

%%% end:list of authors

%% `\KeyWords{}' always has to be placed before `\maketitle'.
 %Do NOT move this preamble from here!

\maketitle

\begin{abstract}
We have measured the clustering of galaxies around active galactic
nuclei (AGN) for which 
single-epoch virial masses of the super-massive black hole (SMBH)
are available to investigate
the relation between the large scale environment of AGNs and the 
evolution of SMBHs.
The AGN samples used in this work were derived from the Sloan Digital 
Sky Survey (SDSS) observations and the galaxy samples were from 
240~deg$^{2}$ S15b data of the Hyper Suprime-Cam 
Subaru Strategic Program (HSC-SSP).
The investigated redshift range is 0.6--3.0, and the masses of the 
SMBHs lie in the range $10^{7.5}$--$10^{10}$~M$_{\solar}$.
The absolute magnitude of the galaxy samples reaches to 
$M_{\lambda 310}$
$\sim$ $-$18 
at rest frame wavelength 
310~nm
for the low-redshift end of the samples.
More than 70\% of the galaxies in the analysis are blue.
We found a significant dependence of the cross-correlation length on 
redshift, which primarily reflects the brightness dependence of the 
galaxy clustering.
At the lowest redshifts
the cross-correlation 
length increases
from 7~$h^{-1}$~Mpc
around 
$M_{\lambda 310}$
$ = $ 
$-19$~mag
to $>$10~$h^{-1}$~Mpc beyond 
$M_{\lambda 310}$
$ = $
$-20$~mag.
No significant dependence of the cross-correlation length on
BH mass was found for whole galaxy samples dominated by blue galaxies, 
while there was an indication of BH mass dependence 
in the cross-correlation with red galaxies.
These results provides a picture of the environment of AGNs studied 
in this paper being enriched with blue starforming galaxies, and 
a fraction of the galaxies are being evolved to red galaxies
along with the evolution of SMBHs in that system.

\end{abstract}

\section{Introduction}

Most galaxies have a supermassive black hole (SMBH) with mass 
greater
than $\sim$10$^{6}$~M$_{\solar}$ at their 
center~\citep{Richstone+98}.
Observations of the local Universe have revealed that the mass 
of the SMBH correlates with the several properties of the bulge 
component of the host galaxy~\citep{Magorrian+98,Ferrarese+00,Gebhardt+00,Ho-07}.
This observational evidence suggests that a SMBH and its host
galaxy co-evolve in a coordinated way in spite of the nine
orders of difference in their physical size scale~\citep{Kormendy+13}.

SMBHs grow through the accretion of gas from their host galaxies
or large-scale environment.
Accretion in a secular mode, which arises through internal dynamical
processes such as bar or disk instability or external processes
driven by galaxy interaction, is one of the mechanisms to deliver the
gas into the SMBH~\citep{Kormendy+04}.
As the mass accretion rate of the secular mode cannot be as high 
as to maintain the activity seen in bright QSOs~\citep{Menci+14}, 
this mode could operate in lower-luminosity AGNs.

A merger of galaxies can induce gravitational torques
which drive inflows of cold gas toward the center of galaxies~\citep[e.g.,][]{Hopkins+08}.
Observational evidence has been obtained that shows the relation
between AGN activity and a galaxy merger~\citep[e.g.,][]{Sanders+88,Treister+12}.
The activity of the bright QSOs could be explained by this model.
Although the accretion can be the most efficient in this mode,
it is also expected that AGN feedback promptly operates and stops
the gas inflows when the mass of the SMBH becomes as 
large
as 
10$^{9}$~M$_{\solar}$~\citep{Fanidakis+13}.

As an alternative process to make a SMBH evolve above 
10$^{9}$~M$_{\solar}$,
quiescent gas accretion from the hot halo~\citep{Keres+09,Fanidakis+13} 
and/or recycled gas from evolving stars~\citep{Ciotti+01,Ciotti+07} has 
been proposed.
According to the predictions of semi-analytical modeling by \citet{Fanidakis+13},
AGNs fueled in the hot-halo mode are located in more massive dark matter haloes
than the AGNs fueled in the merger-driven starburst mode.

The mass of the host dark matter halo that AGNs reside can be inferred from 
the auto-correlation function of AGNs at distance scale $>$1~Mpc.
According to the analysis of auto-correlation of SDSS QSOs by \citet{Ross+09}, 
the halo mass is almost constant at $\sim$2$\times 10^{12}h^{-1}M_{\solar}$ 
in the redshift range from 0.3 to 2.2.
It is also possible to estimate the halo mass from the cross-correlation between
AGNs and galaxies if the auto-correlation functions of galaxies are precisely
obtained.
According to the cross-correlation studies the dark matter halo mass estimated
to be $10^{12}$--$10^{13.5}h^{-1}M_{\solar}$ depending on the detection wavelength
of the AGNs~\citep[e.g.][]{Hickox+09,Krumpe+12}.
Studies on clustering and/or environments of AGNs have also been reported elsewhere
\citep[e.g.][]{Croom+05,Coil+09,Silverman+09,Donoso+10,Allevato+11,Bradshaw+11,
Shen+13,Mountrichas+13,Zhang+13,Georgakakis+14,Krumpe+15,Ikeda+15}.

In spite of a large number of AGN clustering studies published so far,
the studies focused on the relation between the mass of the central BH,
which is the most fundamental property on which the history of mass
accretion is imprinted, and also the properties of galaxies such as color and 
luminosity function around it are very limited.
Since the interaction in the scale of group or cluster of galaxies can
induce concurrent activities of starformation, mass accretion to the central
BH, and transition to the red sequence in the constituent galaxies, 
the contribution of such large-scale phenomena to the evolution of SMBH can be 
inferred from the surrounding galaxies.
\citet{Shen+09} 
measured the clustering of QSOs for samples divided by their 
BH
mass at redshifts 0.4--2.5 and didn't find significant dependence except
for the most massive sample for which marginally larger clustering was found 
in $\sim$2$\sigma$ level.
\citet{Komiya+13} examined dependence of the AGN-galaxy clustering on the 
BH
mass at redshift range from 0.3 to 1.0 using the UKIDSS data 
for the galaxy samples, 
and found that the cross-correlation length increases above $10^{8.2}M_{\solar}$.
\citet{Krolewski+15} also examined the BH mass dependence of the AGN-galaxy 
clustering at redshift $\sim$0.8 using SDSS and WISE data for the galaxy samples,
and found no significant relationship between clustering amplitude and BH mass.
\citet{Krumpe+15} measured clustering of soft X-ray and optically selected 
AGNs at redshifts from 0.16 to 0.36 and detected a weak dependence 
on the mass at significance level of 2.7~$\sigma$, while they didn't detect 
significant dependence on the Eddington ratio.
They conclude that the mass dependence is the origin of the observed weak
X-ray luminosity clustering dependence.
Extending the dataset of \citet{Komiya+13}, \citet{Shirasaki+16} derived, 
for the first time as derived from statistically significant
number of samples,
color and absolute magnitude distributions of galaxies around AGNs, and
found that the increase of the cross-correlation length found by 
\citet{Komiya+13} is due to the increase in the number density of red galaxies.
Those results indicate that the most massive SMBHs are evolved in dark 
matter haloes more massive than the lower mass SMBHs, and the surrounding 
galaxies also evolve in a coordinated way with the SMBHs which are 
mostly located in the center of host halo.
To extend further the study given in \citet{Shirasaki+16} up to redshifts 
$\sim$3, which covers the era of the peak of star formation and mass 
accretion rate to SMBHs, we require a deeper wide multi-band survey.

The Hyper Suprime-Cam Subaru Strategic Program (HSC-SSP) is a multi-band 
imaging survey conducted with the HSC on the 8.2m Subaru Telescope.
The survey consists of three layers: 
Wide (1400~deg$^{2}$, r $\sim$26), 
Deep (27~deg$^{2}$, r$\sim$27), 
and UltraDeep (3.5~deg$^{2}$, r$\sim$28).
The HSC-SSP Wide survey provides the first opportunity to investigate
the environment of AGNs up to redshift three with unprecedented
statistics.
Using the unique dataset of HSC-SSP Wide survey, 
this paper's purpose is
to measure not only the clustering of galaxies around AGNs
but also their color and luminosity distribution as a function of SMBH 
mass at five redshift groups from 0.6 to 3.0.
The results obtained in this work would give us a unified picture of 
evolution of galaxies and SMBHs under the large-scale structure of the 
Universe at their most important stage.

Throughout this paper, we assume a cosmology with $\Omega_{m} = 0.3$, 
$\Omega_{\lambda} = 0.7$, $h = 0.7$ and $\sigma_{8} = 0.8$.
All magnitudes are given in the AB system.
All the distances are measured in comoving coordinates.
The correlation length is presented in unit of $h^{-1}$Mpc.
The group and parameter names that are frequently refereed 
to in the text are summarized in 
Table~\ref{tab:param_desc} 
of 
the Appendix.

\section{Datasets}

\subsection{AGNs}
\label{sec:agn}
The AGN samples used in this paper were drawn from the QSO properties catalog 
of \citet{Shen+11} (S11) and SDSS DR12 Quasar catalog (DR12Q) of \citet{Paris+17}.
We used S11 catalog as a reference for the black hole (BH) mass estimate to be 
consistent with the previous studies~\citep{Komiya+13,Shirasaki+16}, and the BH
masses derived from the DR12Q catalog and spectral measurements were calibrated 
with those derived in S11.

We divided the AGN samples according to their measured redshift and
BH mass.
The mass used in this analysis is based on the H$\beta$, 
Mg~II, or C~IV line width as given in the S11 and DR12Q catalogs.
For AGNs drawn from the DR12Q catalog, the mass was calculated according
to the formula derived by \citet{Mejia-Restrepo+16} as follows:
\begin{equation}
M_{\rm BH,MgII} = 8.05 \times 10^{6}~M_{\solar} \left(\frac{L_{\lambda 300}}{10^{44}~{\rm erg~s^{-1}}}\right)^{0.609} 
            \left( \frac{\rm FWHM(Mg~II)}{\rm km~s^{-1}} \right)^{2},
\label{eq:Mbh_MgII}
\end{equation}
\begin{equation}
M_{\rm BH,CIV} = 5.71 \times 10^{5}~M_{\solar} \left(\frac{L_{\lambda 145}}{10^{44}~{\rm erg~s^{-1}}}\right)^{0.57} 
            \left( \frac{\rm FWHM(C~IV)}{\rm km~s^{-1}} \right)^{2} 
            \left( \frac{L_{\rm p}(\rm C~III])}{L_{\rm p}(\rm C~IV)} \right)^{-2.09},
\label{eq:Mbh_CIV}
\end{equation}
where $L_{\lambda 300}$ and $L_{\lambda 145}$ are the continuum luminosity at 300~nm and 145~nm 
respectively, FWHM(Mg~II) and FWHM(C~IV) are the full width at half-maximum
of the Mg~II and C~IV emission lines respectively,
and $L_{p}(\rm C~III])$ and $L_{p}(\rm C~IV)$ are peak luminosities of the 
corresponding lines.
We measured the continuum luminosities, $L_{\lambda 300}$ and $L_{\lambda 145}$, and the luminosity 
ratio, $\frac{L_{\rm p}(\rm C~III])}{L_{\rm p}(\rm C~IV)}$ directly 
from the SDSS spectra, while the FWHMs of the emission lines were
taken from the DR12Q catalog.
\begin{figure}
  \begin{center}
     \includegraphics[width=0.48\textwidth]{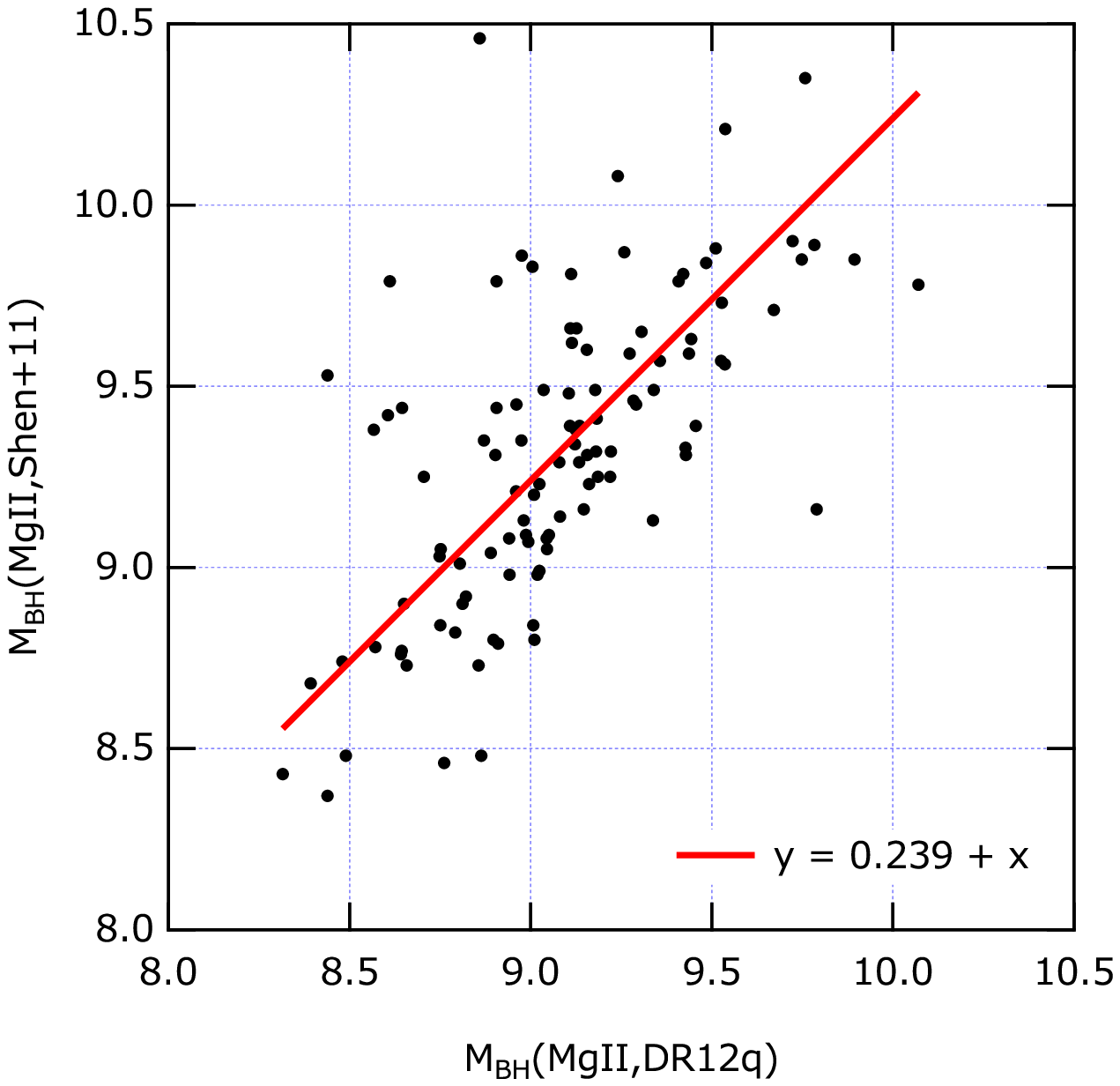} 
     \includegraphics[width=0.48\textwidth]{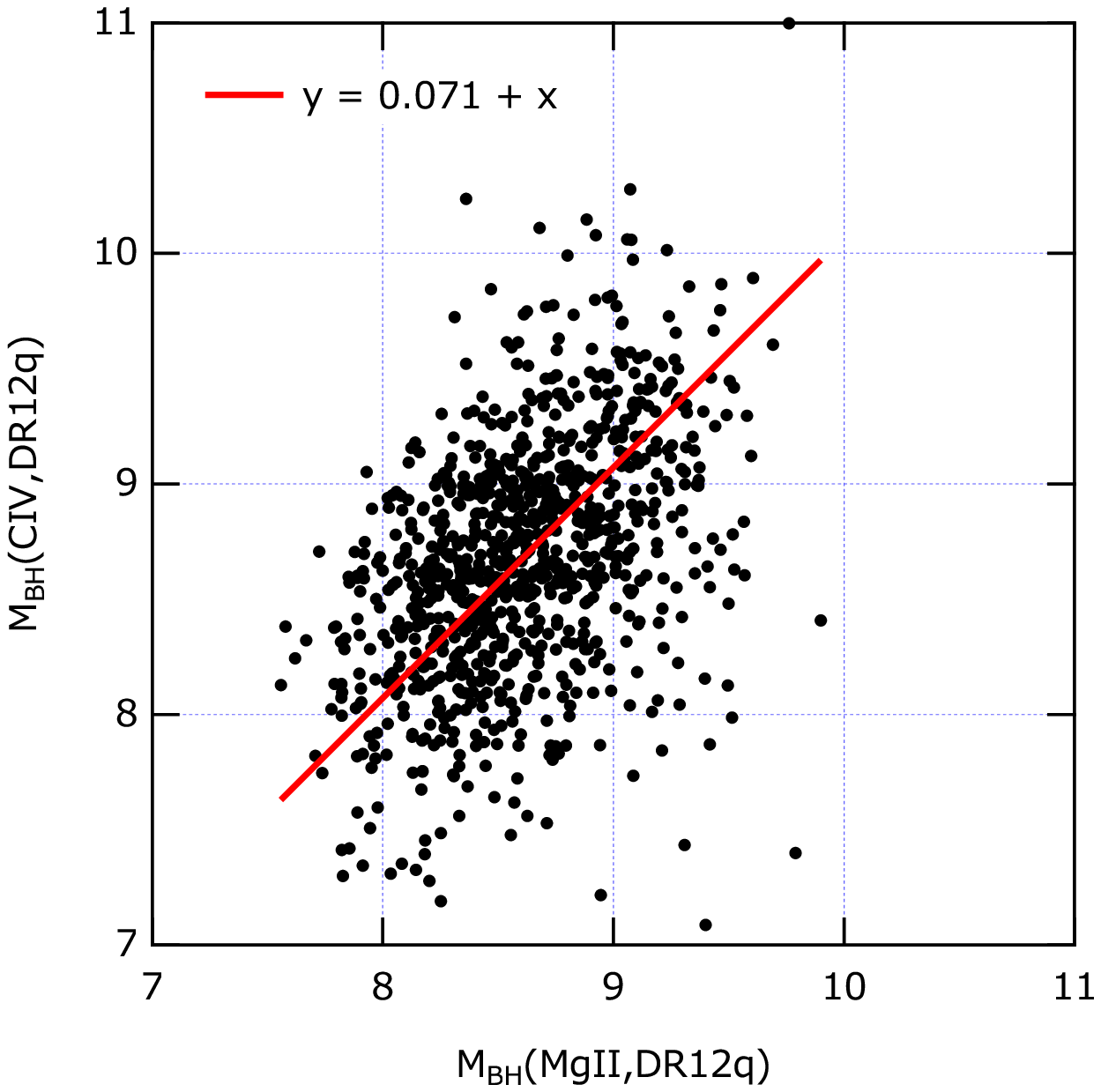} 
  \end{center}
  \caption{Comparisons of BH masses derived by different methods for the same
  object. 
  Left panel shows the comparison between the BH masses as given in
  S11~\citep{Shen+11} and those calculated for the same object in 
  DR12Q~\citep{Paris+17} using 
  equation~(\ref{eq:Mbh_MgII}).
  Right panel shows the comparison between the BH masses calculated for
  an object in DR12Q using equation~(\ref{eq:Mbh_MgII}) and (\ref{eq:Mbh_CIV}).
  The solid lines represent linear expressions fitted to the data points.
}
  \label{fig:AGN_Comp_Mbh}
\end{figure}
To check the consistency between the BH masses derived by 
different relations, we compared them for the same objects.
The left panel in Figure~\ref{fig:AGN_Comp_Mbh} shows the comparison of black hole 
masses given in S11, M$_{\rm BH,MgII}$(S11), and those calculated for the same 
object in DR12Q, M$_{\rm BH,MgII}$(DR12Q), using equation~(\ref{eq:Mbh_MgII}).
We found there is a systematic offset of 0.24 dex and scatter of 0.22, so 
we have corrected for the offset to the mass calculated using 
equation~(\ref{eq:Mbh_MgII}).

The right panel in Figure~\ref{fig:AGN_Comp_Mbh} shows the comparison of
black hole masses calculated for objects in DR12Q using 
equation~(\ref{eq:Mbh_MgII}), M$_{\rm BH,MgII}$(DR12Q), and 
equation~(\ref{eq:Mbh_CIV}), M$_{\rm BH,CIV}$(DR12Q).
Systematic offset of 0.07 dex and scatter of 0.48 dex were found between them.
Thus the offset of M$_{\rm BH,CIV}$(DR12Q) to M$_{\rm BH,MgII}$(S11) was
estimated to be 0.17, and the corresponding correction was made to 
M$_{\rm BH,CIV}$(DR12Q).
As mentioned in \citet{Mejia-Restrepo+16} and apparent from 
Figure~\ref{fig:AGN_Comp_Mbh}, M$_{\rm BH,CIV}$ has a larger uncertainty 
than the mass measured based on other emission lines, so we used M$_{\rm BH,MgII}$,
whenever it was available, as a mass estimator in first priority.
The AGN samples given the C~IV BH mass are mostly 
at $z \ge 2.3$.
The effect of the large uncertainty of the C~IV BH mass is limited only to 
the results for $z \ge 2.0$.

\citet{Coatman+16} recently showed that an empirical correction to C~IV BH masses 
based on the blueshift of C~IV emission line reduces the uncertainty.
To apply the correction requires 
an accurate measure of the AGN systemic redshift.
Thus, we decided to use the relation of equation~(\ref{eq:Mbh_CIV})
which does not require precise spectroscopic measurements.

The systematic offset and scatter between $M_{\rm BH, H\beta}$ and 
$M_{\rm BH, Mg~II}$ given in S11 catalog are 0.009 dex and 0.38 dex
respectively as shown in Figure~10 of \citet{Shen+11}.
Thus, no correction to $M_{\rm BH, H\beta}$ is required.

For AGNs appearing in both catalogs, we used the redshift and
black hole mass of the S11 catalog.
The redshift range of the AGNs was chosen to be 0.6--3.0, so 
that it overlaps with the 
highest redshift bin ($z$=0.6--1.0) of the previous 
studies~\citep{Komiya+13,Shirasaki+16} to cross check between 
both the results and 
extend up to the limit of sensitivity of this analysis.
The BH mass range was set to $10^{7}$--$10^{11}$M$_{\solar}$.
We selected 6166 AGNs which are in the redshift and BH mass range
and are located within the footprint of HSC-SSP Wide survey.
\subsection{Galaxies}
\label{sec:galaxy}

The galaxy sample was collected from the Hyper Suprime-Cam Subaru
Strategic Program (HSC-SSP) S15b Wide survey dataset.
HSC-SSP is a three-layered, multi-band ($grizy$ plus 4 narrow-band 
filters) imaging survey with the Hyper Suprime-Cam (HSC)~\citep{Miyazaki+12,Miyazaki+17} 
on the 8.2m Subaru Telescope.
The total area and the depth of observation will be 1400 deg$^{2}$ 
with $r \sim 26$ (Wide layer), 
27 deg$^{2}$ with $r \sim 27$ (Deep layer), and
3.5 deg$^{2}$ with $r \sim 28$ (UltraDeep layer).

\begin{table}
  \tbl{Summary of the survey area}{%
  \begin{tabular}{cccc}
      \hline
Field name & Center coordinates & $S_{i}$$^{a}$ & $S_{griz}$$^{b}$ \\
           &                    &  deg$^{2}$  & deg$^{2}$ \\
      \hline
XMM-LSS    & 02$^{h}$18$^{m}$ $-$04$^{\circ}$30' & 51.8 & 48.9 \\
GAMA09H    & 09$^{h}$00$^{m}$ $+$01$^{\circ}$00' & 53.0 & 41.9 \\
WIDE12H    & 11$^{h}$58$^{m}$ $+$00$^{\circ}$00' & 32.5 & 24.6 \\
GAMA15H    & 14$^{h}$32$^{m}$ $+$00$^{\circ}$00' & 38.6 & 35.4 \\
HECTOMAP   & 16$^{h}$24$^{m}$ $+$43$^{\circ}$30' &  9.6 &  9.0 \\
VVDS       & 22$^{h}$24$^{m}$ $+$01$^{\circ}$00' & 52.0 & 47.8 \\
AEGIS      & 20$^{h}$58$^{m}$ $+$52$^{\circ}$30' &  1.9 &  1.8 \\
      \hline
  \end{tabular}}\label{tab:survey_param}
  \begin{tabnote}
$^{a}$ Effective area of $i$-band detected samples.
$^{b}$ Effective area of four-band detected samples.
  \end{tabnote}
\end{table}

We used the dataset derived from Wide layer.
The observed locations and effective area are summarized in 
Table~\ref{tab:survey_param}.
The typical depths of the observation are 26.8, 26.4, 26.4, 25.5, 24.7 for
$g$, $r$, $i$, $z$, and $y$ bands, respectively.
The detail of survey itself is described in \citet{Aihara+17a},
and the content of the S15b dataset is in \citet{Aihara+17b}.
The S15b dataset were analyzed through the HSC pipeline
(version 4.0.1) developed by the HSC software team 
using codes from the Large Synoptic
Survey Telescope (LSST) software pipeline~\citep{Ivezic+08,Axelrod+10}.
The photometric and astrometric calibrations are made based on
data obtained from the Panoramic Survey Telescope and
Rapid Response System (Pan-STARRS) 1 imaging 
survey~\citep{Magnier+13,Schlafly+12,Tonry+12}.

The photometric magnitude used in this work is a CModel magnitude.
The galactic reddening was corrected according to the dust maps derived
by \citet{Schlegel+98}.

The analysis performed in this paper is based on two galaxy samples;
one is the $i$-band detected sample which is drawn from all sources
selected by the criteria for $i$-band as described below and measured 
to be brighter than 27 magnitude in $i$-band regardless of the detection 
in the other four bands;
the other is the four-band detected sample which is selected by 
enforcing the same criteria for $griz$-bands 
except for the magnitude cut adapted only to $i$-band data,
regardless of the detection in $y$-band.
Since the observations in $y$-band are shallower than the others,
the detection in $y$-band was not required to avoid the bias to
redder galaxies.
For the cross-correlation analysis, we used the $i$-band detected sample.
When we measure the distribution of galaxy color and 
luminosity around AGNs,
we used the four-band detected sample.
The term ''detected'' used here means that the source 
satisfies the criteria defined below and has a non-NULL magnitude.
The criteria used to select $i$-band detected samples are:
\begin{verbatim}
   iflags_pixel_edge is not True
   AND iflags_pixel_interpolated_any is not True
   AND iflags_pixel_saturated_any is not True
   AND iflags_pixel_cr_any is not True
   AND iflags_pixel_bad is not True
   AND icmodel_flux_flags is not True
   AND icentroid_sdss_flags is not True
   AND detect_is_tract_inner is True
   AND detect_is_patch_inner is True
   AND deblend_nchild = 0
\end{verbatim}
where
\verb|iflags_pixel_edge| is true if the source is near the edge of 
the frame;
\verb|iflags_pixel_interpolated_any| is true if any pixels in the source
footprint have been interpolated due to saturation or cosmic rays;
\verb|iflags_pixel_saturated_any| is true if any pixels are saturated;
\verb|iflags_pixel_cr_any| is true if any pixels are masked as cosmic rays;
\verb|iflags_pixel_bad| is true if any pixels are masked for non-functioning
or in severely vignetted areas;
\verb|icmodel_flux_flags| is false if the cmodel measurement fails;
\verb|icentroid_sdss_flags| is false if the SDSS centroiding algorithm
fails;
\verb|detect_is_tract_inner| and \verb|detect_is_patch_inner| 
are true if the source is in the inner region of a tract and patch, which
is used to select a source of primary detection;
\verb|deblend_nchild| is the number of children this source was deblended
into.
Similar flag checks were performed for the other bands.

In addition to the above criteria, we adapted a criterion:
\begin{verbatim}
    flags_pixel_bright_object_center is not True
\end{verbatim}
to remove the galaxy samples
in bright source masks only for those located at $\ge$2~Mpc from the AGN.
The bright source masks implemented in this data release are over-conservative 
in the choice of radius, so it has a drawback that decreases the significance of 
the clustering.
Thus we did not adapt the bright source mask to the galaxy samples located at 
$<$2~Mpc from the AGN.
We found that there is a non-negligible number of false
detections in the deblended sources, especially at 
larger magnitudes.
To remove the false detections, we selected only the deblended sources which 
are brighter than 27 mag in $i$-band and also brighter than $m_{i,\rm top} + 6$ mag, 
where $m_{i,\rm top}$ is an $i$-band magnitude brightest in the deblended sources 
which belong to the same parent.
To avoid saturation in HSC photometry, no magnitude data brighter than 20 mag 
were used for any of the five bands.

\subsection{AGN dataset selection}
\label{sec:agn_selection}

In the analysis of this paper, we treated each AGN and its surrounding 
galaxies as a set.
Hereafter, we refer to the unit of the dataset as an AGN dataset.
In this section we describe the criteria to include the AGN datasets
for analysis.

For each AGN dataset, we measured the surface density of galaxies in annuli
spaced by 0.2~Mpc out to 10~Mpc from the position of AGN.
We kept only those AGN in which $>$60\% of the area of all annuli at $\ge$2~Mpc
around it and $>$80\% of the area at $<$2Mpc were included in the survey 
footprint and not masked by bright sources.
By this selection, among the original 6166 AGN datasets, 346 datasets were 
removed for 
the $i$-band detected sample and 585 were removed for 
the four-band detected sample.
Thus 5820 and 5581 AGN datasets passed this selection for $i$-band
and four-band detected samples, respectively.

The spatial uniformity of the galaxy samples in the AGN field was also
examined to identify the AGN datasets that are significantly
contaminated by nearby galaxies and stellar groups or showing non-uniformity
for any other reason.
For this purpose, we calculated 
two 
parameters for the radial 
number density distribution of galaxies, $\chi^{2}$ and
$\sigma_{\rm max}$, where $\chi^{2}$ is a square sum
of the deviation from the number density distribution fitted to the
observed data using equation~(\ref{eq:model}), which will be derived
in section~\ref{sec:cc_length},
$\sigma_{\rm max}$ is a maximum deviation from the density distribution.
Adapted criteria for those parameters are:
$\chi^{2}/n \le 3.0$ and $\sigma_{\rm max} \le 5$.
By this selection, 268 (236) AGN datasets were removed from the 5820 (5581)
AGN datasets for the $i$-band (four-band) detected sample.
Thus 5552 (5345) AGN datasets passed all the above selections.

\begin{figure}
  \begin{center}
  \includegraphics[width=0.8\textwidth]{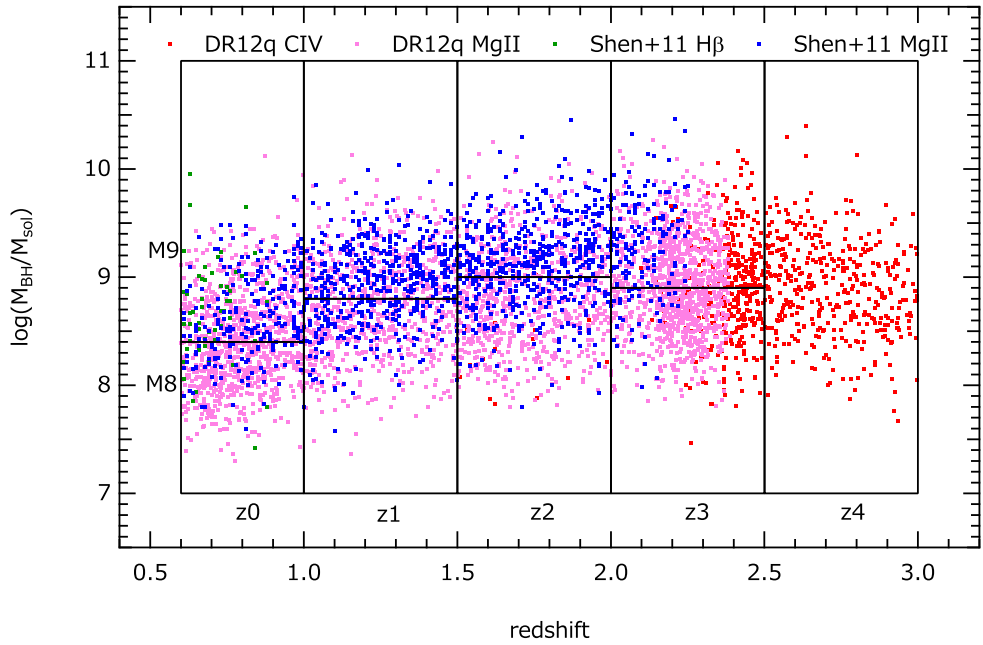} 
  \end{center}
  \caption{Distribution of redshift and BH mass of 5552 AGNs which are used
in cross-correlation analysis using $i$-band detected galaxy samples.}
  \label{fig:AGN_M-z}
\end{figure}

\begin{figure}
  \begin{center}
  \includegraphics[width=0.8\textwidth]{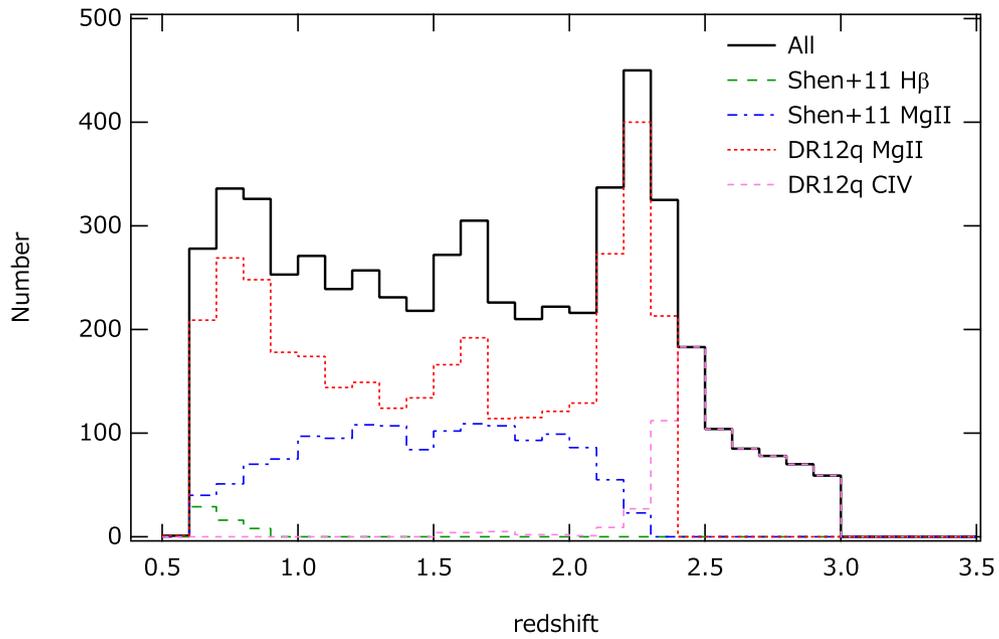} 
  \end{center}
  \caption{Distribution of BH mass for the samples shown in 
    Figure~\ref{fig:AGN_M-z}.}
  \label{fig:hist_AGN_M}
\end{figure}

\begin{figure}
  \begin{center}
  \includegraphics[width=0.8\textwidth]{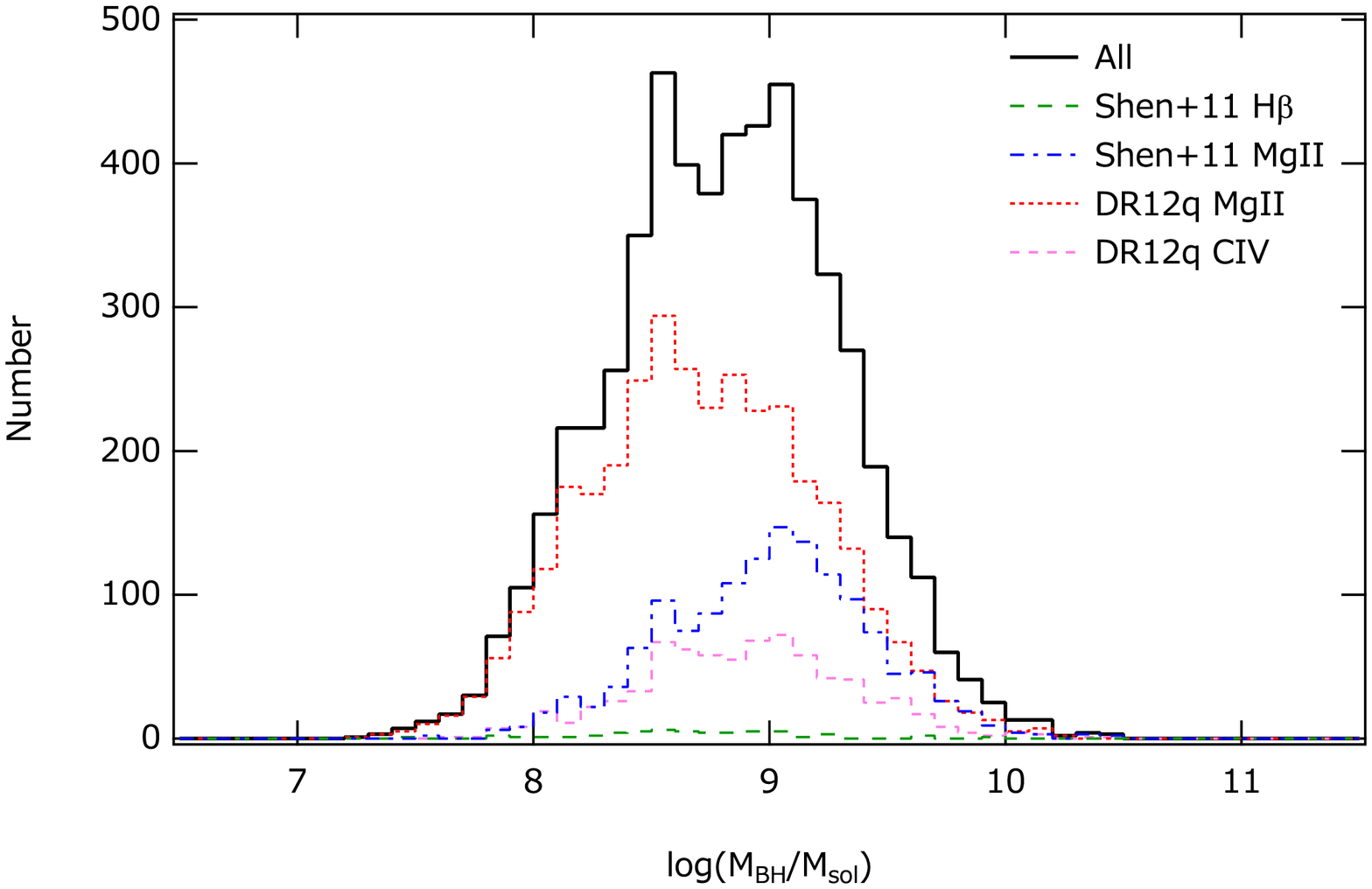} 
  \end{center}
  \caption{Distribution of redshift for the samples shown in
    Figure~\ref{fig:AGN_M-z}.}
  \label{fig:hist_AGN_z}
\end{figure}

Figure~\ref{fig:AGN_M-z} shows the mass vs redshift distribution of the AGNs 
that were selected according to the conditions described 
above and used in the cross-correlation analysis in this paper.
Figures~\ref{fig:hist_AGN_M} and \ref{fig:hist_AGN_z} show histograms of redshift
and black hole mass, respectively.
We divided the redshift range into five groups as shown in Figure~\ref{fig:AGN_M-z},
which we call z0, z1, z2, z3 and z4 for $z =$ 0.6--1.0, 1.0--1.5,
1.5--2.0, 2.0--2.5, and 2.5--3.0, respectively.
For each redshift group, except for z4, the BH mass range was divided into two 
groups M8 and M9 so that each group has a similar number of AGNs rather than 
being divided at the same mass.
This is because the effect of sample variance becomes dominant when the 
number of AGN samples is small and it is difficult to homogenize the samples
among different mass groups by the division at constant mass for all the redshifts.
The mass group was divided at $\log{(M_{\rm BH}/M_{\solar})}$ = 8.4, 8.8, 9.0, 
and 8.9 for z0, z1, z2 and z4 redshift groups, respectively, which allows us
to make the statistical uncertainties even for both mass groups.
The mass dependence for the z4 redshift group was not examined, since the number 
of samples in the z4 group is too small to do so.


Table~\ref{tab:samples} shows the number of AGN datasets which passed all the 
selection criteria described above for each dataset.
The $i$-band detected galaxy samples were used for cross-correlation analysis,
and the four-band detected samples were used for deriving color distribution
and luminosity function of clustering galaxies, and they were also used
to measure the dependence of cross-correlation length on galaxy luminosity.
The redshift-matched samples were used for examining the BH mass 
dependence at each redshift.
\begin{table}
  \tbl{Number of AGN samples used in the analysis for each dataset.}{%
  \begin{tabular}{crrrrrr}
      \hline
            & \multicolumn{3}{c}{$i$-band detected} 
            & \multicolumn{3}{c}{four-band detected} \\
            & \multicolumn{1}{c}{all} 
            & \multicolumn{2}{c}{z-matched$^{a}$} 
            & \multicolumn{1}{c}{all} 
            & \multicolumn{2}{c}{z-matched} \\
 redshift group & M8+M9 & M8 & M9 & M8+M9 & M8 & M9 \\ 
      \hline
      z0$^{b}$    & 1194 & 482 & 482   & 1181 & 491 & 491 \\
      z1$^{b}$    & 1216 & 506 & 506   & 1212 & 500 & 500 \\
      z2$^{b}$    & 1235 & 534 & 534   & 1204 & 508 & 508 \\
      z2$^{c}$    &      &     &       & 1202 & 503 & 503 \\
      z3$^{c}$    & 1511 & 640 & 640   & 1385 & 581 & 581 \\
      z4$^{c}$    &  396 &     &       &  363 &     &     \\
      \hline
    \end{tabular}}\label{tab:samples}
\begin{tabnote}
$^{a}$ redshift-matched samples.
$^{b}$ $M_{\lambda 310}$ based samples for the four-band detected samples.
$^{c}$ $M_{\lambda 220}$ based samples for the four-band detected samples.
\end{tabnote}
\end{table}

\section{Analysis method}\label{sec:analysis}

\subsection{Cross-correlation length}
\label{sec:cc_length}

The cross-correlation function between AGNs and galaxies was 
calculated using the method described in our previous
papers~\citep{Shirasaki+11,Komiya+13,Shirasaki+16}.
The analysis method is briefly described here.

We assumed the power-law form of the cross-correlation function,
\begin{equation}
   \xi(r) = \left( \frac{r}{r_{0}} \right)^{-\gamma},
   \label{eq:cc_func}
\end{equation}
where $r_{0}$ is a cross-correlation length and $\gamma$ is a
power-law index fixed to 1.8, which is a typical value found
in previous studies on AGN-galaxy cross-correlation 
studies~\citep[e.g.,][]{Hickox+09,Coil+09,Krumpe+15}.
The projected cross-correlation function $\omega(r_{p})$ is expressed
in an analytical form as:
\begin{equation}
   \omega(r_{p}) 
      = r_{p} \left( \frac{r_{0}}{r_{p}} \right)^{\gamma} 
      \frac{\Gamma(\frac{1}{2})\Gamma(\frac{\gamma-1}{2})}{\Gamma(\frac{\gamma}{2})},
   \label{eq:omega_1}
\end{equation}
where $r_{p}$ is a projected distance from AGN and $\Gamma$ is the 
Gamma function.
$\omega(r_{p})$ is related to the observed average surface number density of 
galaxies $n(r_{p})$ as:
\begin{equation}
   \omega(r_{p}) = \frac{n(r_{p}) - n_{\rm bg}}{\bar{\rho_{0}}},
   \label{eq:omega_2}
\end{equation}
where $n_{\rm bg}$ is an average surface number density
of
background galaxies 
integrated along the line of sight, and $\bar{\rho_{0}}$ is an average
of the number density of galaxies at AGN redshifts.
From equations~(\ref{eq:omega_1}) and (\ref{eq:omega_2}), the observed 
average surface 
number
density 
of galaxies
around AGNs can be expressed as:
\begin{equation}
   n(r_{p}) = C(\gamma) \bar{\rho_{0}} r_{p} \left( \frac{r_{0}}{r_{p}} \right) ^{\gamma} 
            + n_{\rm bg},
   \label{eq:model}
\end{equation}
where $C(\gamma) = \Gamma(\frac{1}{2}) \Gamma(\frac{\gamma-1}{2}) / \Gamma(\frac{\gamma}{2})$.
We fitted the model function of equation~(\ref{eq:model}) to the surface 
number
density derived by averaging over a group of 
AGN 
datasets, 
then obtained best fit parameters
for the cross-correlation length $r_{0}$ and the background density $n_{\rm bg}$.

In calculating $n(r_{p})$, we masked the central 10'' around each 
AGN to avoid blending problems with the AGN.
Around a bright source there is a region where the number density of galaxy 
selected by the criteria described in section~\ref{sec:galaxy} is significantly 
reduced due to blending with the bright source and/or increase in the
background noise level.
That region needs to be counted as a dead 
region
in calculating
an effective area
around AGNs.

To estimate the dead region, we adapted a bright source flag to 
remove the galaxies
near bright sources, and the area of the masked region were calculated using the
random catalog which was created avoiding the masked region with number density 
100 arcmin$^{-2}$.
The random catalog we used is the one included in the S15b 
database.
The bright source mask was applied only for those which are located $\ge$2~Mpc
from the AGN as explained in section~\ref{sec:galaxy}.
To estimate the dead region at area of $<$2~Mpc, we calculated the correction
factor from the average of the ratio of number density derived using masked 
($n_{\rm masked}(r)$) and un-masked ($n_{\rm unmasked}(r)$) galaxy samples as follows:
\begin{equation}
   c(r) =  \frac{n_{\rm unmasked}(r)}{n_{\rm masked}(r)}
\end{equation}
\begin{equation}
   n_{\rm corrected}(r<{\rm 2Mpc}) = 
           \frac{n_{\rm unmasked}(r<{\rm 2Mpc})}{\langle c(r) \rangle_{\rm r=2-10Mpc}}
\end{equation}

To estimate $\bar{\rho_{0}}$ in equation~(\ref{eq:model}),
$\rho_{0}$ for each AGN dataset was estimated from the luminosity 
function which 
was
derived by parametrizing the luminosity
functions in the literature.
The luminosity function is expressed with the Schechter 
function~\citep{Schechter+76}:
\begin{equation}
   \phi(M) = 0.4 \times \ln{10}~\phi_{*} 
     10^{-0.4 (\rm M-M_{*})(\alpha + 1)} \exp{\{-10^{-0.4(\rm M-M_{*})}\}}.
   \label{eq:schechter}
\end{equation}
The parameters $\phi_{*}$, $M_{*}$, and $\alpha$ were parametrized as a 
function of redshift $z$ at rest-frame wavelengths 150~nm, 280~nm, 
SDSS $u'$, $g'$, and $r'$ band so that they approximated the parameters derived 
in literature~\citep{Gabasch+04,Gabasch+06,Dahlen+05,Dahlen+07,Parsa+16}
for redshift $z$ = 0.5--3.0.
In total 40 sets of luminosity function parameters from the literature
were used to determine the parametrization.
Then they were interpolated as a function of wavelength.

\begin{figure}
  \begin{center}
     \includegraphics[width=0.7\textwidth]{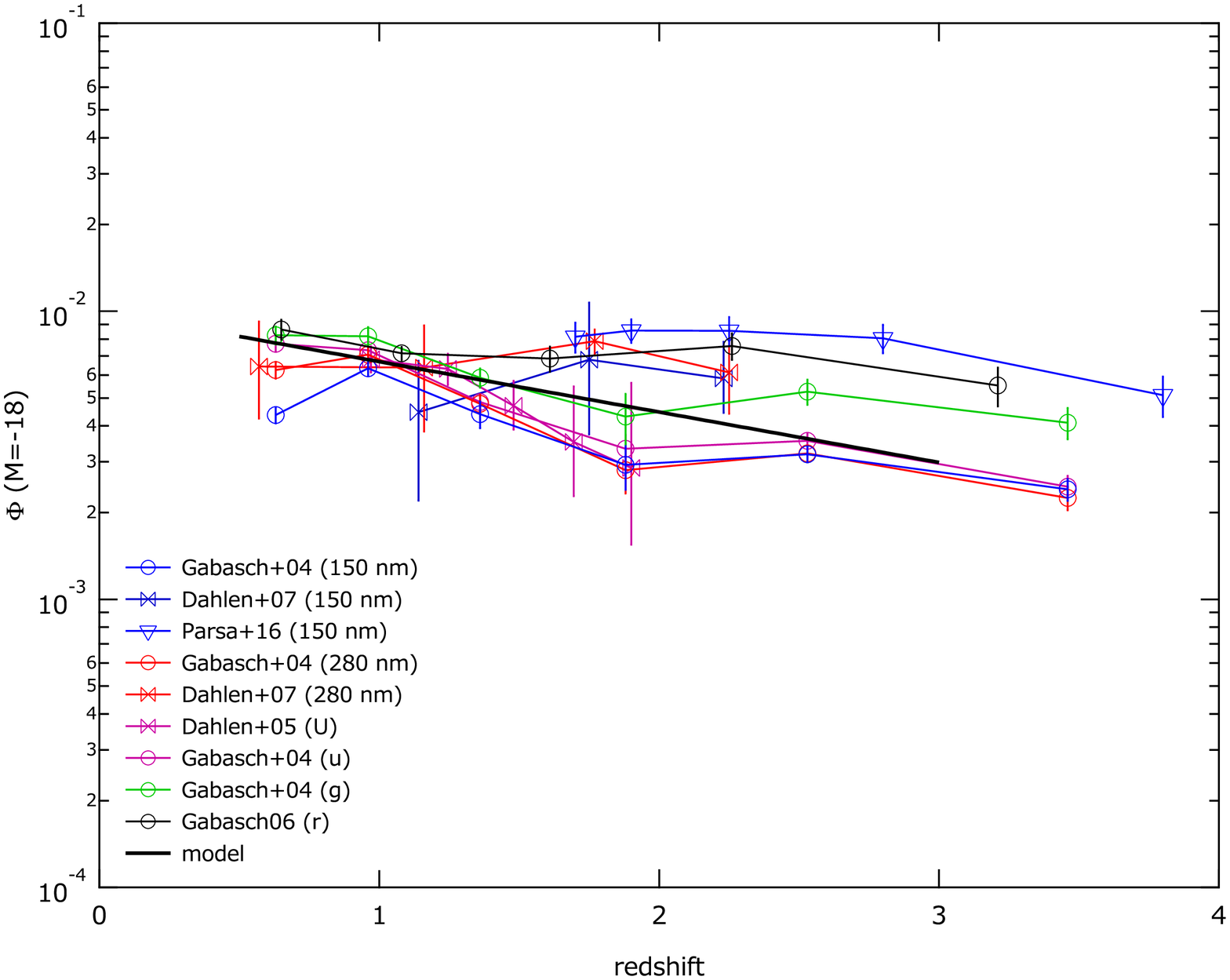} 
  \end{center}
  \caption{$\phi_{-18}$ parameters of equation~(\ref{eq:phi18})
      derived from literature
      \citep{Gabasch+04,Gabasch+06,Dahlen+05,Dahlen+07,Parsa+16}.
      Our parametrization is shown with a thick line.}
  \label{fig:Phi18}
\end{figure}

\begin{figure}
  \begin{center}
     \includegraphics[width=0.7\textwidth]{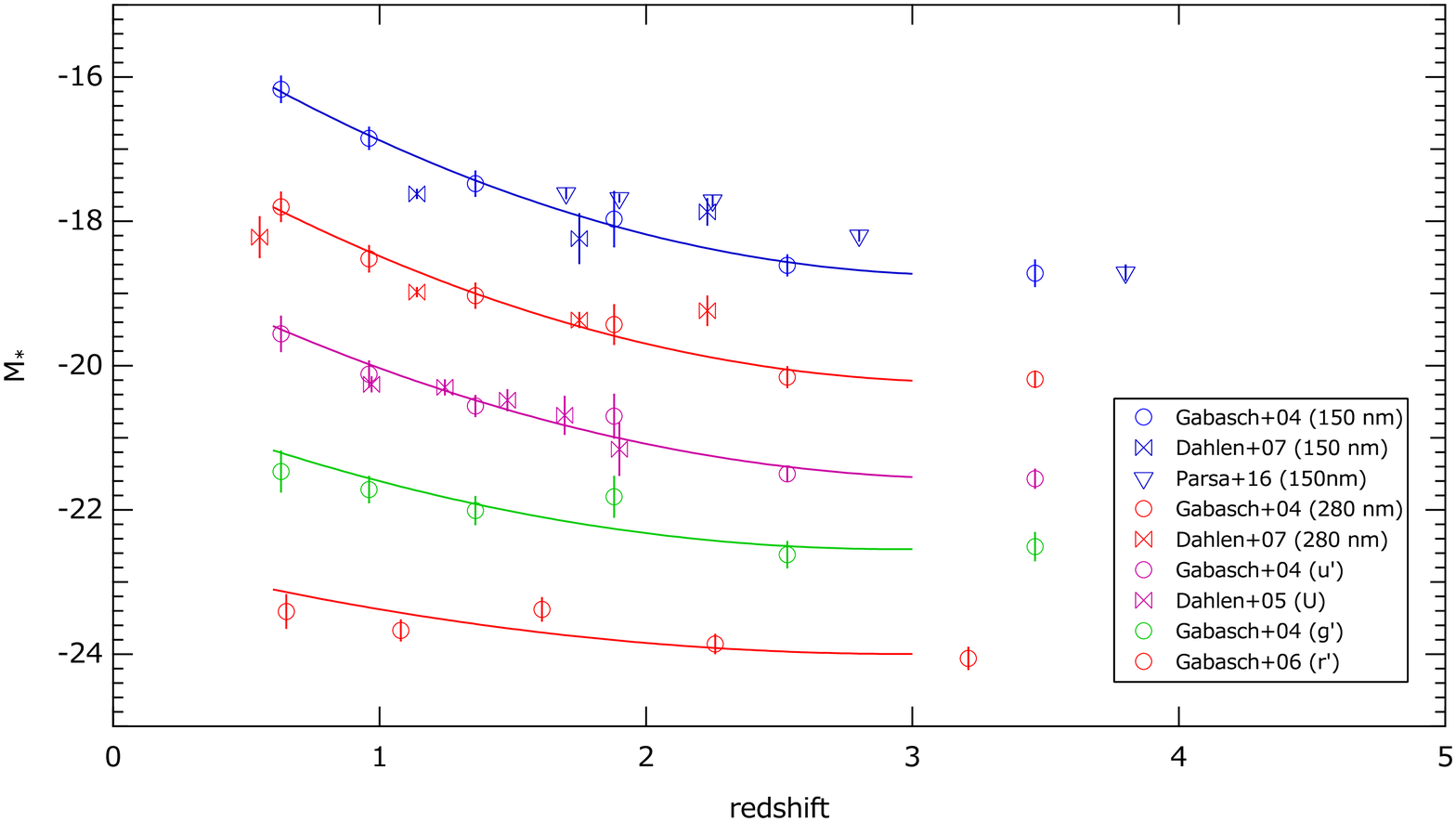} 
  \end{center}
  \caption{$M_{*}$ parameter of equation~(\ref{eq:Mstar})
      derived from literature 
      \citep{Gabasch+04,Gabasch+06,Dahlen+05,Dahlen+07,Parsa+16}.
      Our parametrization is shown with solid lines.}
  \label{fig:Mstar}
\end{figure}

\begin{figure}
  \begin{center}
     \includegraphics[width=0.55\textwidth]{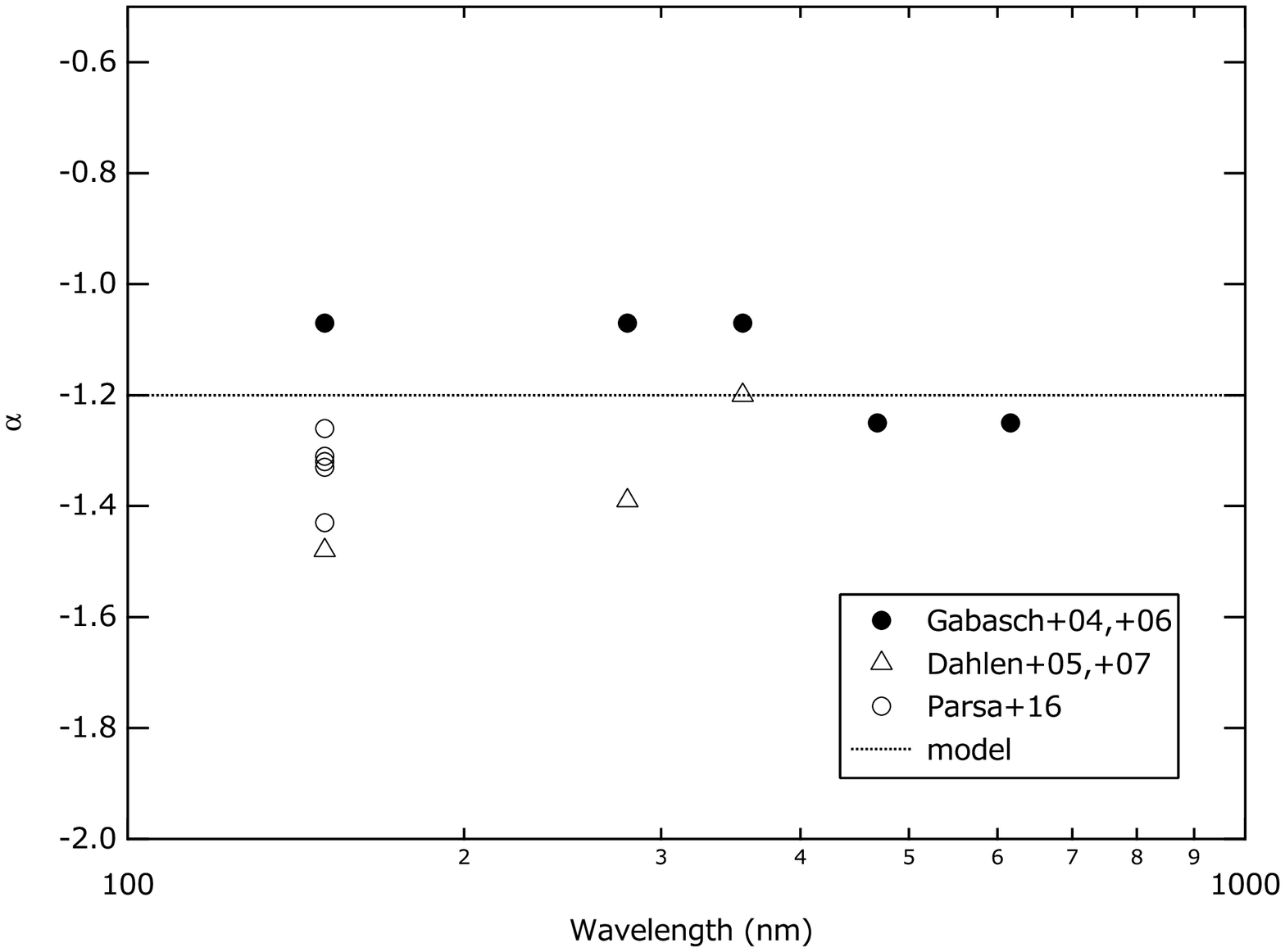} 
  \end{center}
  \caption{$\alpha$ parameter of equation~(\ref{eq:alpha})
      derived from literature
      \citep{Gabasch+04,Gabasch+06,Dahlen+05,Dahlen+07,Parsa+16}.
      Our parametrization is shown with a dotted line.}
  \label{fig:alpha}
\end{figure}

The redshift parametrization was performed with the following functions:
\begin{equation}
   \phi_{*} = \frac{\phi_{-18}}
              { 0.4 \times \ln{10} \times
                10^{0.4 (\rm 18+M_{*})(\alpha + 1)} \exp{\{-10^{0.4(\rm 18+M_{*})}\}} },
   \label{eq:phistar}
\end{equation}
\begin{equation}
   \phi_{-18}  = 10^{(a_{0} + a_{1} z)},
   \label{eq:phi18}
\end{equation}
\begin{equation}
   M_{*} = b_{0} + b_{1} z + b_{2} z^{2},
   \label{eq:Mstar}
\end{equation}
\begin{equation}
   \alpha = c_{0} + c_{1} z,
   \label{eq:alpha}
\end{equation}
where $\phi_{-18}$ represents luminosity density at $M=-18$.
We used $\phi_{-18}$ for the parametrization instead of using $\phi_{*}$,
because $\phi_{*}$ correlates with $M_{*}$ parameter and is strongly
affected by the uncertainty of $M_{*}$.
$\phi_{-18}$, on the other hand, correlates with $M_{*}$ more weakly than
$\phi_{*}$, and its dependence on redshift and wavelength is rather small.
The coefficients of equations~(\ref{eq:phi18})--(\ref{eq:alpha}) used in
this analysis are summarized in Table~\ref{tab:lf_coeff}.
The comparison of these parametrization with the parameters in literature
is shown in Figures~\ref{fig:Phi18}--\ref{fig:alpha}.

The root mean square (RMS) of the difference between parameters from the literature
and those calculated by parametrization of equations~(\ref{eq:phistar})--(\ref{eq:alpha}) 
are: 0.15, 0.28~mag, and 0.14 for $\log{(\phi_{-18})}$, $M_{*}$, and $\alpha$, 
respectively.
The error of $\log{(\phi_{-18})}$ by 0.15 corresponds to a systematic error of
$r_{0}$ by $\sim$20\% for $\gamma = 1.8$.

The error of $M_{*}$ significantly affects the luminosity 
densities at the bright end of $M<M_{*}$, as the 
slope
of the luminosity function 
becomes steeper there.
Thus, comparison was made between the number densities at $M<M_{*}$ 
calculated using the parameters from the literature $\rho_{\rm lit}$
and those calculated with the above parametrization $\rho_{\rm par}$.
According to the comparison, RMS of the relative error
$(\rho_{\rm lit}-\rho_{\rm par})/\rho_{\rm par}$ is
0.36 for 40 sets of luminosity functions of the literature, which
corresponds to an upper deviation of cross-correlation length of 
$\delta r_{0} = 0.28 \times r_{0}$, that is 28\% of the cross correlation
length, and a lower deviation of $\delta r_{0} = 0.16 \times r_{0}$, i.e. 16\%,
for $\gamma = 1.8$.
When the comparison is made by extending to a lower luminosity side as $M>-17$,
the RMS of the relative error is 0.16 for 20 sets of luminosity functions
of $z<1.5$, which corresponds to upper and lower deviation of 10\% and 8\%
for $r_{0}$, respectively.
\begin{table}
  \tbl{The coefficients of equations~(\ref{eq:phi18})--(\ref{eq:alpha}) 
used to calculate luminosity function in this analysis.}{%
    \begin{tabular}{cccc}
\hline
parameter      & unit    & wavelength band & coefficients \\
\hline
$\phi_{-18}$ & mag$^{-1}$Mpc$^{-3}$ & any & $a_{0} = -2.0$,  $a_{1} = -0.175$ \\
$M_{*}$      & mag & 150~nm & $b_{0}=-16.82$, $b_{1}=-2.437$, $b_{2}=0.378$ \\
$M_{*}$      & mag & 280~nm & $b_{0}=-17.57$, $b_{1}=-2.265$, $b_{2}=0.351$ \\
$M_{*}$      & mag & $u'$     & $b_{0}=-18.40$, $b_{1}=-1.932$, $b_{2}=0.294$ \\
$M_{*}$      & mag & $g'$     & $b_{0}=-20.38$, $b_{1}=-1.470$, $b_{2}=0.250$ \\
$M_{*}$      & mag & $r'$     & $b_{0}=-21.60$, $b_{1}=-0.936$, $b_{2}=0.157$ \\
$\alpha$     &     & any    & $c_{0}=-1.2$, $c_{1}=0.0$  \\
\hline    
    \end{tabular}}
  \label{tab:lf_coeff}
  \begin{tabnote}
  \end{tabnote}
\end{table}

Then $\rho_{0}$ is calculated as an integral of the product of the 
luminosity function and a detection efficiency DE($m$) for apparent
magnitude $m$.
The detection efficiency was modeled with the following function:
\begin{equation}
   {\rm DE}(m) = \left\{
   \begin{array}{ll}
     1  & ( m < m_{\rm th} )  \\
     \exp{( -(m-m_{\rm th})^2/\sigma_{m}^{2} ) }  & ( m \ge m_{\rm th}), \\
     \exp{( -(m-m_{\rm th})^2/\sigma_{m}^{2} ) } \times
     \exp{( -(m-m_{\rm th2})^2/\sigma_{m2}^{2} ) }
      & ( m \ge m_{\rm th2} \ge m_{\rm th} ), \\
   \end{array}
   \right.
   \label{eq:DE}
\end{equation}
where 
$m_{\rm th}$ and $m_{\rm th2}$ represent the threshold magnitudes for a decrease 
in the detection efficiency, $\sigma_{m}$ and $\sigma_{m2}$ represent the 
attenuation widths.
Two attenuation functions which have different threshold magnitudes and
attenuation widths were required to fit the data accurately.

These parameters were obtained for each AGN dataset by fitting a
model function to the observed magnitude distribution, $N_{\rm obs}(m)$
at $m =$ 21--27.
We assumed a power law form for the model function of the magnitude distribution 
which is expected for an ideal observation of 100\% detection efficiency at
any magnitude: 
\begin{eqnarray}
   N_{\rm org}(m)  =  c \times 10^{a (m-23)},
   \label{eq:mag_dist_org}
\end{eqnarray}
Then the function $N_{\rm org}(m) {\rm DE}(m)$ is fitted to the observed
magnitude distribution $N_{\rm obs}(m)$.
The residuals between them are within the statistical error.

Using the ${\rm DE}(m)$ obtained from the fitting, $\rho_{0}$ is
calculated as:
\begin{equation}
   \rho_{0} = \int_{m_{\rm low} - {DM}}^{m_{\rm upp}-{DM}} 
        \phi(M) {\rm DE}(M+{DM}) dM,
   \label{eq:rho0}
\end{equation}
where $DM$ represents the distance modulus for the AGN redshift, and
$m_{\rm low}$ and $m_{\rm upp}$ represent the magnitude range 
of the galaxy samples.

\subsection{Color and absolute magnitude distributions for galaxies}
\label{sec:ana_color}

The color ($D$) and absolute 
magnitude ($M$) distributions for 
galaxies around AGNs were
derived by subtracting the distribution in an offset region ($n_{\rm off}$)
from that at an central region of the AGN field ($n_{\rm cen}$) as follows
\citep{Shirasaki+16}:
\begin{equation}
   n(D) = 
   n_{\rm cen}(D) - 
   n_{\rm off}(D)
   \label{eq:n_D}
\end{equation}
\begin{equation}
   n(M) = 
   n_{\rm cen}(M) - 
   n_{\rm off}(M)
   \label{eq:n_M}
\end{equation}
The offset region in this work is defined as an annular region
at a projected distance from 7 to 9.8~Mpc from the AGN, and the
central region is from 0.2 to 2~Mpc.

To calculate the color and absolute magnitude for the same rest frame 
bandpass at different redshifts, we performed SED fitting using
the EAZY software developed by \citet{Brammer+08} and calculated
color and magnitude at fixed bandpasses.
For the redshift z0, z1, and z2 groups ($z=$0.6--2.0), 
the color is defined as $D_{1} = M_{\lambda 270} - M_{\lambda 380}$ and the absolute 
magnitude is $M_{\lambda 310}$, where $M_{\lambda 270}$, $M_{\lambda 380}$ and $M_{\lambda 310}$ represent 
absolute magnitudes at wavelengths 270, 380, and 310~nm, 
respectively.
For the redshift z2, z3, and z4 groups ($z=$1.5--3.0), the color is
defined as $D_{2} = M_{\lambda 165} - M_{\lambda 270}$ and the absolute magnitude is
$M_{\lambda 220}$, where $M_{\lambda 165}$, $M_{\lambda 270}$, and $M_{\lambda 220}$ represent 
absolute magnitudes at wavelengths 165, 270, and 210~nm, 
respectively.
For redshift z2 group, we calculated the distributions for both
definitions for comparison between them.

Although EAZY can be used to estimate photometric redshift (photo-z), it was
used here to interpolate the observed SEDs of all the sources assuming 
its redshift to be the same as the AGN redshift.
This is because the photo-z is not well constrained at
redshifts explored in this work, as the structure around 400~nm,
which is primarily used to determine the photo-z, is out of the
observed passband.
For this reason, photo-z wasn't used in our analysis and the clustering
of galaxies was measured by stacking the galaxy distribution around AGNs 
and subtracting the DC offset, which is a contribution
from foreground/background galaxies, from the distribution.

\section{Results}

\subsection{Cross-correlation function}

First we present the cross-correlation functions 
derived by using the $i$-band detected 
galaxy samples for each redshift group.
To test also for the dependence on galaxy luminosity,
two galaxy samples were constructed with different threshold 
magnitudes.
The threshold magnitudes were chosen to be $M_{90\%}$ and $M_{50\%}$,
where $M_{90\%}$ ($M_{50\%}$) represents the absolute
magnitude where averaged detection efficiency estimated
as described in section~\ref{sec:cc_length} 
is 90\% (50\%).
The absolute magnitude is measured in $i$-band observer frame
as calculated with $m_{i} - DM$, where $m_{i}$ is $i$-band
apparent magnitude and $DM$ is distance modulus for the AGN redshift.
The values of $M_{50\%}$ and $M_{90\%}$ are summarized in the
second and third columns of Table~\ref{tab:M50_M90}.

\begin{table}
  \tbl{Absolute magnitude threshold $M_{50\%}$$^{a}$ and $M_{90\%}$$^{b}$
for each magnitude definition and each redshift group.}{%
  \begin{tabular}{crrrrrr}
      \hline
   & \multicolumn{2}{c}{$m_{i}-DM$ } 
   & \multicolumn{2}{c}{$M_{\lambda 310}$}
   & \multicolumn{2}{c}{$M_{\lambda 220}$} \\
redshift
   &  $M_{50\%}$   & $M_{90\%}$    &  $M_{50\%}$   &  $M_{90\%}$   &  $M_{50\%}$   &  $M_{90\%}$    \\
   &  mag   & mag    &  mag   &  mag   &  mag   &  mag    \\
      \hline
z0 & -17.61 & -18.40 & -17.82 & -18.80 &        &         \\
z1 & -18.77 & -19.49 & -19.12 & -20.01 &        &         \\
z2 & -19.65 & -20.34 & -20.17 & -21.06 & -19.97 & -20.85  \\
z3 & -20.40 & -21.09 &        &        & -20.74 & -21.55  \\
z4 & -20.93 & -21.62 &        &        & -21.36 & -22.20  \\
      \hline
  \end{tabular}}
  \label{tab:M50_M90}
  \begin{tabnote}
$^{a}$ absolute magnitudes where the detection efficiency is 50\% for a given redshift group,
$^{b}$ absolute magnitudes where the detection efficiency is 90\% for a given redshift group.
  \end{tabnote}
\end{table}

\begin{figure}
  \begin{center}
  \includegraphics[width=0.4\textwidth]{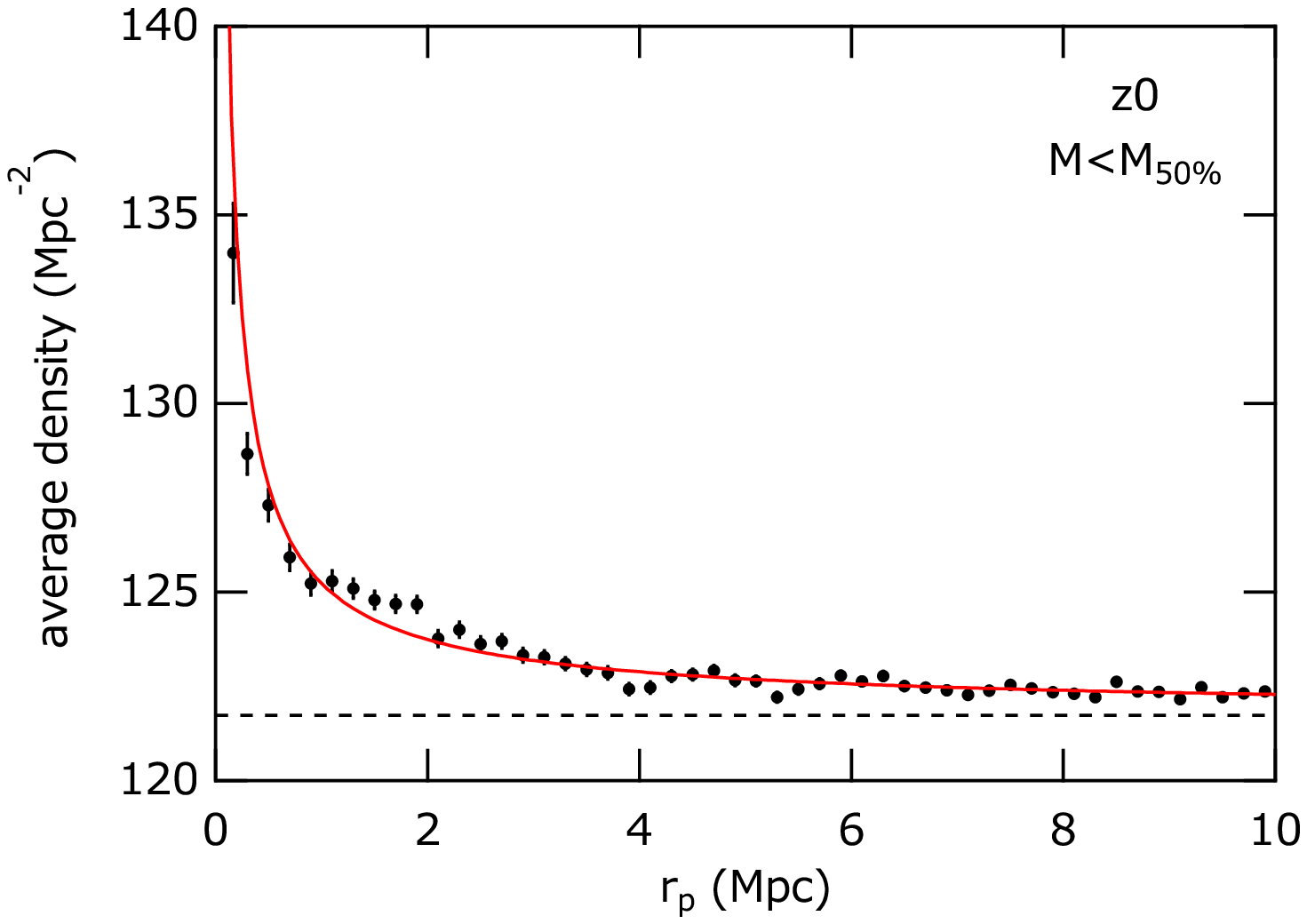} 
  \includegraphics[width=0.4\textwidth]{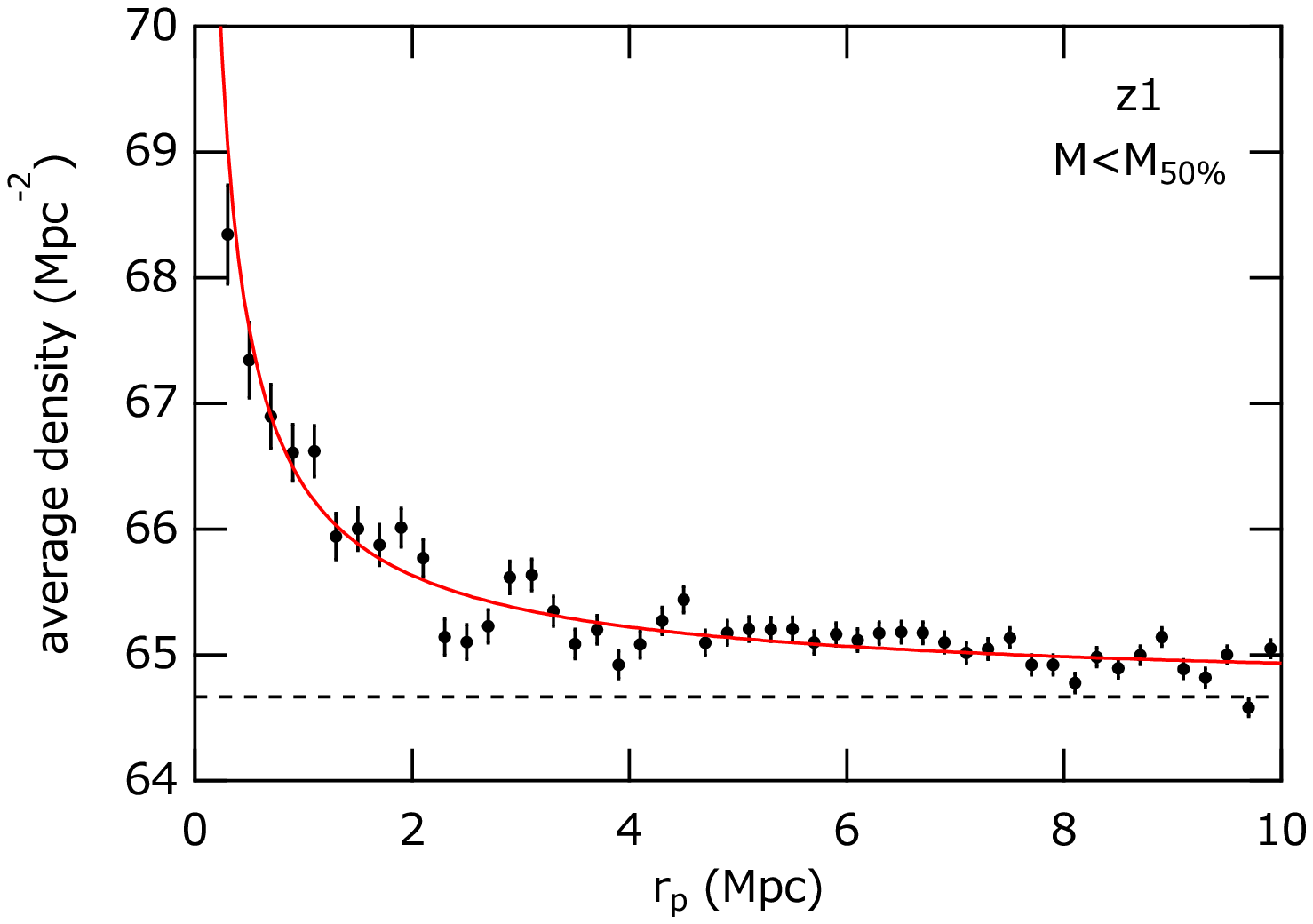} 
  \includegraphics[width=0.4\textwidth]{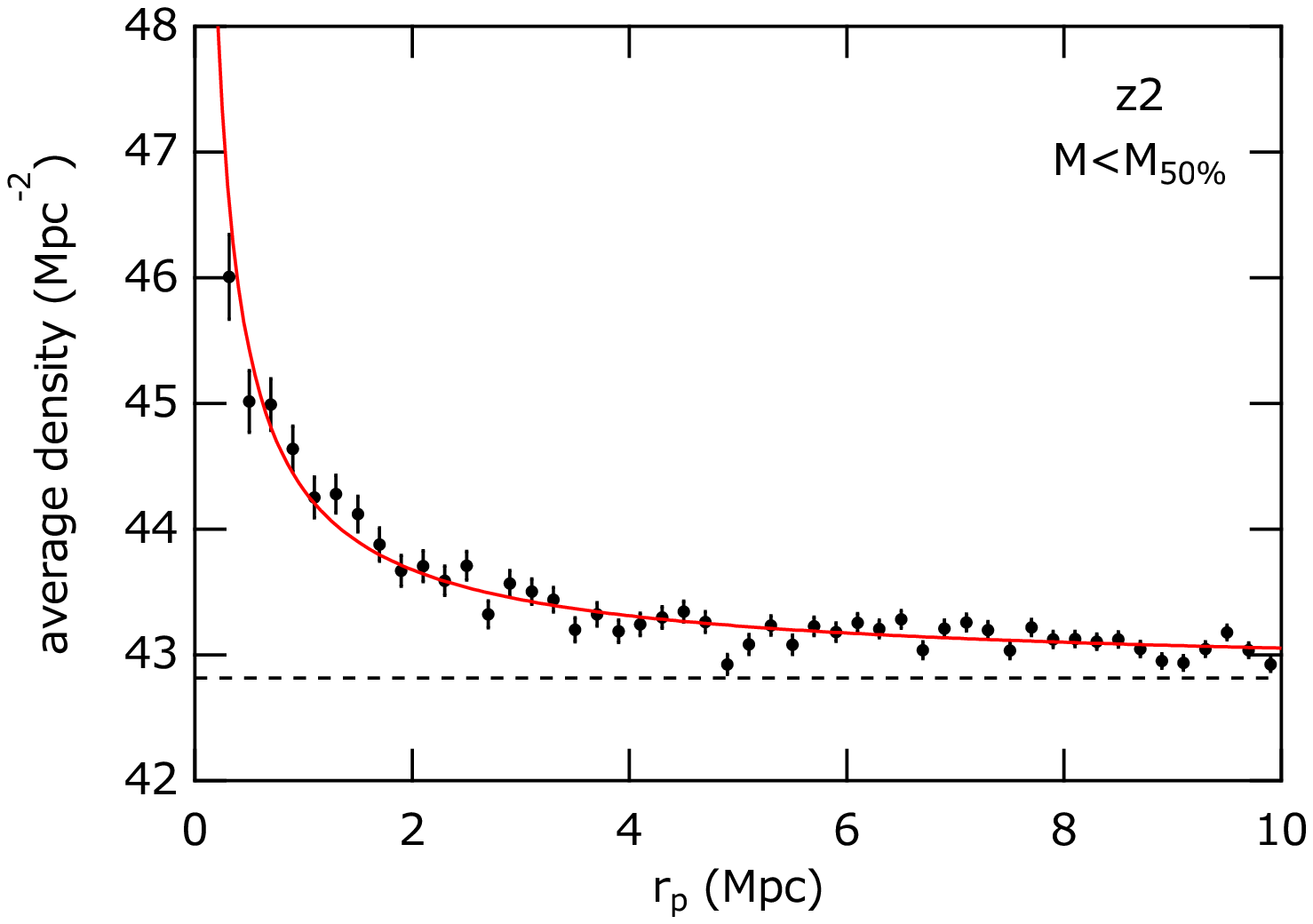} 
  \includegraphics[width=0.4\textwidth]{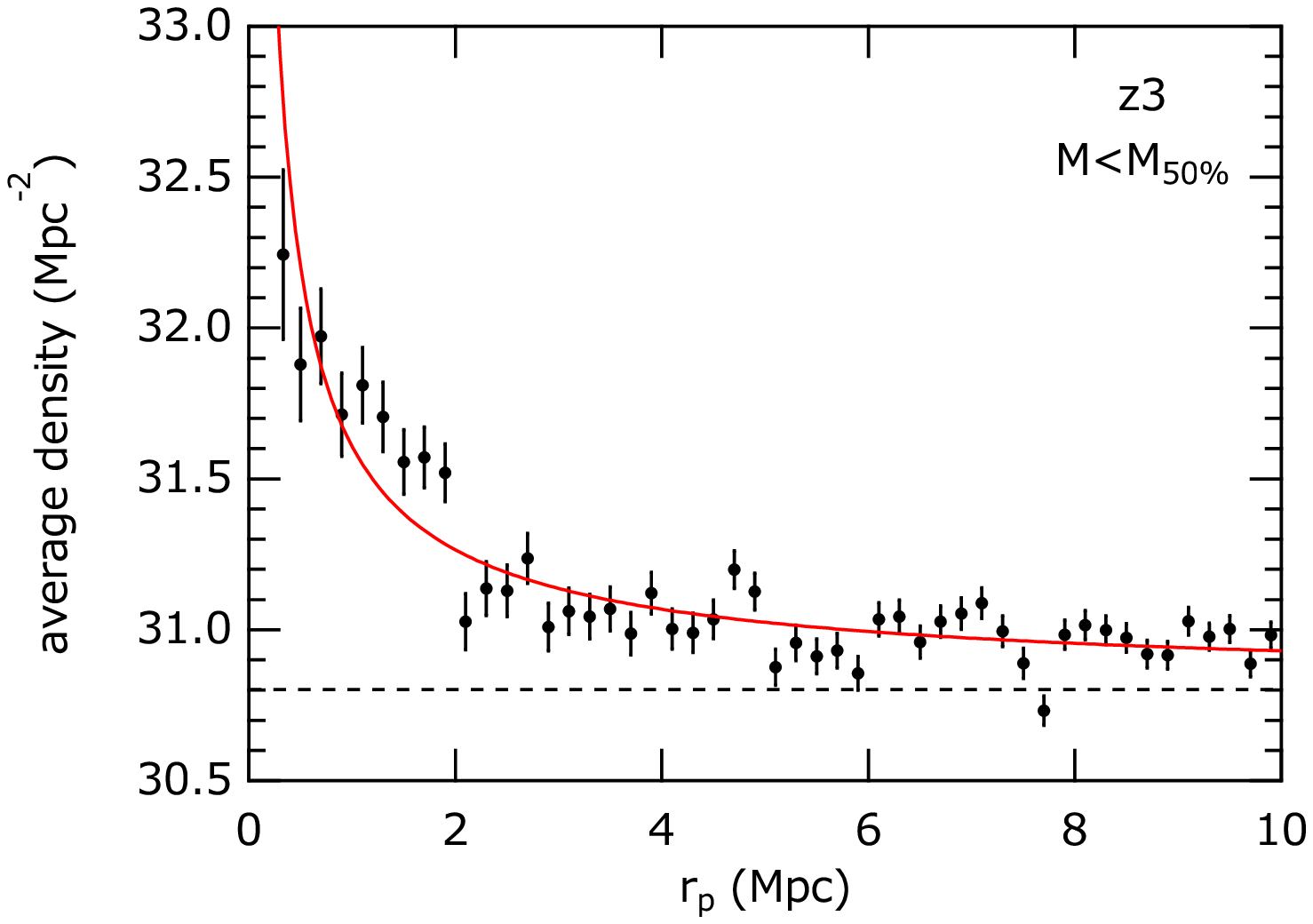} 
  \includegraphics[width=0.4\textwidth]{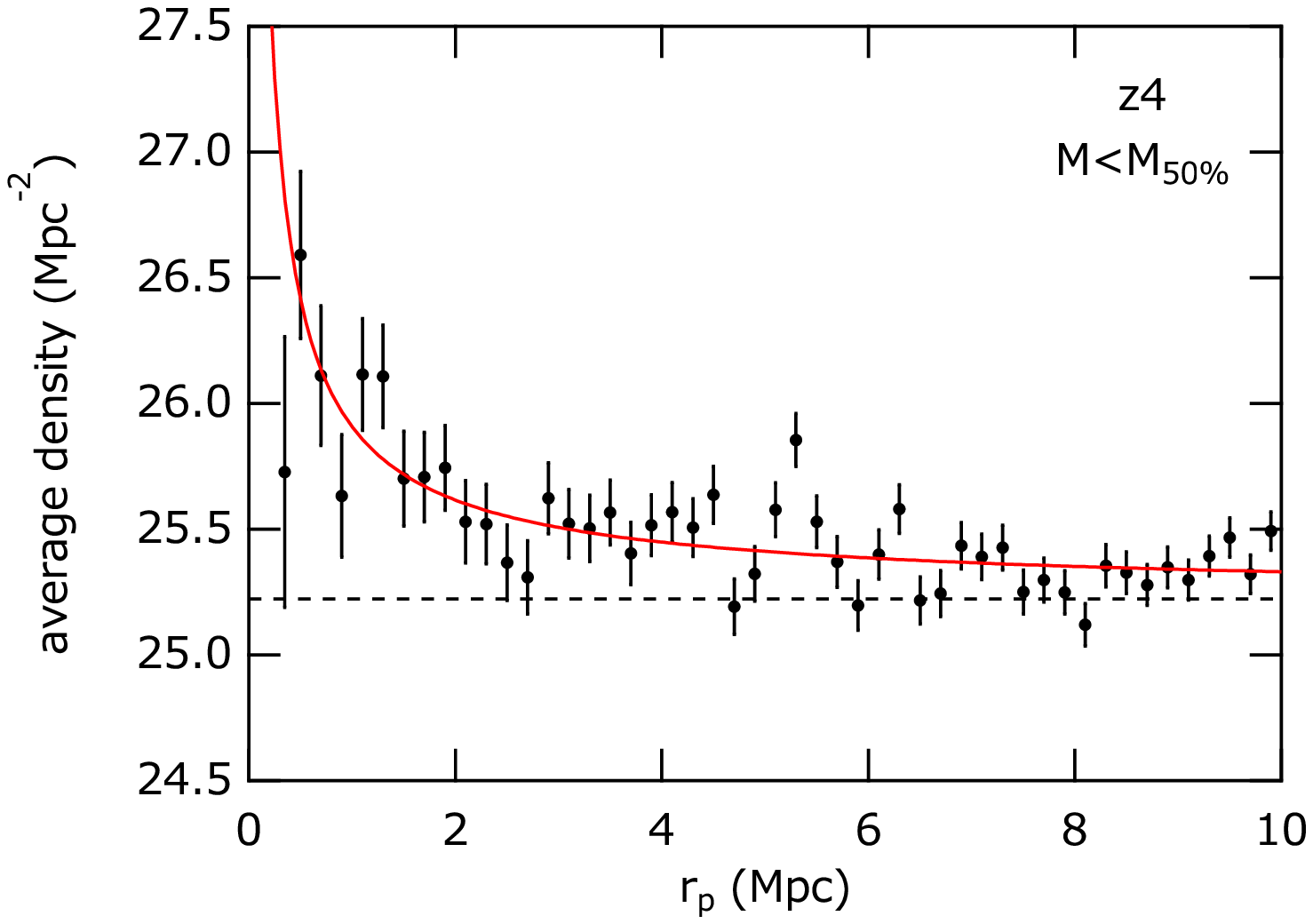} 
  \end{center}
  \caption{Radial distribution of galaxy surface density around AGNs
  for each redshift group. $i$-band detected galaxy samples
  with absolute magnitude brighter than $M_{50\%}$ are used, where
  $M_{50\%}$ represents the magnitude where detection efficiency 
  is 50\%}
  \label{fig:density_1}
\end{figure}

\begin{figure}
  \begin{center}
    \includegraphics[width=0.32\textwidth]{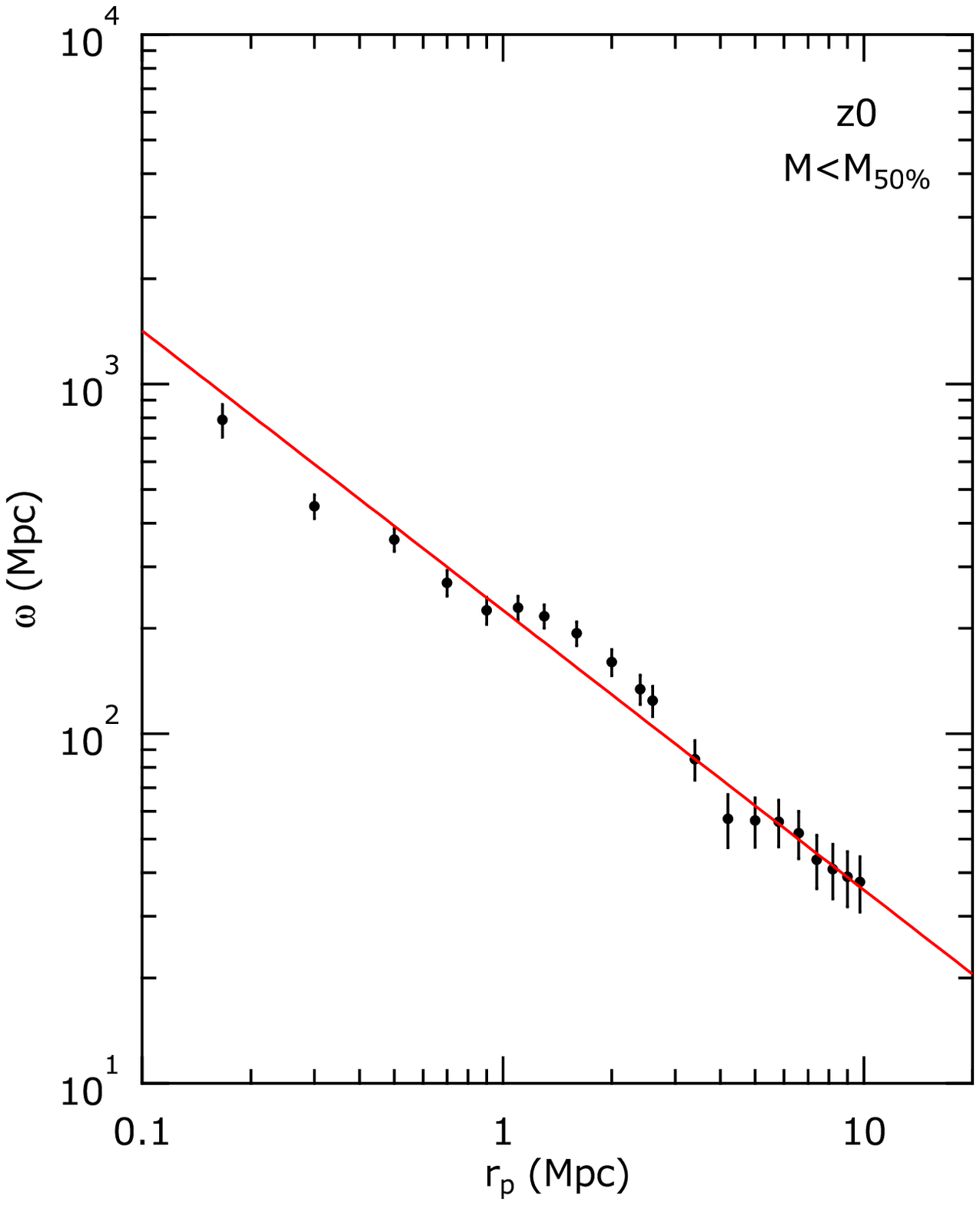} 
    \includegraphics[width=0.32\textwidth]{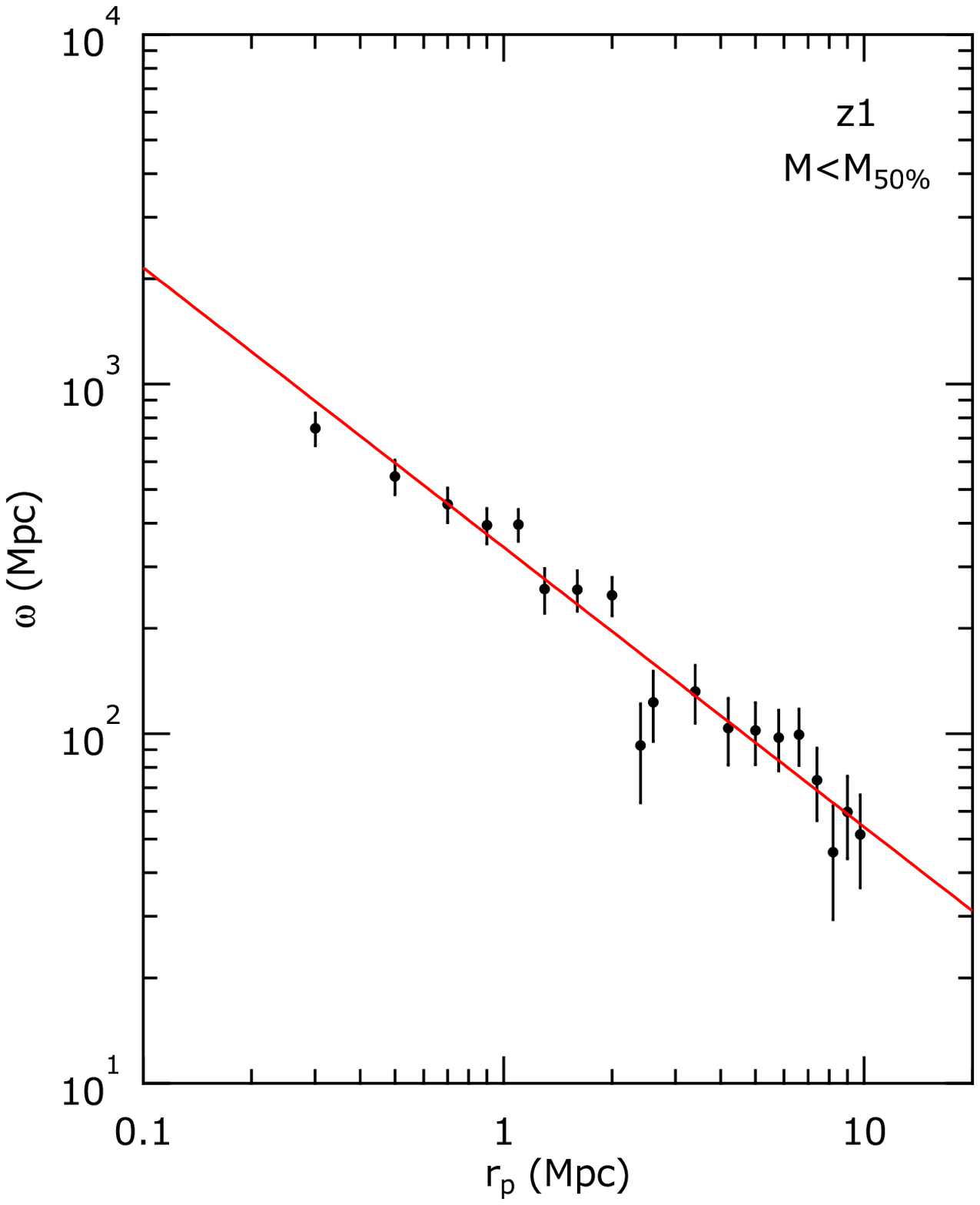} 
    \includegraphics[width=0.32\textwidth]{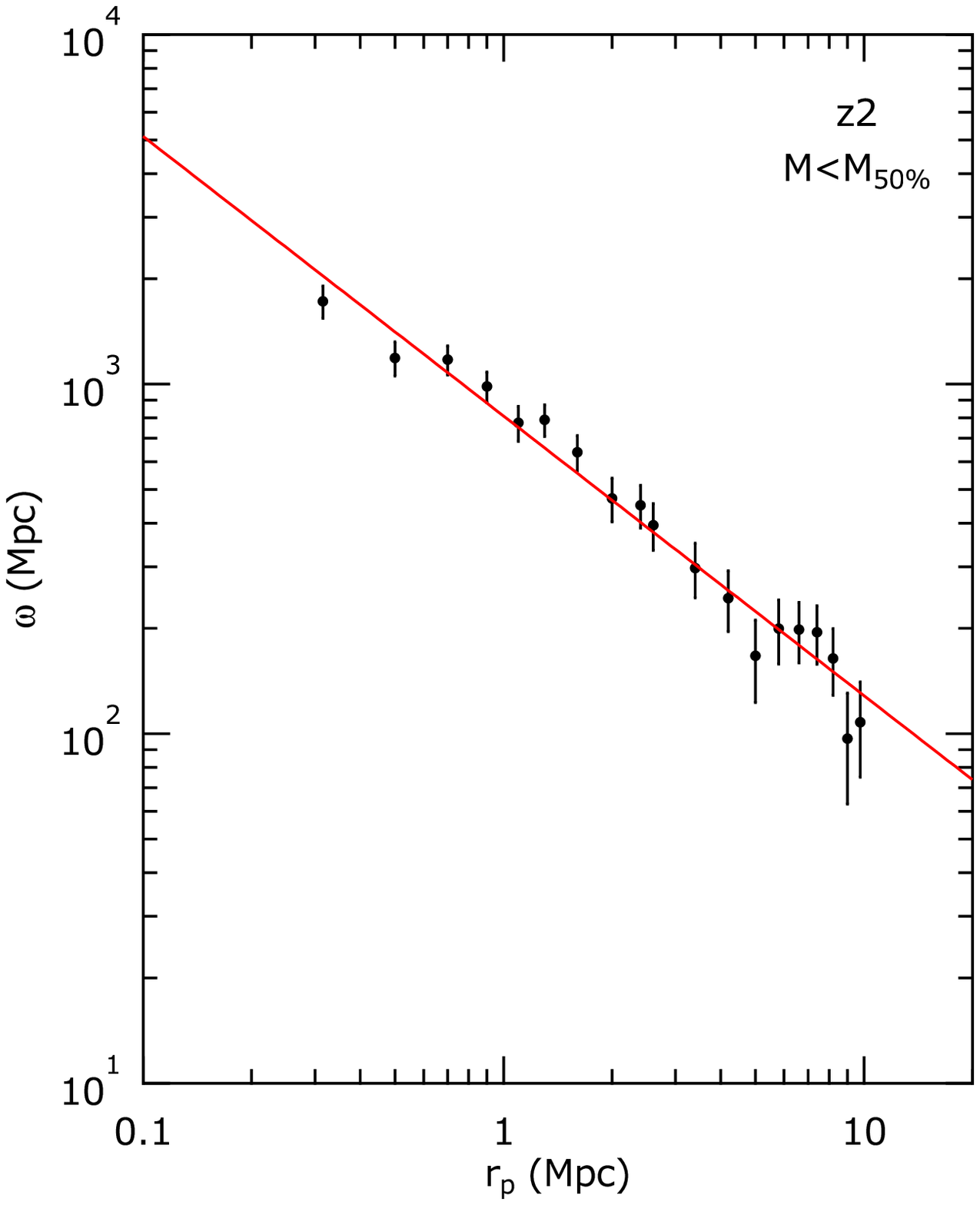} 
    \includegraphics[width=0.32\textwidth]{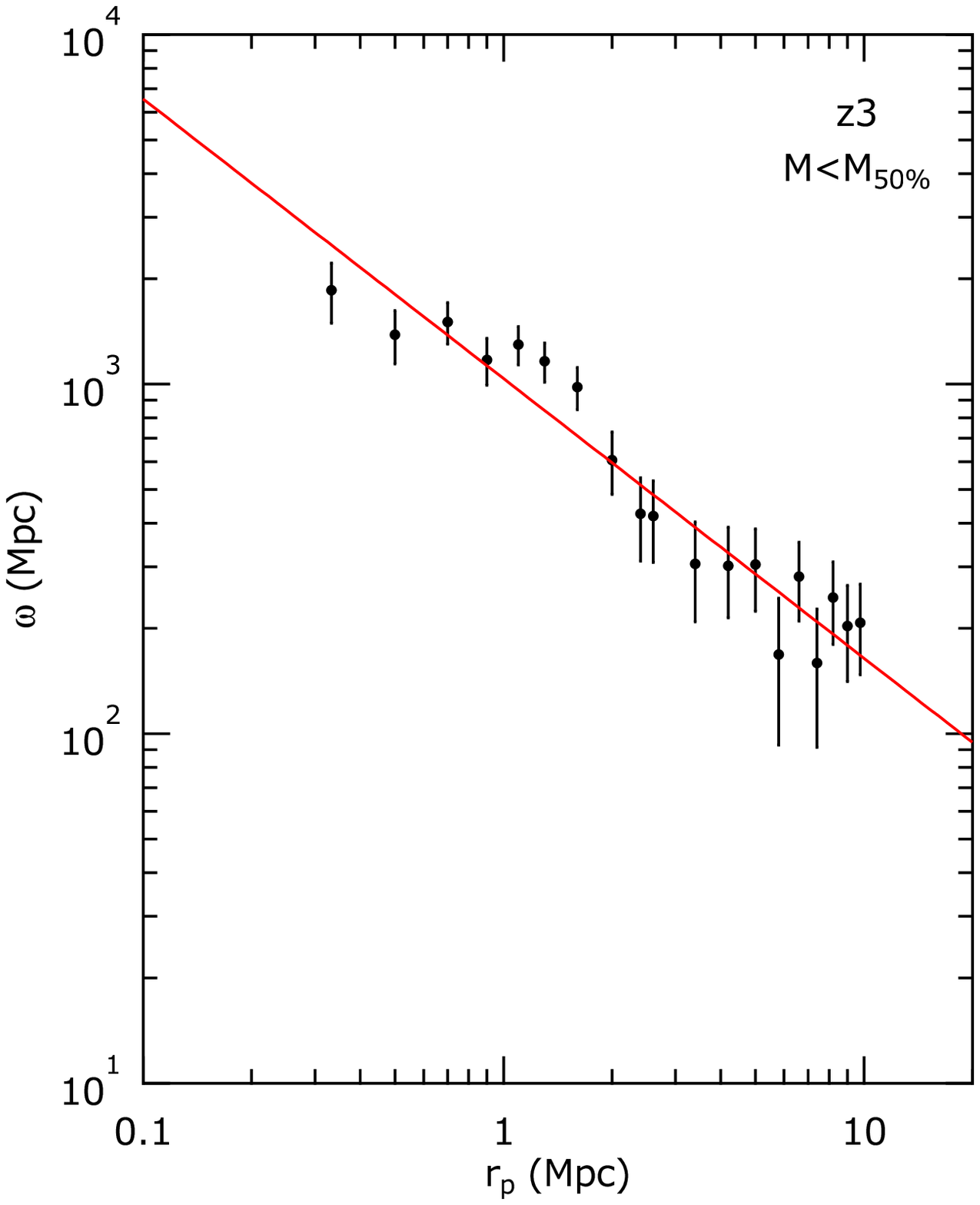} 
    \includegraphics[width=0.32\textwidth]{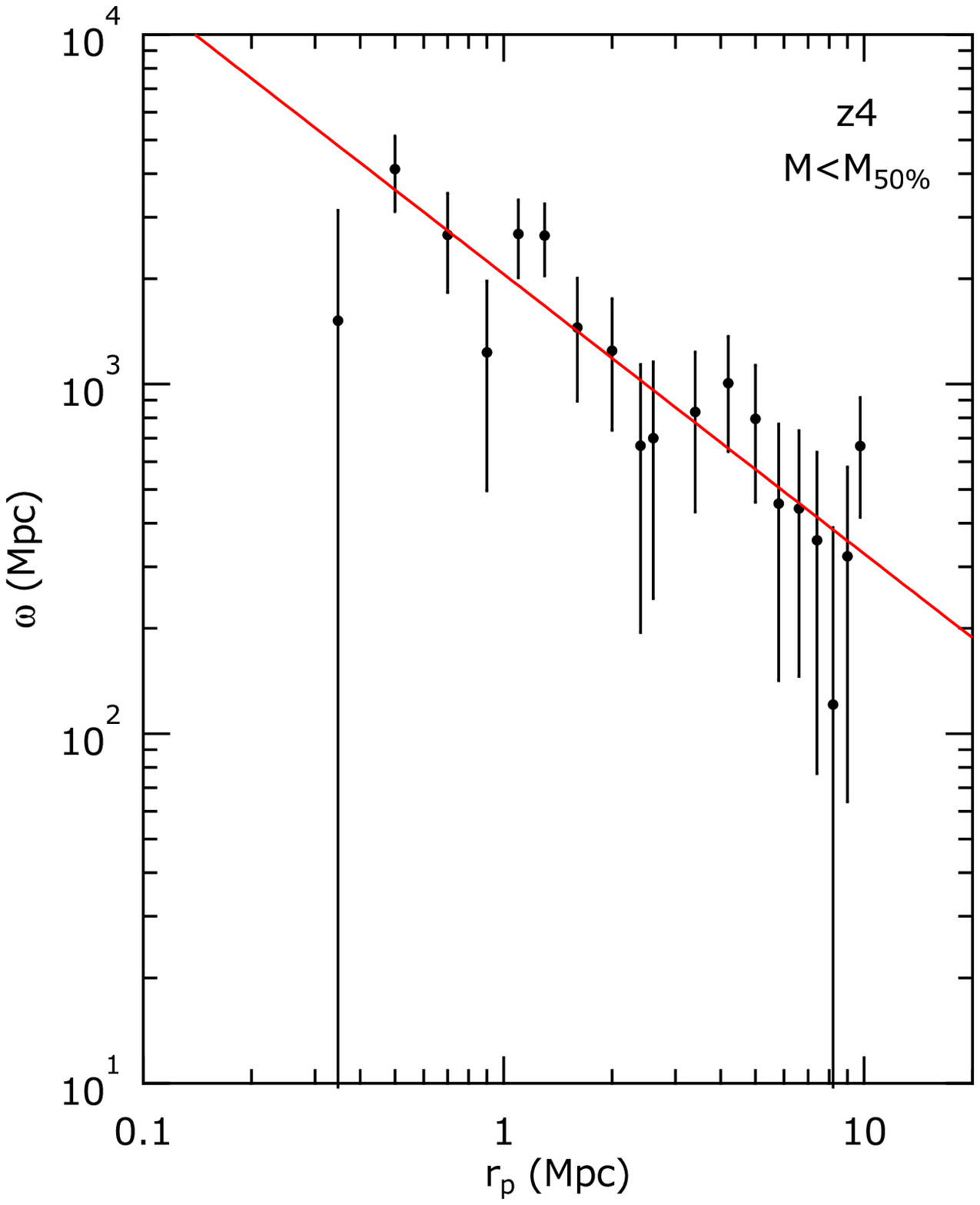} 
  \end{center}
  \caption{Projected cross-correlation functions derived from
    data shown in Figure~\ref{fig:density_1}. Solid lines represent
    the power law model fitted to the data.}
  \label{fig:cc_function}
\end{figure}

Figure~\ref{fig:density_1} shows
the distributions of average number density of galaxies with 
absolute magnitude brighter than $M_{50\%}$ for five redshift
groups.
The model functions of equation~(\ref{eq:model}) fitted to the
observations are also plotted in the same figure.
The corresponding projected cross-correlation functions calculated 
using equation~(\ref{eq:omega_2}) are shown in
Figure~\ref{fig:cc_function}.
The estimated fitting parameters are summarized in upper parts
of Table~\ref{tab:fit_density} for each threshold magnitude.

The statistical error of $r_{0}$ was estimated by the jackknife
method.
Jackknife resamplings were made by omitting, in turn, each of
the AGN datasets.
Then $r_{0}$ was calculated for each jackknife resampling and
its error was estimated from their variance:
\begin{equation}
\sigma_{r_{0}}^{2} = \frac{N-1}{N} \sum_{i=1}^{N} 
   (r_{0,i} -  \bar{r}_{0} )^{2},
\end{equation}
where $r_{0,i}$ is a cross-correlation length obtained for the
i-th jackknife sample, $\bar{r}_{0}$ is the average of
$r_{0,i}$, and $N$ is the number of jackknife samples.

\begin{figure}
  \begin{center}
    \includegraphics[width=0.6\textwidth]{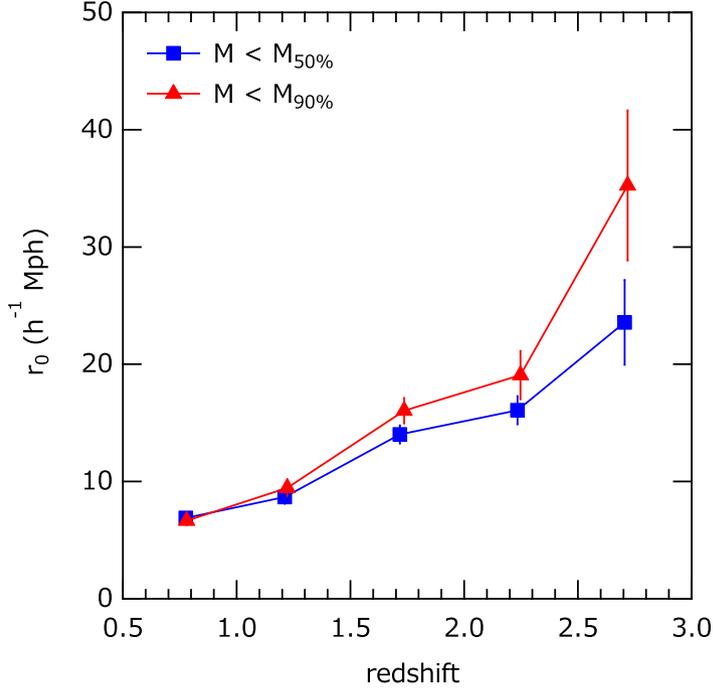} 
  \end{center}
  \caption{Cross-correlation length as a function of redshift.
  Squares and triangles represent the cross-correlation
  length derived using galaxies brighter than $M_{50\%}$
  and $M_{90\%}$ respectively, where
  $M_{50\%}$ ($M_{90\%}$) represents the magnitude where detection efficiency 
  is 50\% (90\%).}
  \label{fig:r0_z}
\end{figure}

The cross-correlation lengths are plotted as a function of redshift
in Figure~\ref{fig:r0_z} for each threshold sample.
The result shows that the cross-correlation length increases
as the redshift increases, and also is larger for more luminous galaxy 
samples.
Therefore, it is important for examining the BH mass dependence
after matching the distribution of redshift 
and galaxy luminosity for different BH mass groups.

To reduce the effect of redshift dependence in the comparisons between
the different BH mass groups, we constructed 
redshift-matched samples for each redshift group.
The redshift-matched samples were constructed by selecting the
same number of AGN datasets for each 
$\mathnormal{\Delta} z =$ 0.02 bin.

\begin{figure}
  \begin{center}
    \includegraphics[width=0.40\textwidth]{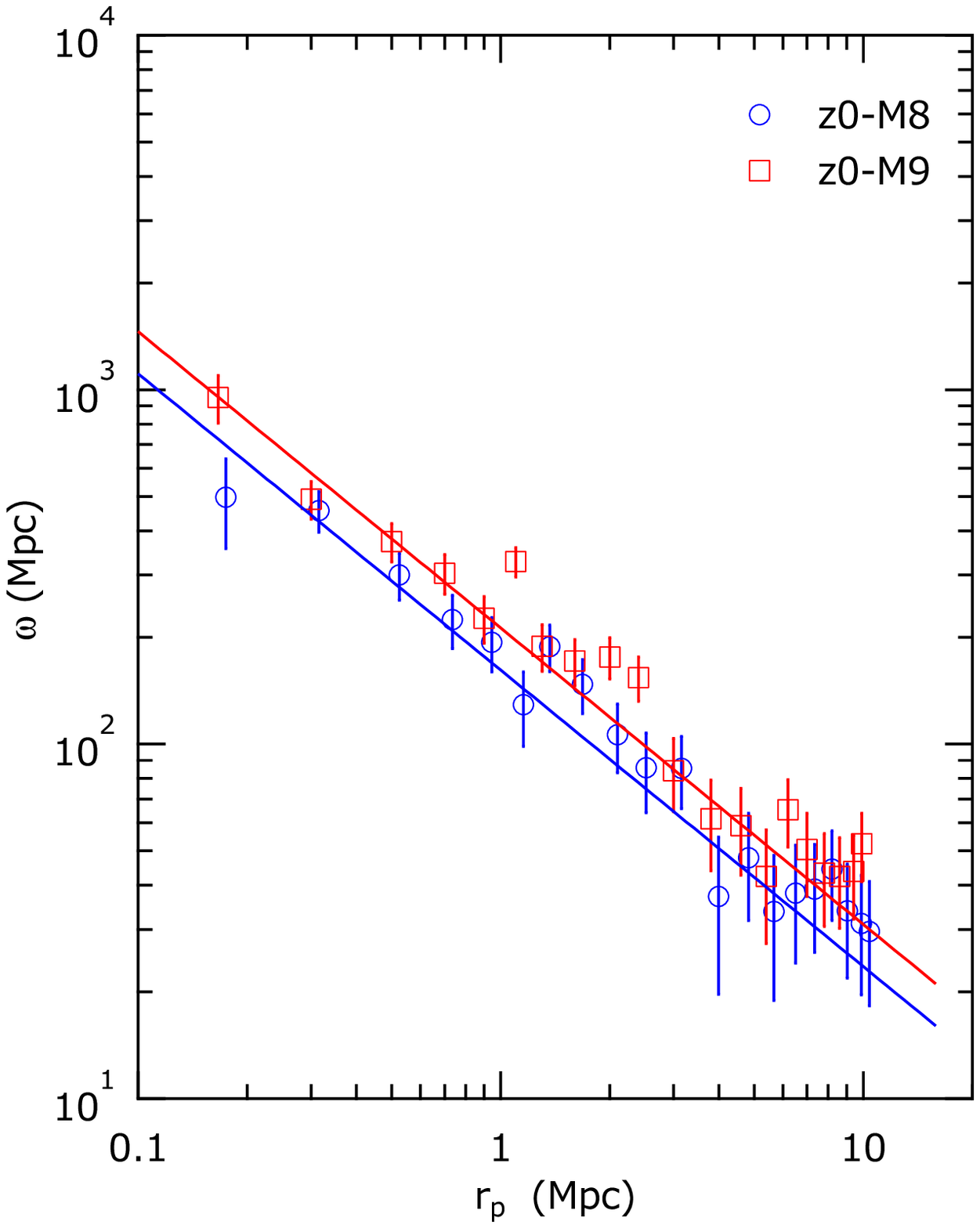} 
    \includegraphics[width=0.40\textwidth]{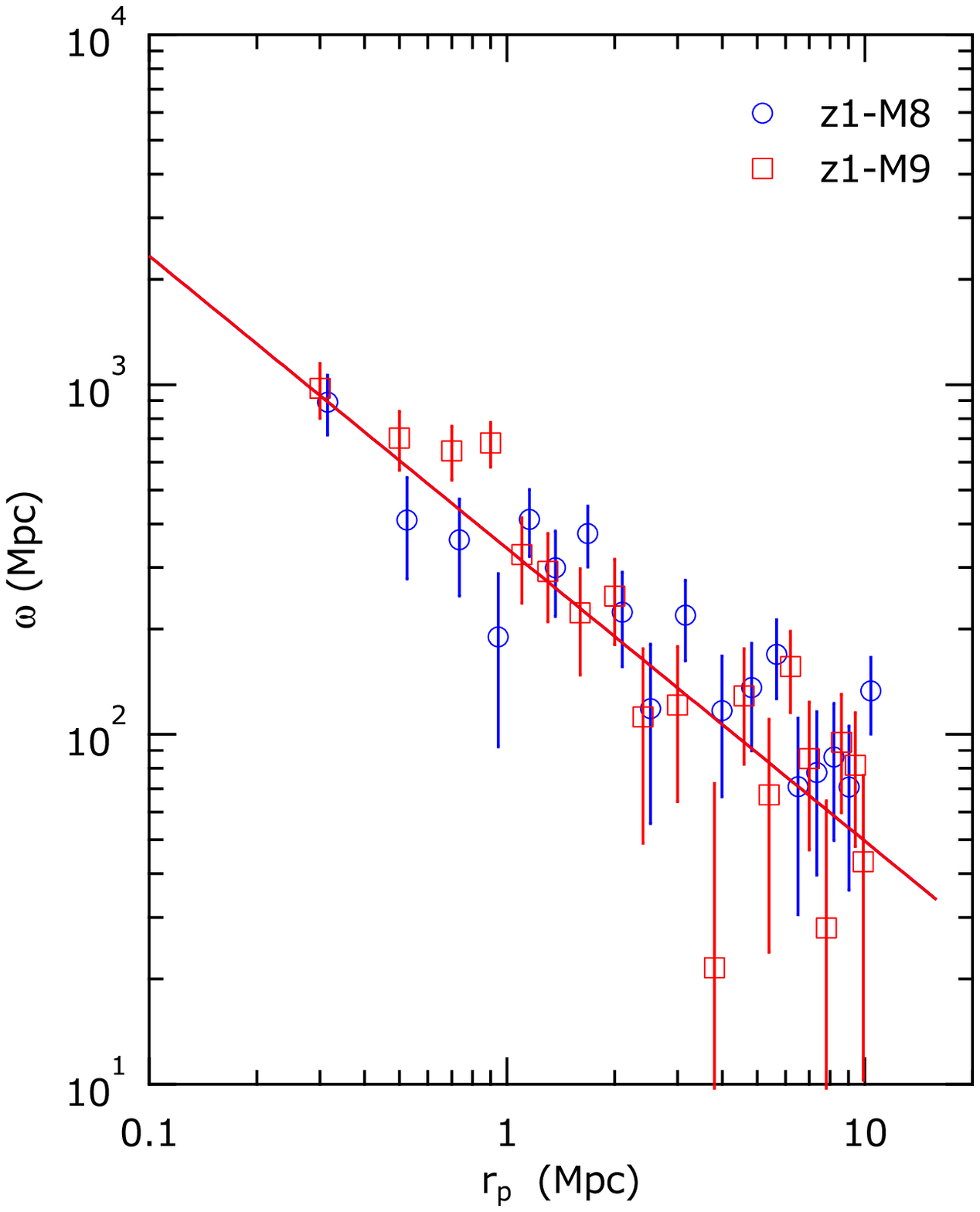} 
    \includegraphics[width=0.40\textwidth]{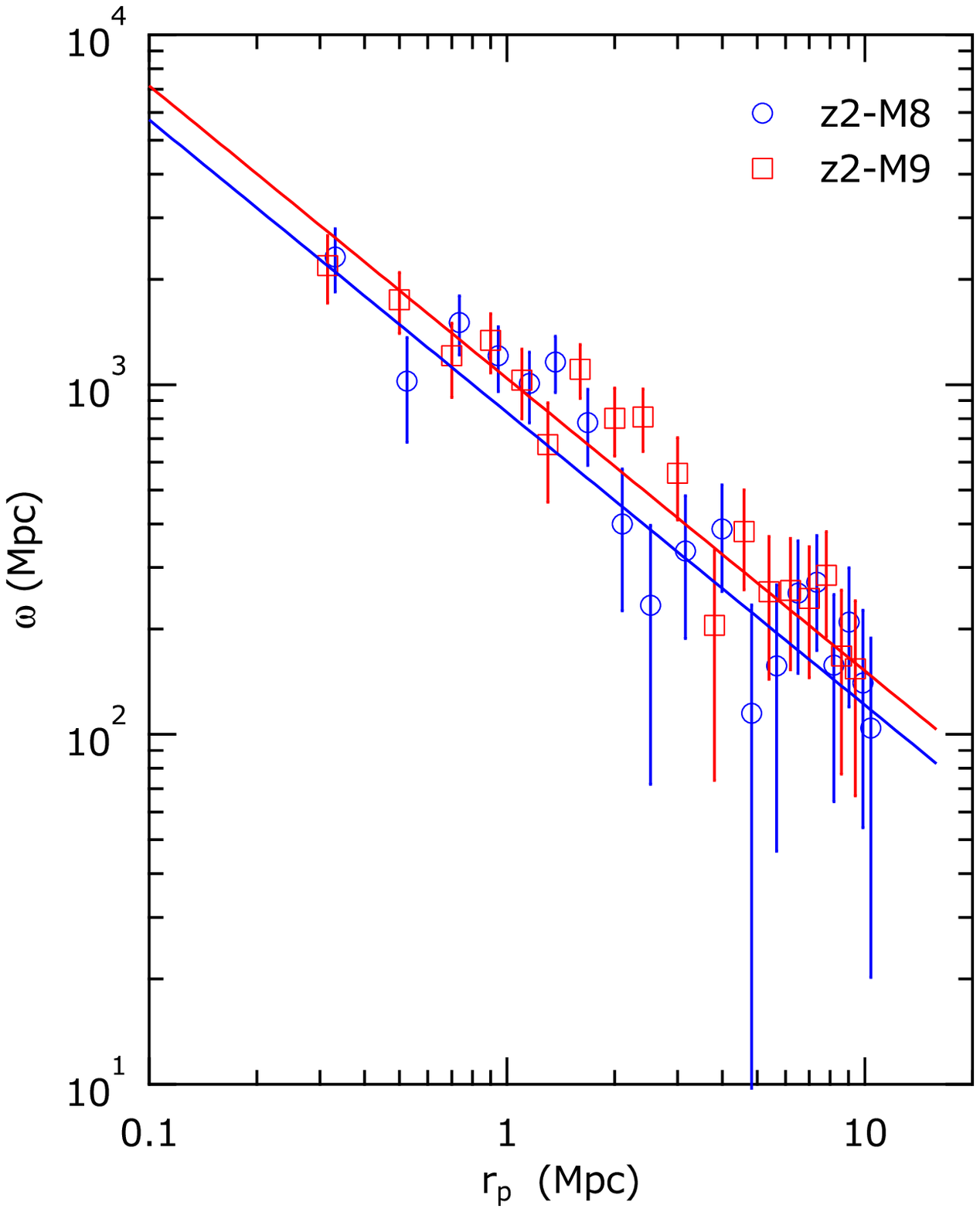} 
    \includegraphics[width=0.40\textwidth]{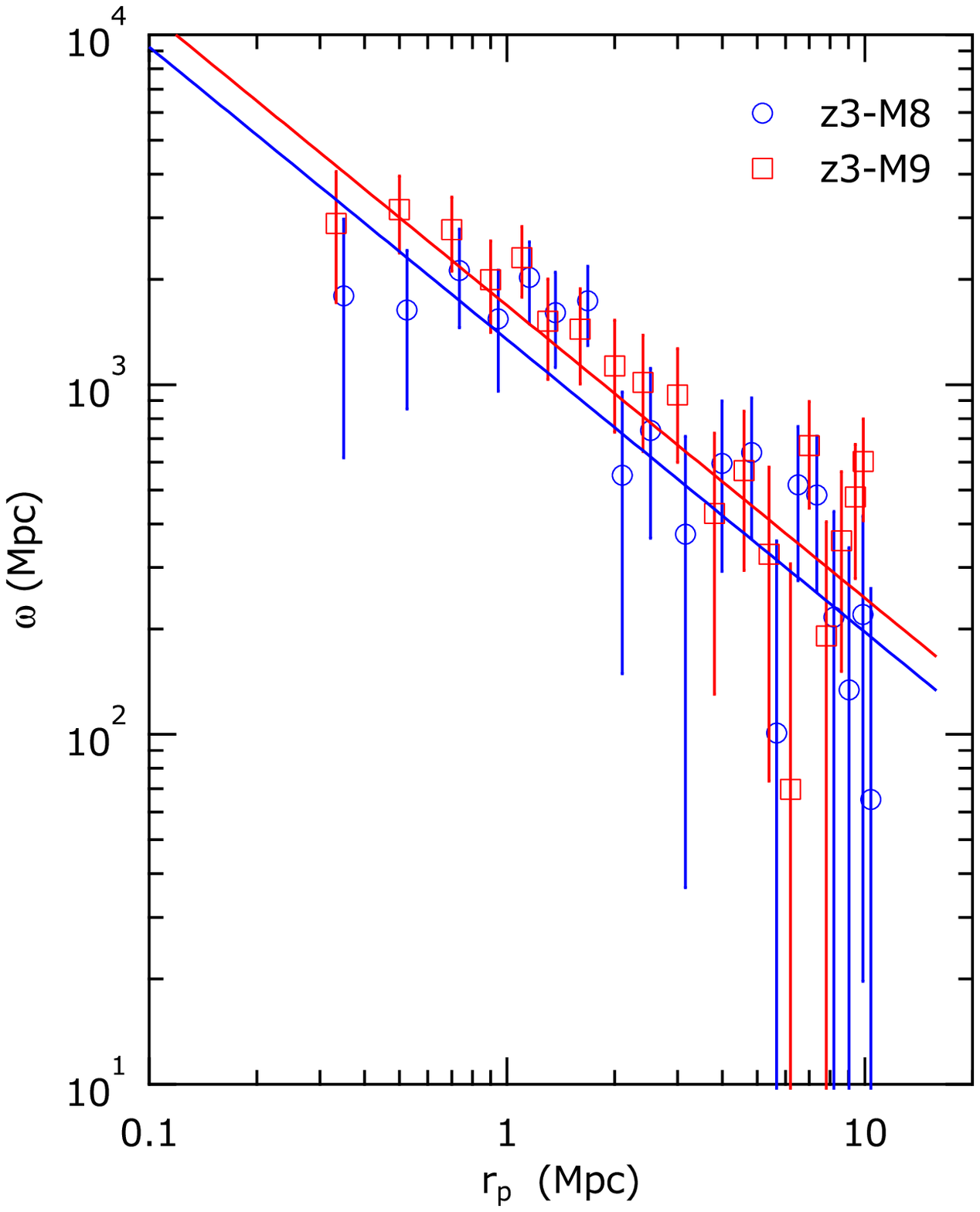} 
  \end{center}
  \caption{Comparisons of projected cross-correlation functions 
     derived for mass groups of M8 (circles) and M9 (squares). 
     Each panel represents the result for redshift 
     z0, z1, z2, and z3.  Redshift-matched AGN samples 
     and galaxy samples brighter than $M_{90\%}$ are used.
     Solid lines represent the power law model fitted to the 
     data points.}
  \label{fig:cc_function_mass}
\end{figure}

The projected cross-correlation functions calculated for the redshift-matched 
samples are shown in Figure~\ref{fig:cc_function_mass} for
four redshift groups z0, z1, z2 and z3.
The mass dependence for the z4 group was not examined due to poor statistics.
In this figure, the circles represent lower BH mass groups (M8)
and the squares represent higher BH mass groups (M9).
Galaxy samples brighter than $M_{90\%}$ were used.
The estimated fitting parameters are summarized in the bottom part 
of Table~\ref{tab:fit_density}.

\begin{figure}
  \begin{center}
    \includegraphics[width=0.6\textwidth]{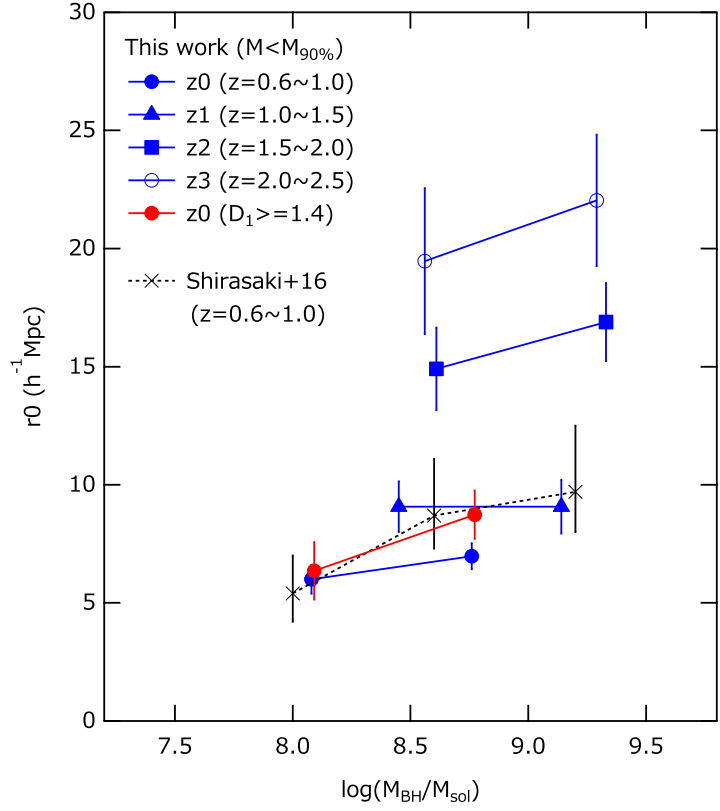} 
  \end{center}
  \caption{Cross-correlation lengths plotted as a function 
           of
           average BH mass obtained for data shown in
           Figure~\ref{fig:cc_function_mass} 
           (blue solid markers).
           Note that the galaxy sample for these datasets
           are dominated by blue galaxies.
           The result obtained by \citet{Shirasaki+16},
           for which galaxy samples are dominated by red 
           galaxies contrary to the current dataset, is
           shown with crosses.
           The result obtained using the data selected by color 
           to construct a dataset comparable with the result of
           \citet{Shirasaki+16} is shown with open circles.
}
  \label{fig:r0_MBH}
\end{figure}

The cross-correlation lengths are plotted as a function of
BH mass in Figure~\ref{fig:r0_MBH}.
From these results, no significant difference is seen 
between the two mass groups.

For comparison with the previous results \citep{Shirasaki+16}
derived using UKIDSS and SDSS data for galaxy samples, 
the result obtained for red galaxy samples ($D_{1}\ge1.4$) as described 
in the next subsection are also plotted.
In the work of \citet{Shirasaki+16}, they calculated 
cross-correlation functions between
AGNs and galaxies, which were mostly red galaxies, at redshifts $<$ 1.0, 
and obtained the result that the cross-correlation length depends 
on BH mass but does not depend on redshift.
Our current result at higher redshifts indicates its strong 
dependence on redshift and not on BH mass.
The different behavior partly comes from the difference in the properties 
of galaxy samples; that is, the galaxy samples in the previous work
are dominated by red-type galaxies and mostly dimmer than $L_{*}$, while 
those in the current work are dominated by blue-type galaxies and
mostly consist of galaxies brighter than $L_{*}$ at higher redshifts.
As shown in the following subsection the clustering of galaxies 
significantly increases above $L_{*}$, which introduces redshift dependence
to the cross-correlation length.

\begin{table}
  \tbl{Fitting result of cross-correlation function.}{%
  \begin{tabular}{llrcccrrr}
      \hline
      BH mass 
 & Redshift 
 & $n_{\rm AGN}$$^{\rm a}$ 
 & $\langle \log{M_{\rm BH}/M_{\solar}} \rangle$$^{\rm b}$
 & $\langle z \rangle$$^{\rm c}$
 & $r_0$$^{\rm d}$
 & $\langle n_{\rm bg} \rangle$$^{\rm e}$
 & $\langle \rho_{0} \rangle$$^{\rm f}$ \\
 $\log{(M_{\rm BH}/M_{\solar})}$ & & & 
 & & ${h^{-1}{\rm Mpc}}$ & Mpc$^{-2}$ & $10^{-3}$Mpc$^{-3}$ \\
\hline
\multicolumn{8}{c}{$M<M_{90\%}$ galaxy samples}\\
7.0--11.0 (M8+M9) & 0.6--1.0 (z0) & 1194 & 8.45 & 0.78 &  6.66$\pm$0.36 & 76.29$\pm$0.03 & 11.2  \\
7.0--11.0 (M8+M9) & 1.0--1.5 (z1) & 1216 & 8.80 & 1.22 &  9.46$\pm$0.69 & 42.53$\pm$0.02 &  2.85 \\
7.0--11.0 (M8+M9) & 1.5--2.0 (z2) & 1235 & 8.94 & 1.74 & 16.05$\pm$1.09 & 28.73$\pm$0.02 &  0.886\\
7.0--11.0 (M8+M9) & 2.0--2.5 (z3) & 1511 & 8.96 & 2.25 & 19.07$\pm$2.06 & 20.67$\pm$0.01 &  0.299\\
7.0--11.0         & 2.5--3.0 (z4) &  396 & 8.90 & 2.72 & 35.24$\pm$6.41 & 16.89$\pm$0.02 &  0.0968\\
\multicolumn{8}{c}{$M<M_{50\%}$ galaxy samples}\\
7.0--11.0 (M8+M9) & 0.6--1.0 (z0) & 1194 & 8.45 & 0.78 &  6.89$\pm$0.36 & 121.73$\pm$0.04 & 15.5 \\
7.0--11.0 (M8+M9) & 1.0--1.5 (z1) & 1216 & 8.80 & 1.22 &  8.67$\pm$0.58 &  64.67$\pm$0.03 & 4.92 \\
7.0--11.0 (M8+M9) & 1.5--2.0 (z2) & 1235 & 8.94 & 1.74 & 14.02$\pm$0.77 &  42.82$\pm$0.02 & 1.85 \\
7.0--11.0 (M8+M9) & 2.0--2.5 (z3) & 1511 & 8.96 & 2.25 & 16.08$\pm$1.21 &  30.80$\pm$0.02 & 0.777\\
7.0--11.0         & 2.5--3.0 (z4) &  396 & 8.90 & 2.72 & 23.58$\pm$3.62 &  25.22$\pm$0.03 & 0.332\\
\multicolumn{8}{c}{Redshift-matched AGN samples \& $M<M_{90\%}$ galaxy samples}\\
7.0-- 8.4 (M8) & 0.6--1.0 (z0) & 482 & 8.08 & 0.77 & 6.00$\pm$0.61 & 75.42$\pm$0.04 & 11.4 \\
8.4--11.0 (M9) & 0.6--1.0 (z0) & 482 & 8.76 & 0.77 & 6.98$\pm$0.55 & 76.44$\pm$0.04 & 11.4 \\
7.0-- 8.8 (M8) & 1.0--1.5 (z1) & 506 & 8.45 & 1.22 & 9.07$\pm$1.08 & 42.03$\pm$0.03 & 2.85 \\
8.8--11.0 (M9) & 1.0--1.5 (z1) & 506 & 9.14 & 1.22 & 9.07$\pm$1.14 & 43.05$\pm$0.03 & 2.86 \\
7.0-- 9.0 (M8) & 1.5--2.0 (z2) & 534 & 8.61 & 1.74 &14.91$\pm$1.75 & 28.67$\pm$0.03 & 0.890 \\
9.0--11.0 (M9) & 1.5--2.0 (z2) & 534 & 9.33 & 1.74 &16.89$\pm$1.65 & 28.95$\pm$0.03 & 0.887 \\
7.0-- 8.9 (M8) & 2.0--2.5 (z3) & 640 & 8.56 & 2.25 &19.47$\pm$3.10 & 20.75$\pm$0.02 & 0.299 \\
8.9--11.0 (M9) & 2.0--2.5 (z3) & 640 & 9.29 & 2.25 &22.04$\pm$2.78 & 20.61$\pm$0.02 & 0.300 \\
      \hline
  \end{tabular}}\label{tab:fit_density}
  \begin{tabnote}
$^{a}$number of AGN datasets, 
$^{b}$average of logarithm of BH mass,
$^{c}$average redshift,
$^{d}$cross-correlation length and its error,
$^{e}$average of projected number density of background galaxies,
$^{f}$average of the averaged number density of galaxies at the AGN redshift,
  \end{tabnote}
\end{table}

\subsection{Galaxy luminosity dependence of cross-correlation length}
\label{sec:result_cc_luminosity}

As larger clustering was indicated for more luminous galaxies as shown
in Figure~\ref{fig:r0_z}, we calculated cross-correlation lengths
for flux-limited samples derived from the 
four-band
detected samples
to examine the dependence on the luminosity of galaxies more closely.
To calculate the absolute magnitude for the same rest frame 
bandpass at different redshifts, we performed SED fitting using
the EAZY software developed by \citet{Brammer+08} and calculated
the magnitude at a fixed bandpass.
The observed magnitudes at $g$, $r$, $i$, $z$, and $y$ bands are used in
the SED fitting.
The flux-limited galaxy samples were constructed based on 
absolute magnitude $M_{\lambda 310}$ for z0, z1 and z2 and $M_{\lambda 220}$ for z2, z3 
and z4, where $M_{\lambda 310}$ and $M_{\lambda 220}$ represents the absolute magnitude at 
wavelength 310~nm and 220~nm, respectively.

\begin{figure}
  \begin{center}
    \includegraphics[width=0.6\textwidth]{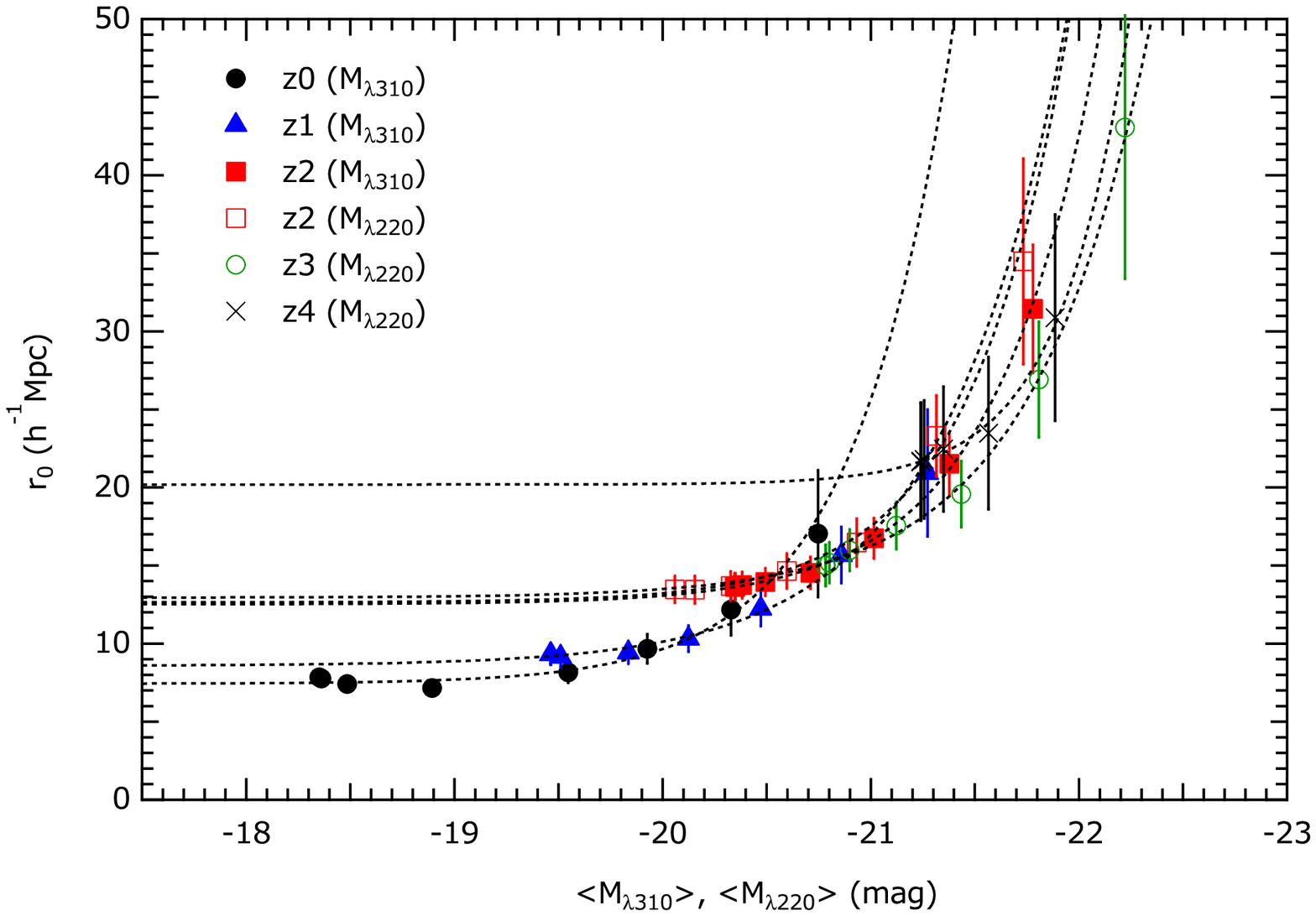} 
  \end{center}
    \caption{Cross-correlation length as a function of 
     average
     absolute 
     magnitude of galaxies measured in $M_{\lambda 310}$ or $M_{\lambda 220}$.
     The cross-correlation lengths were calculated using the galaxy 
     samples which are brighter than given threshold magnitudes.
     Dotted lines represent functions of equation~(\ref{eq:r0_vs_M})
     fitted to the data points.
     }
    \label{fig:r0_M}
\end{figure}

\begin{figure}
  \begin{center}
    \includegraphics[width=0.6\textwidth]{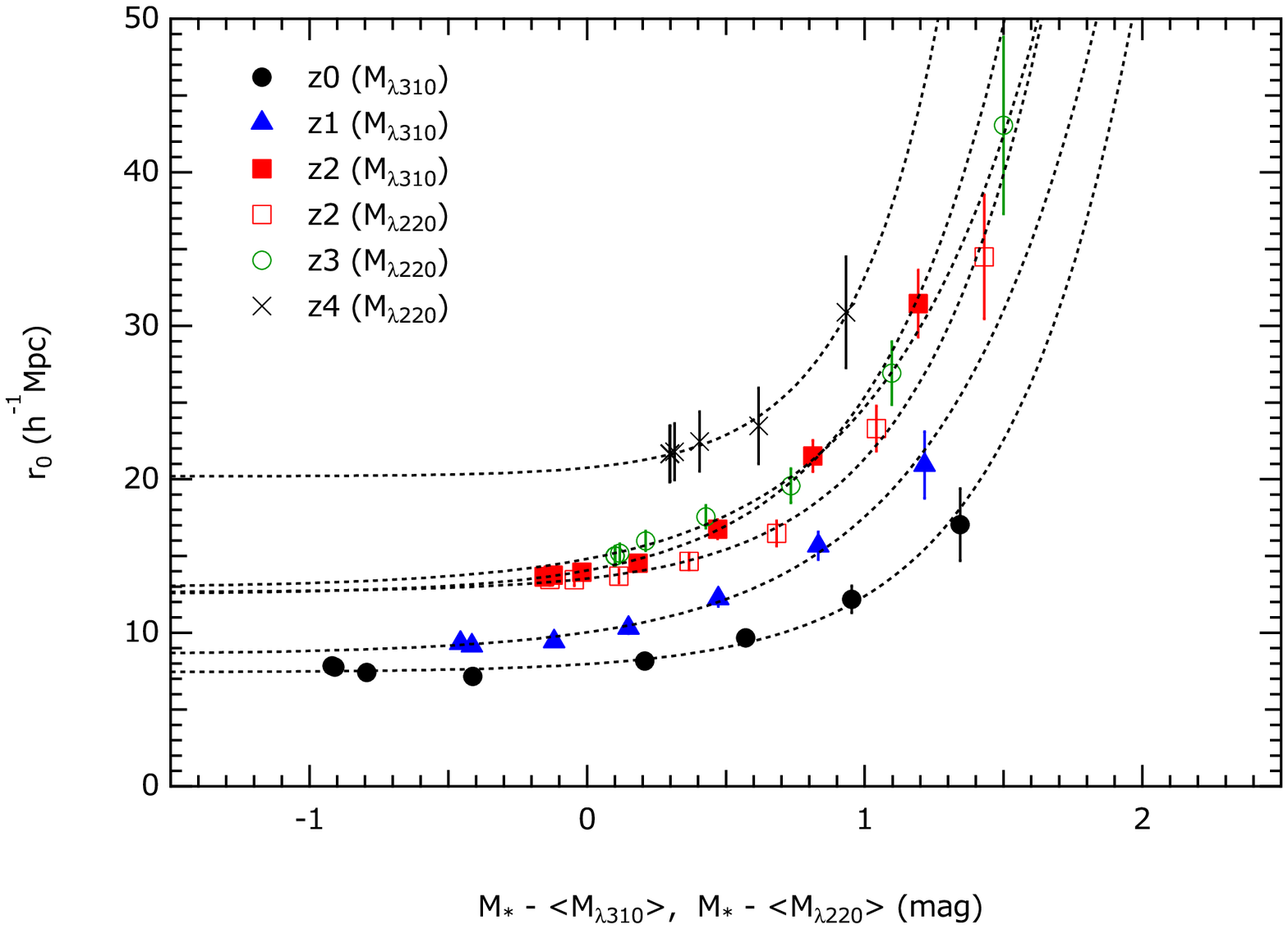} 
  \end{center}
    \caption{Cross-correlation length as a function of 
             $M_{*} - \langle M_{\lambda 310} \rangle$ or 
             $M_{*} - \langle M_{\lambda 220} \rangle$.
 $\langle M_{\lambda 310} \rangle$ and $\langle M_{\lambda 220} \rangle$ are the 
 average absolute magnitudes of galaxies measured at retsframe
 wavelengths of 310~nm and  220~nm, respectively, and 
 the cross-correlation lengths were calculated by using the galaxy 
 samples which are brighter than the threshold magnitude.
%%
%% The data points plotted in Figure~\ref{fig:r0_M} were shift by
%% $-M_{*}$ and flipped in the direction of horizontal-axis.
     Dotted lines represent functions of equation~(\ref{eq:r0_vs_M})
     fitted to the data points.
    }
    \label{fig:r0_dM}
\end{figure}

Figure~\ref{fig:r0_M} shows the cross-correlation
length as a function of average absolute magnitude of the galaxy sample 
brighter than the threshold magnitude in $M_{\lambda 310}$ or $M_{\lambda 220}$, 
and Figure~\ref{fig:r0_dM} shows it as a function of difference from the 
characteristic magnitude 
$M_{\rm *}$
of the parametrized luminosity function derived in
section~\ref{sec:cc_length}.
In these figures, dotted lines represent an empirical formula
expressed as:
\begin{equation}
  r_{0} = r_{0,\rm min} + \exp{\left( -\frac{
   M
   +20}{\sigma_{M}} \right) },
  \label{eq:r0_vs_M}
\end{equation}
where $r_{0,\rm min}$ represents an asymptotic cross-correlation length
at low luminosities, $M$
represents $M_{\lambda 310}$ or $M_{\lambda 220}$, and $\sigma_{M}$
corresponds to a slope of the curve at larger luminosities.

From these figures, it is apparent that the cross-correlation length 
rapidly increases at magnitudes brighter than 
$M_{\rm *}$, 
and it also increases with increasing redshift when
compared for the same $M_{*}-M$.
$r_{0,\rm min}$ corresponds to the cross-correlation length calculated 
for the whole galaxy samples of each redshift group.
Due to the difference in the detection threshold of galaxy luminosity,
the average luminosity for all the galaxy
samples is larger for the higher redshift groups than for the 
lower ones.
Thus the difference of $r_{0,\rm min}$ is due to the difference 
in galaxy luminosity as well as the difference in redshift.

\subsection{Color distributions of galaxies around AGNs}
\label{sec:result_color}

\begin{figure}
  \begin{center}
    \includegraphics[width=0.32\textwidth]{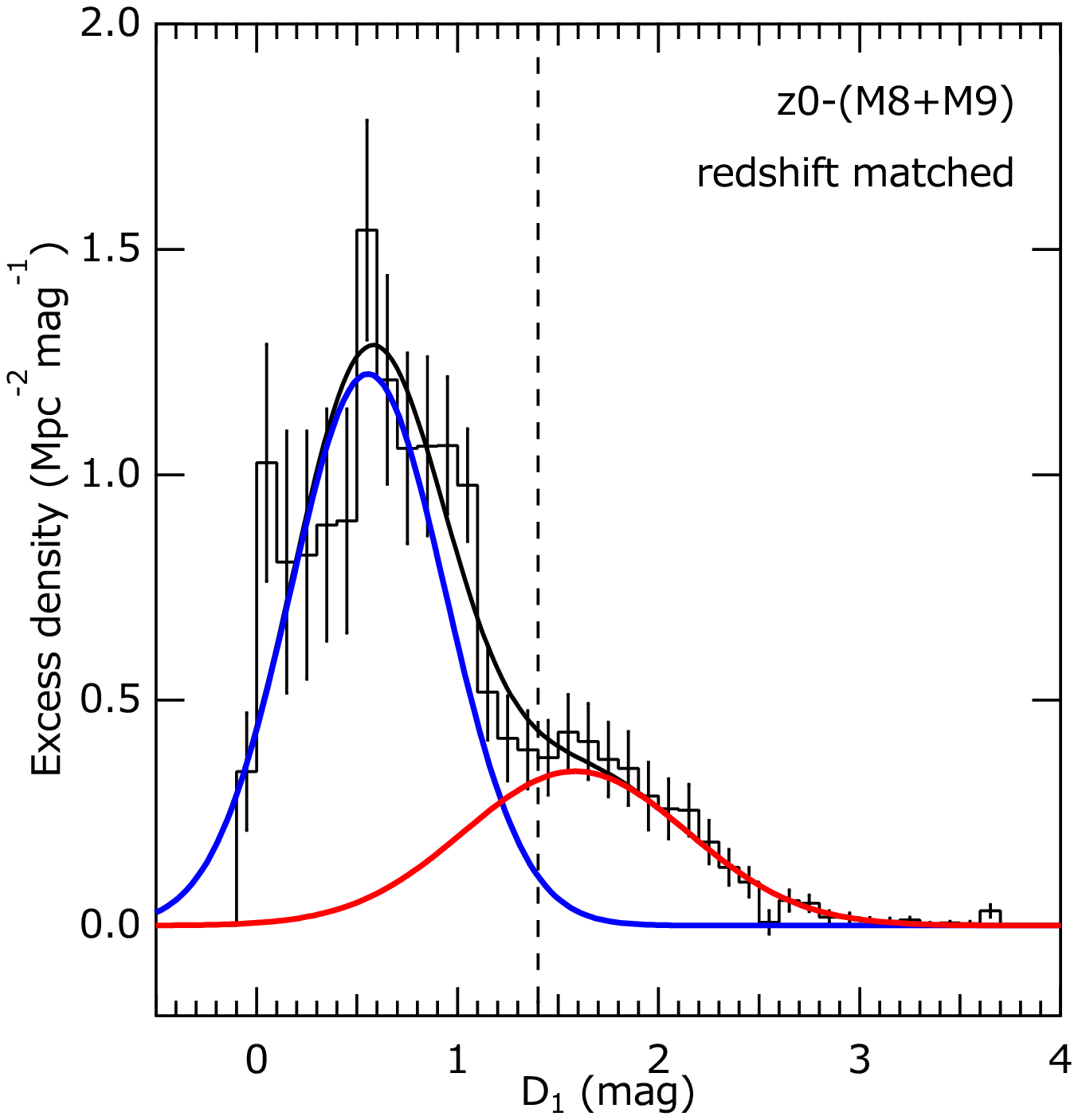} 
    \includegraphics[width=0.32\textwidth]{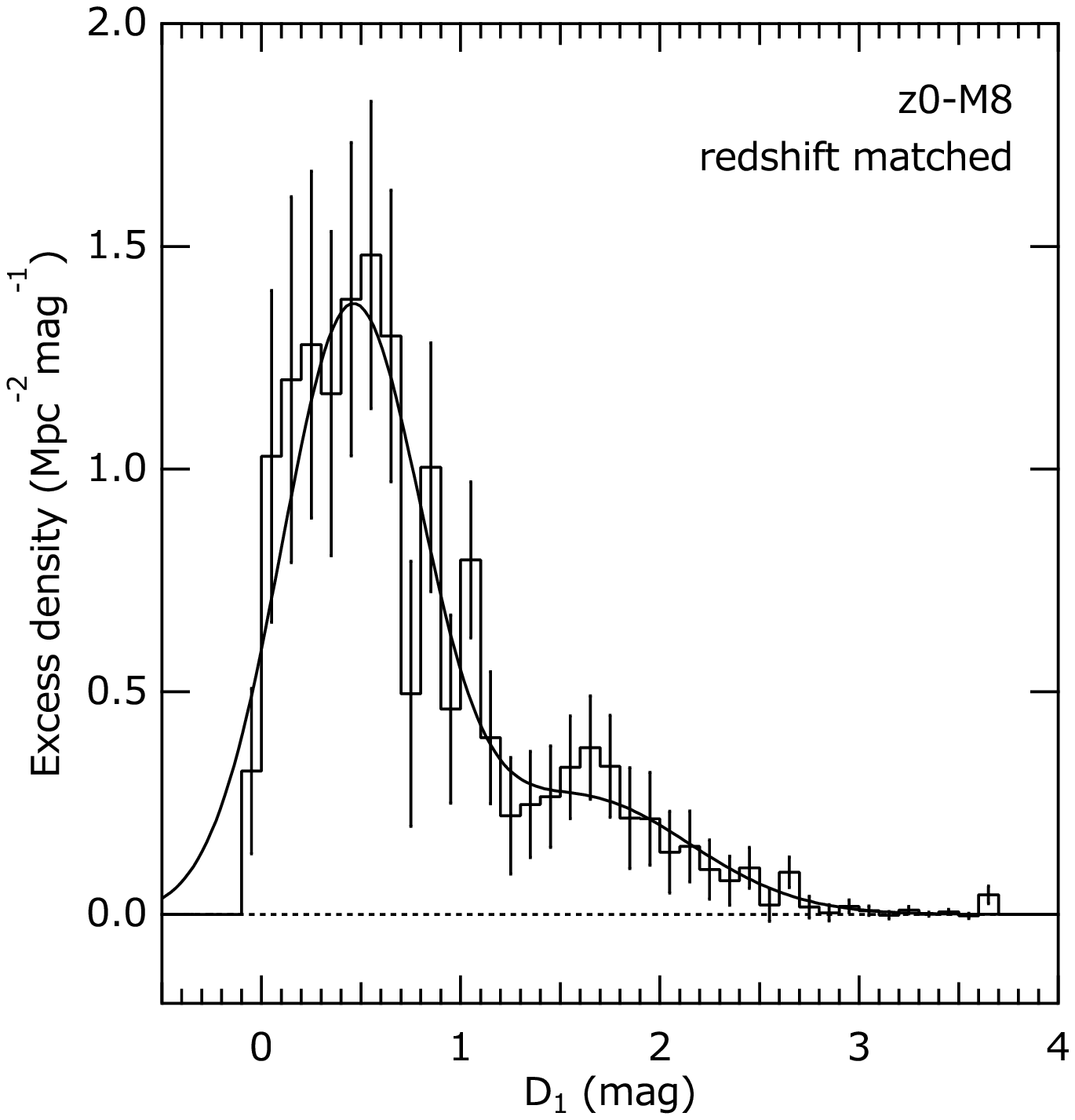} 
    \includegraphics[width=0.32\textwidth]{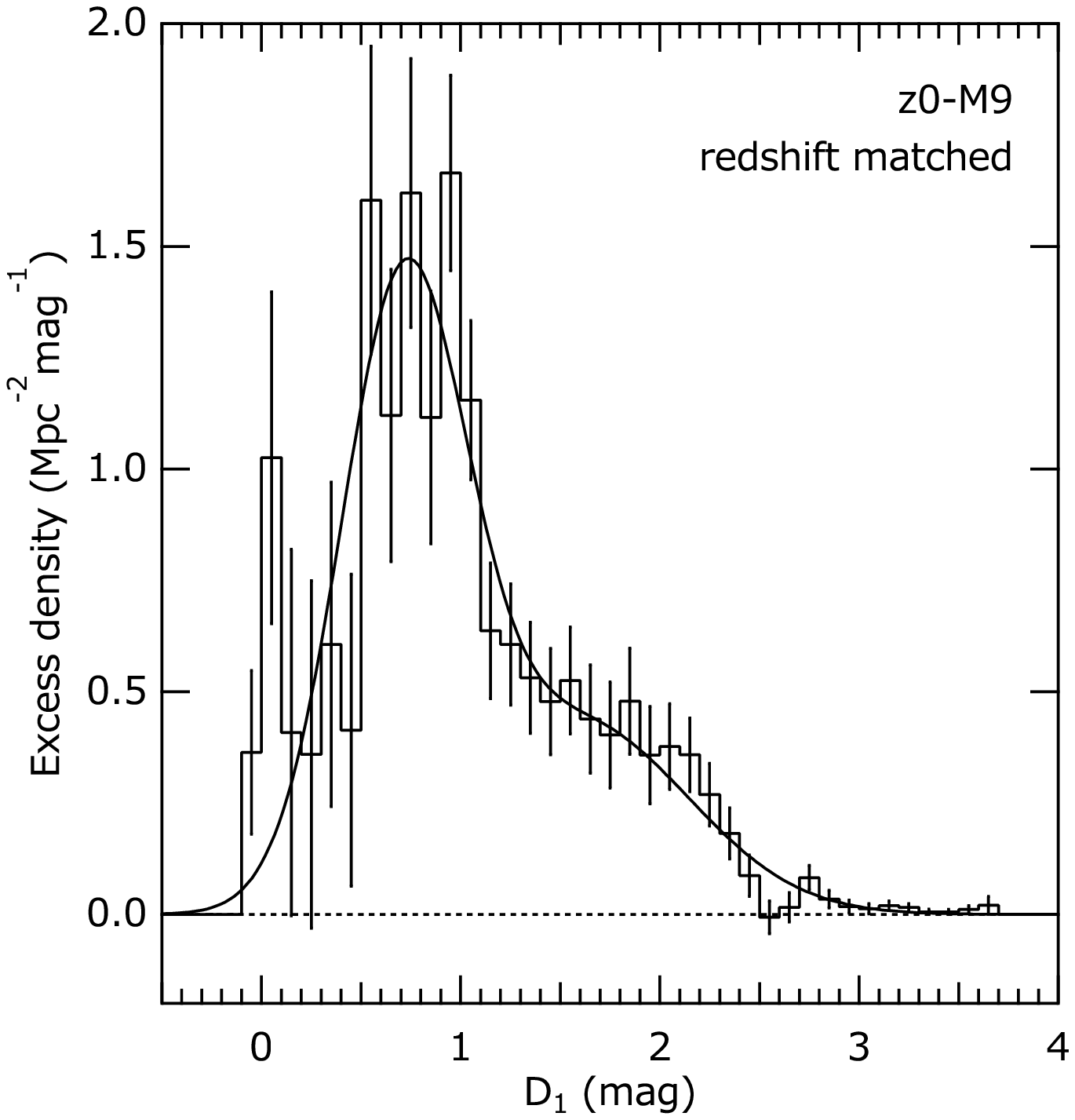} 
    \includegraphics[width=0.32\textwidth]{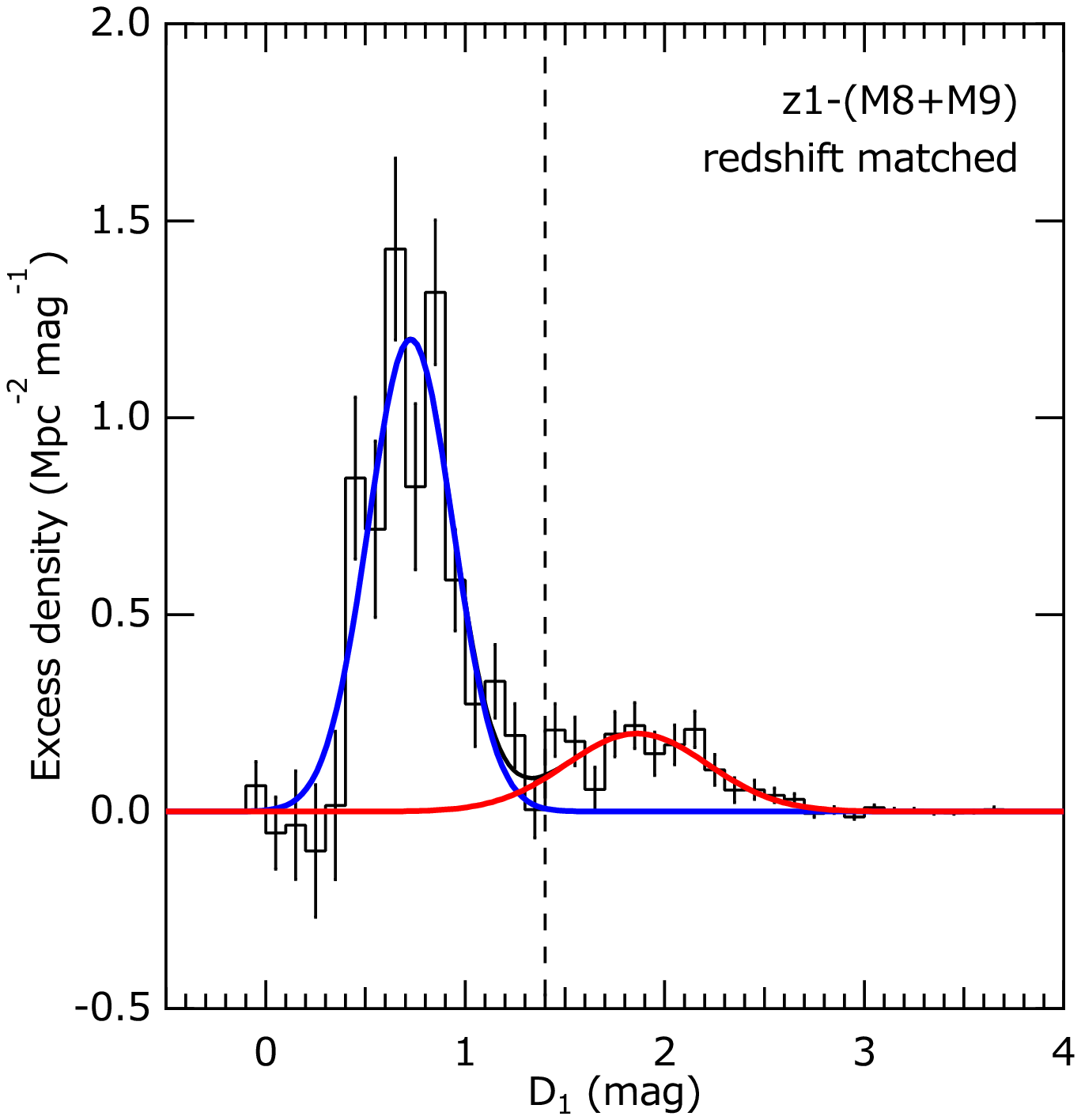} 
    \includegraphics[width=0.32\textwidth]{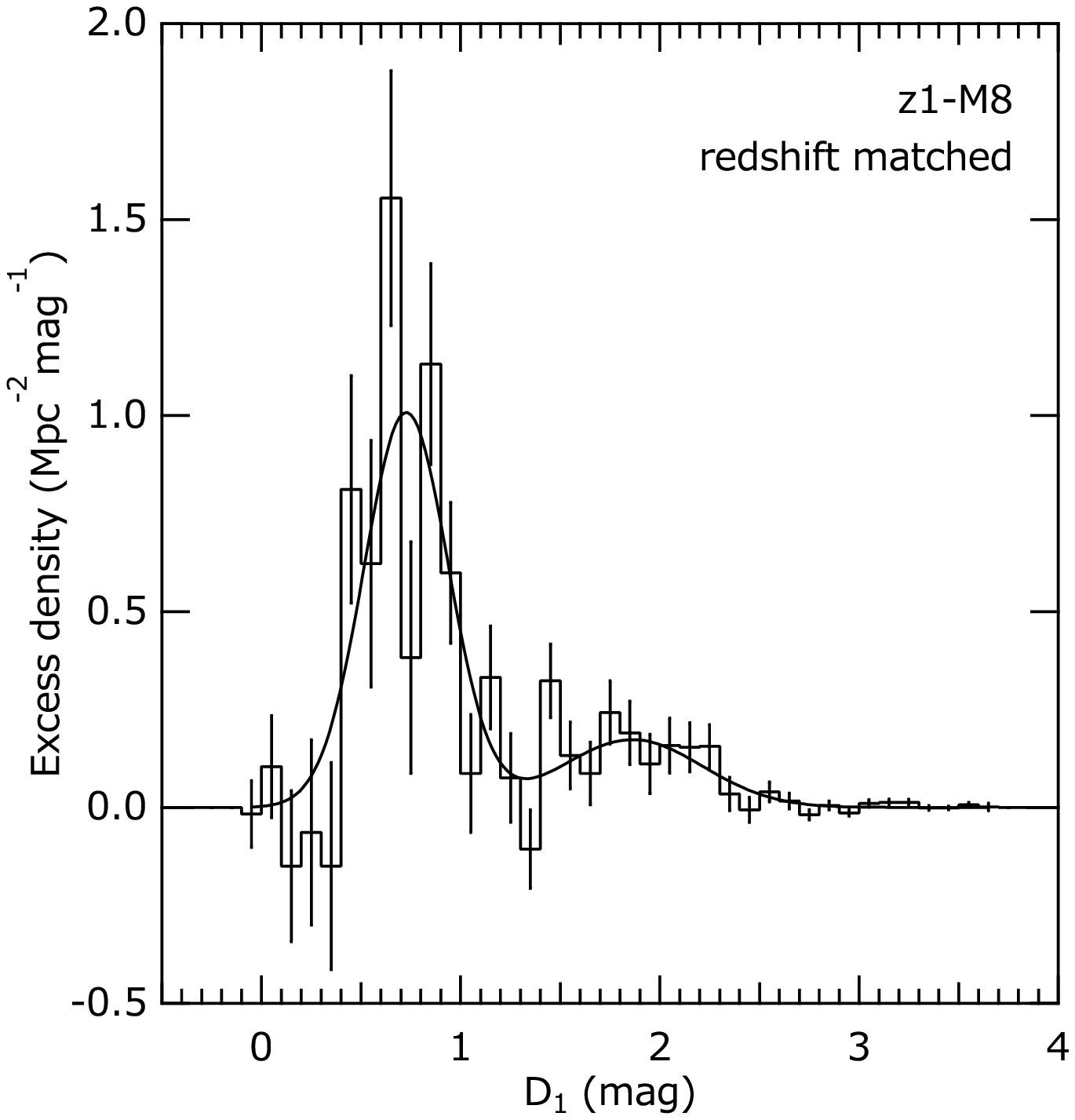} 
    \includegraphics[width=0.32\textwidth]{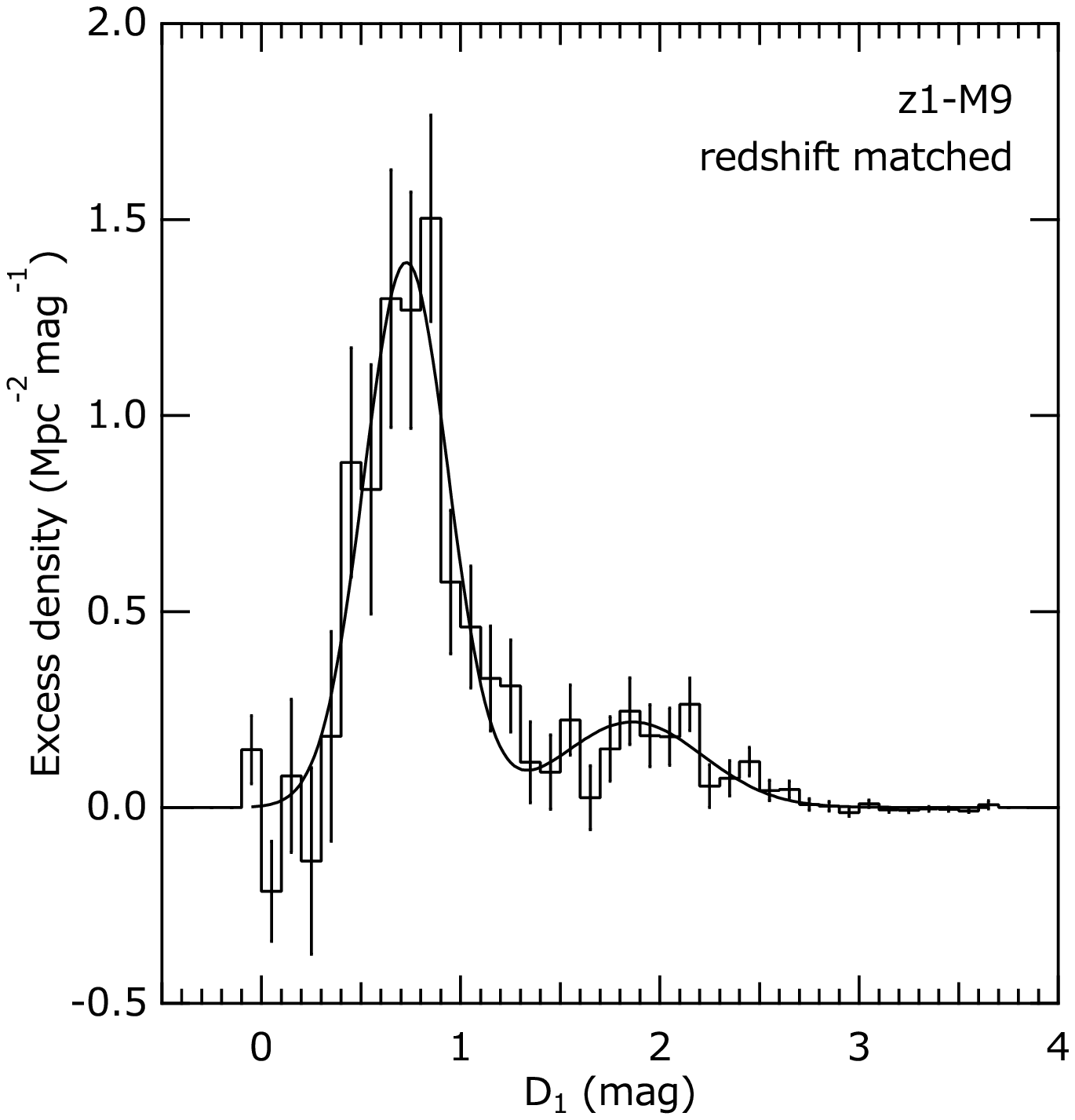} 
    \includegraphics[width=0.32\textwidth]{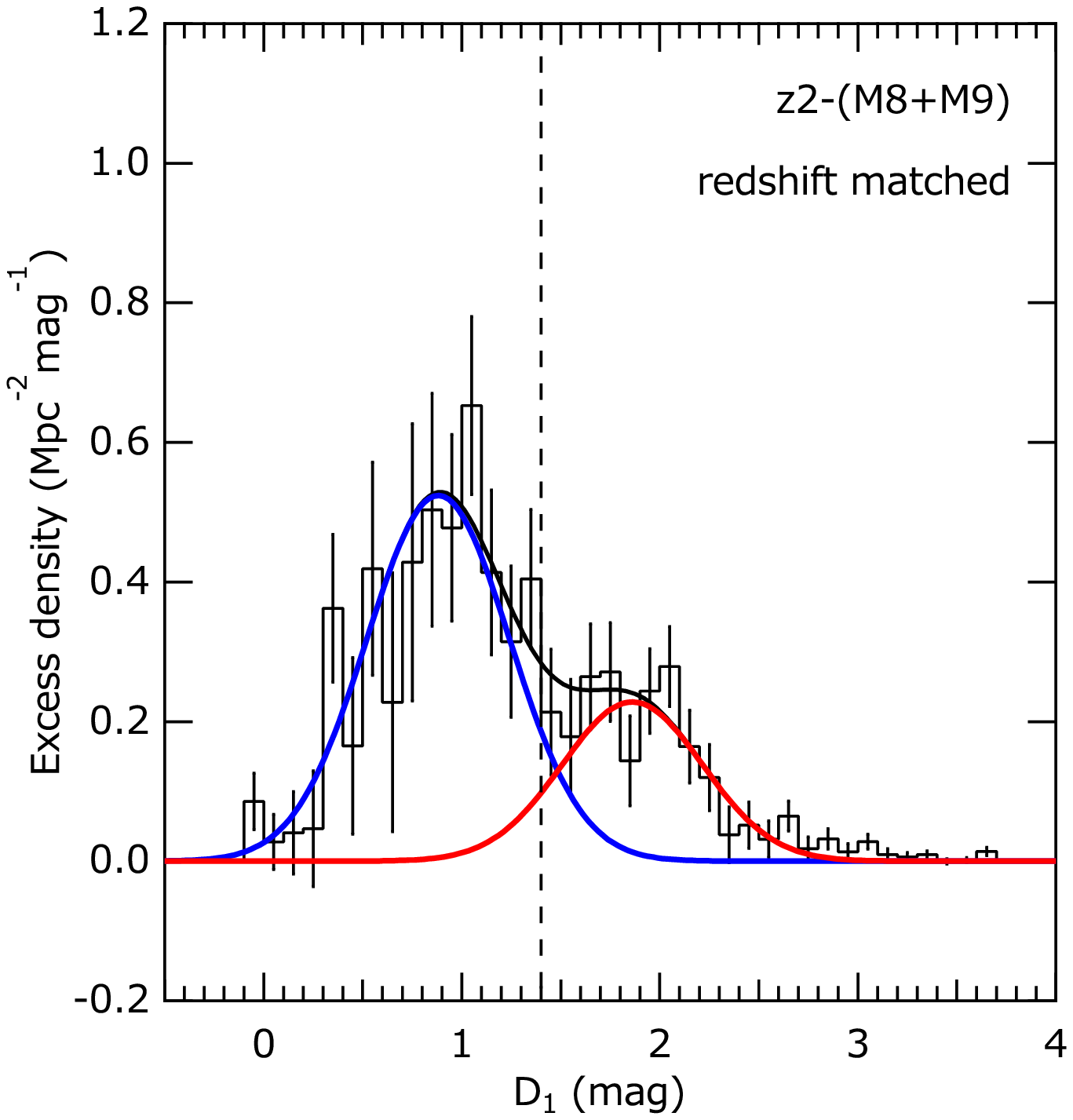} 
    \includegraphics[width=0.32\textwidth]{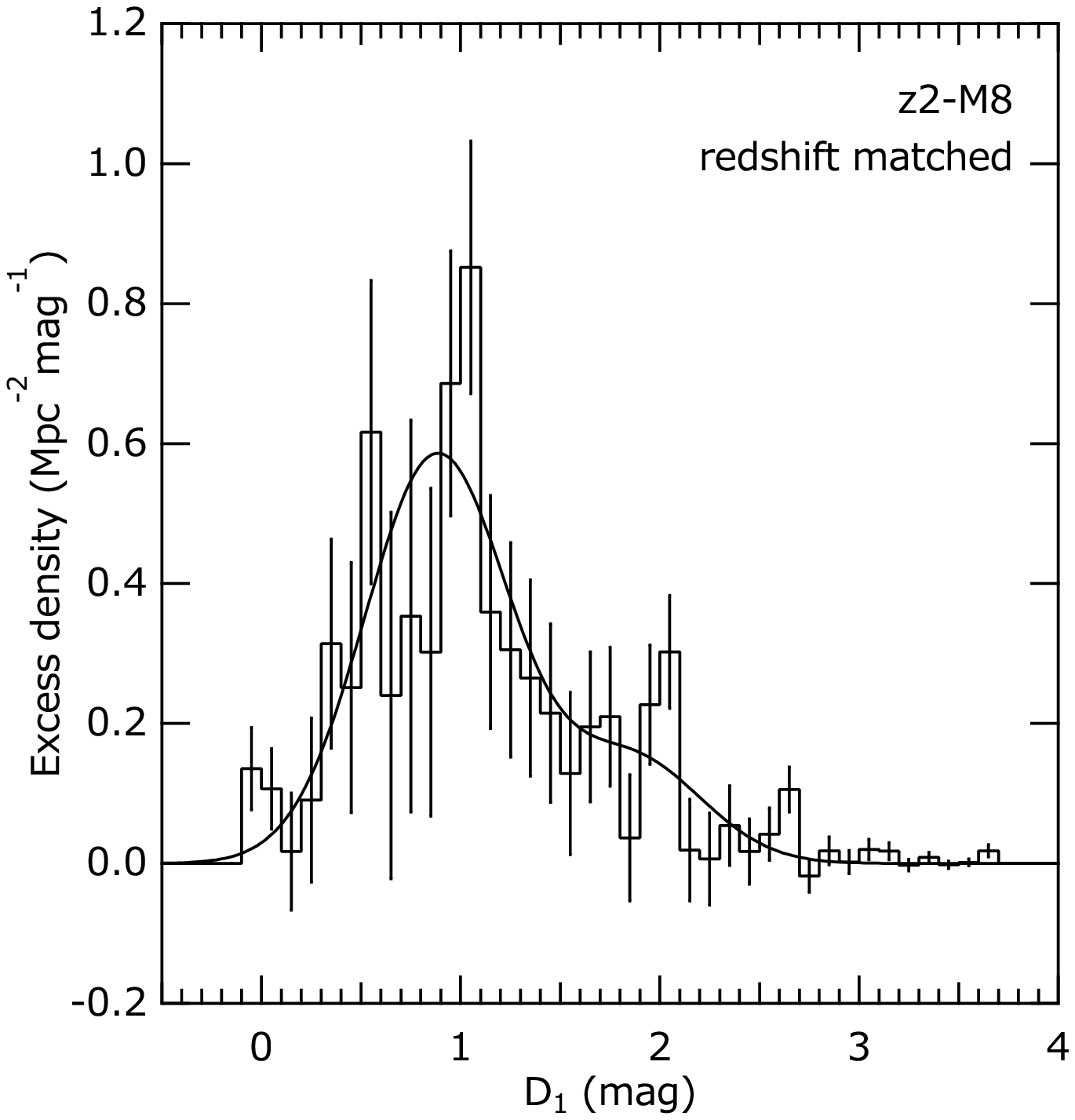} 
    \includegraphics[width=0.32\textwidth]{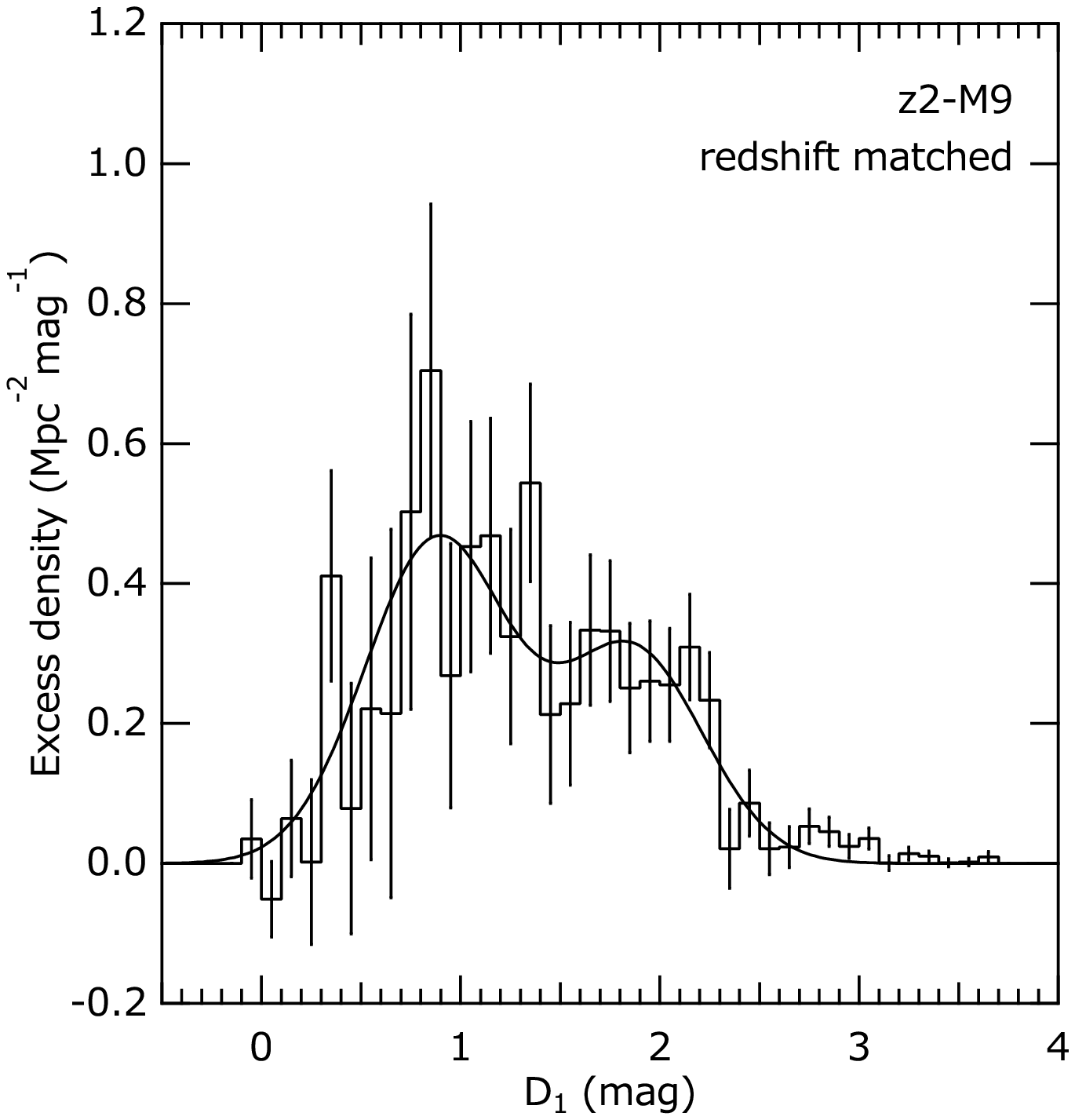} 
  \end{center}
  \caption{Color ($D_{1}$) distributions derived for redshift-matched 
     AGNs and $< M_{50\%}$ galaxy samples 
     calculated as
     shown in equation~(\ref{eq:n_D}). 
     The panels in the left column are the distributions for
     combined mass groups (M8+M9), and on the right of each 
     row are the distributions for individual
     mass groups of M8 and M9.
     The top three panels are those for redshift z0, and the middle and 
     bottom panels are for z1 and z2, respectively.
     The solid lines represent the results of double Gaussian fitting.
     The dashed vertical lines represent the boundary defining blue and
     red galaxies in this paper.}
  \label{fig:hist_D1}
\end{figure}

\begin{figure}
  \begin{center}
    \includegraphics[width=0.32\textwidth]{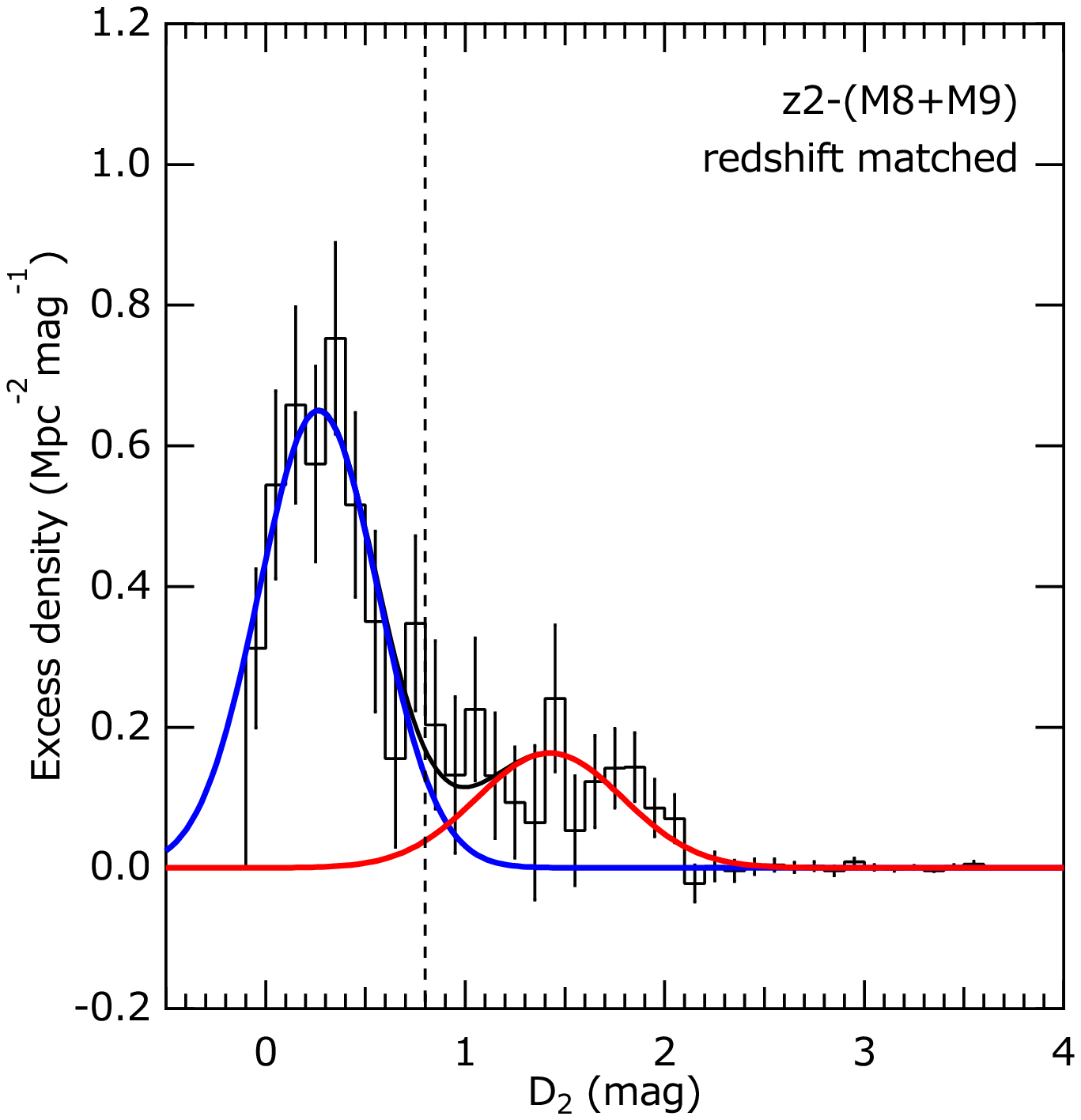} 
    \includegraphics[width=0.32\textwidth]{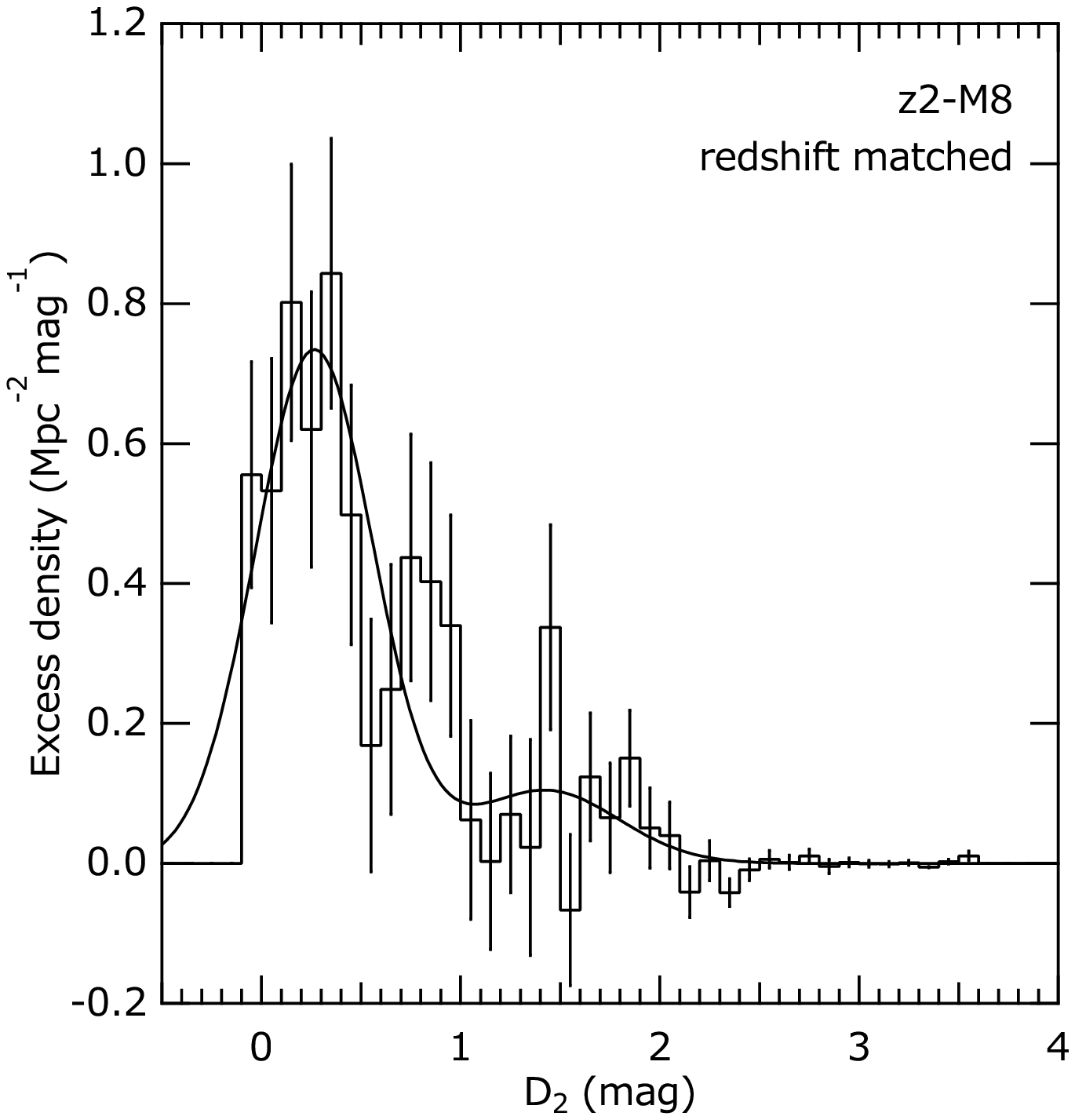} 
    \includegraphics[width=0.32\textwidth]{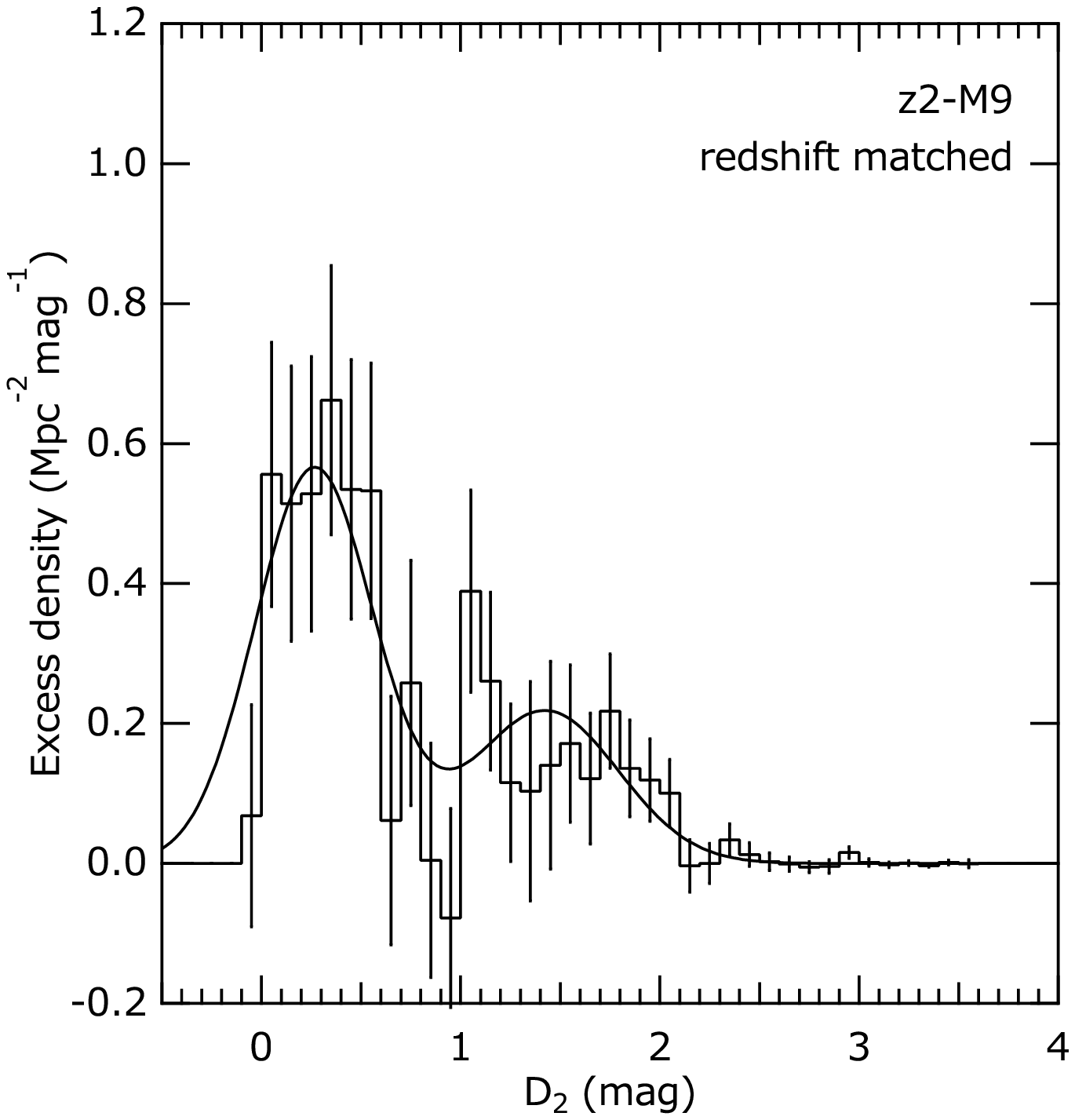} 
    \includegraphics[width=0.32\textwidth]{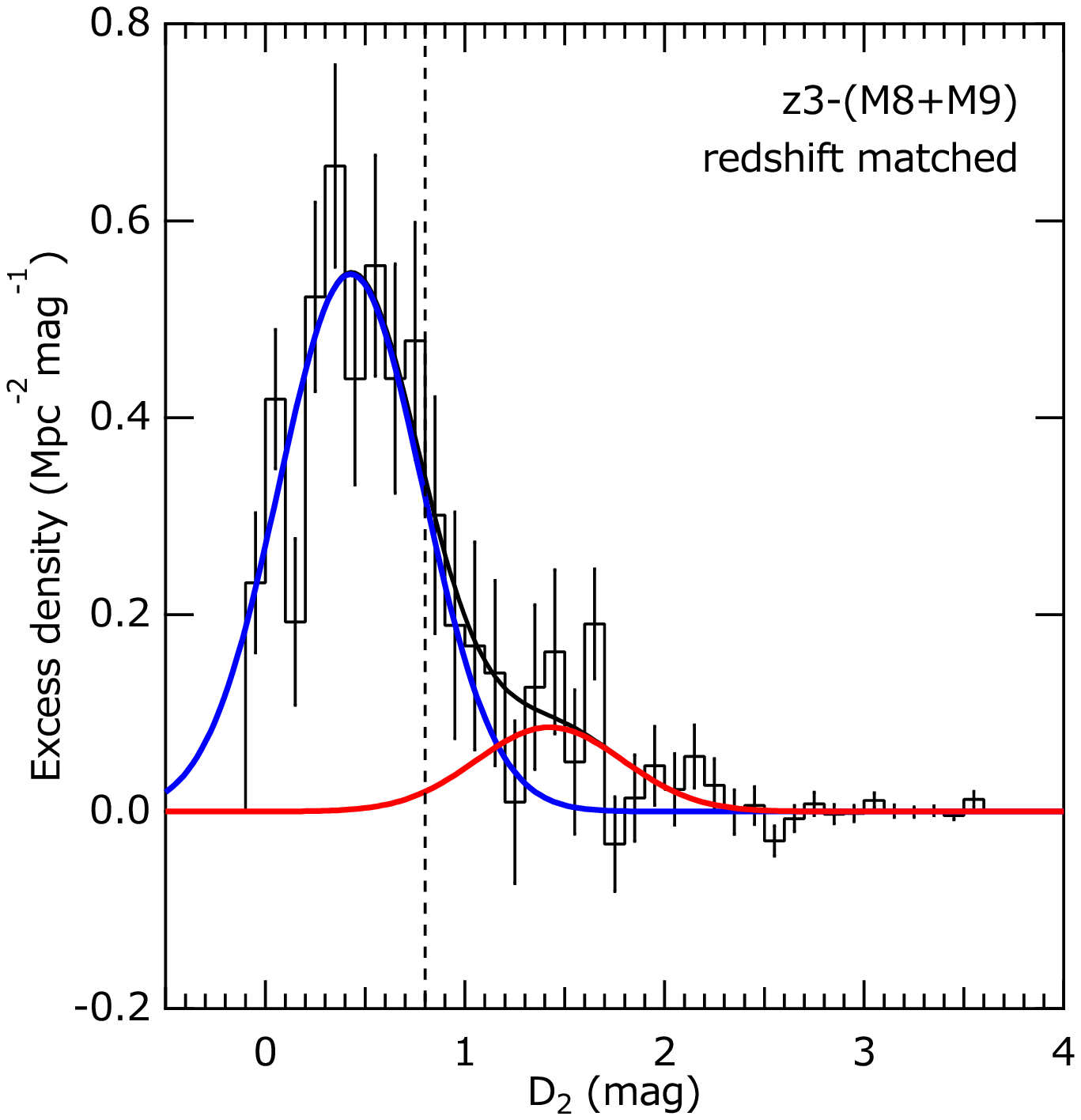} 
    \includegraphics[width=0.32\textwidth]{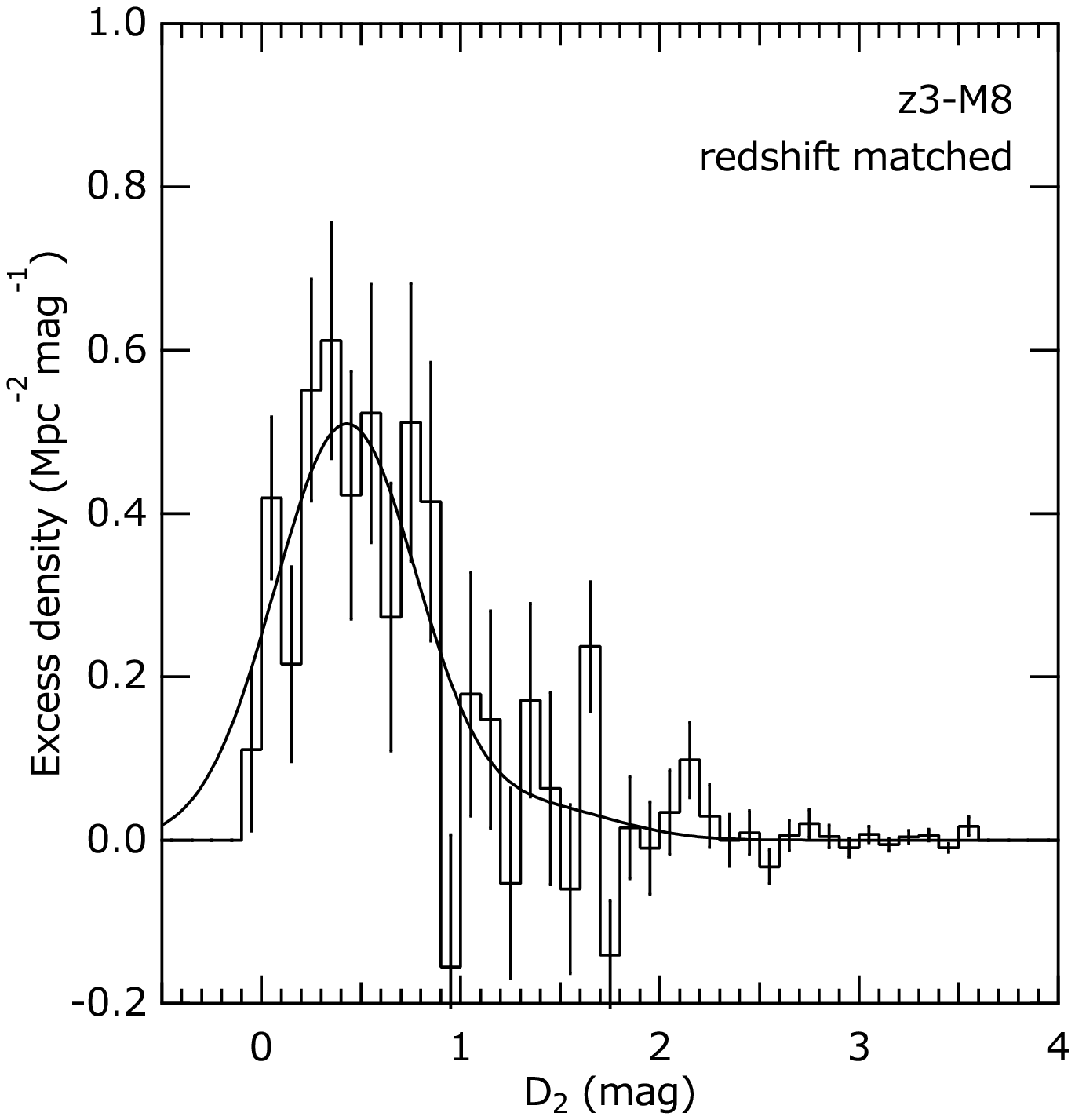} 
    \includegraphics[width=0.32\textwidth]{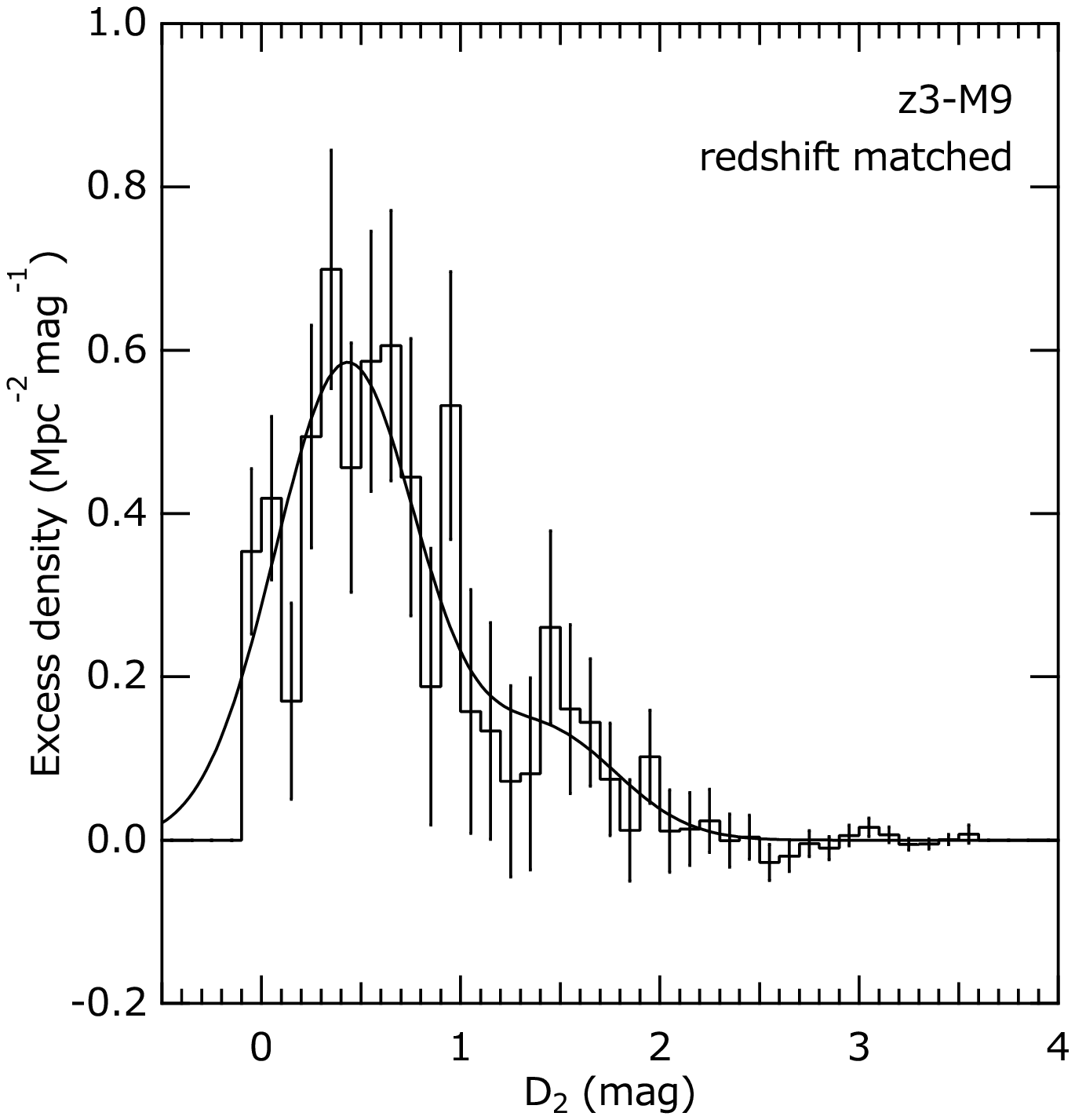} 
  \end{center}
   \includegraphics[width=0.32\textwidth]{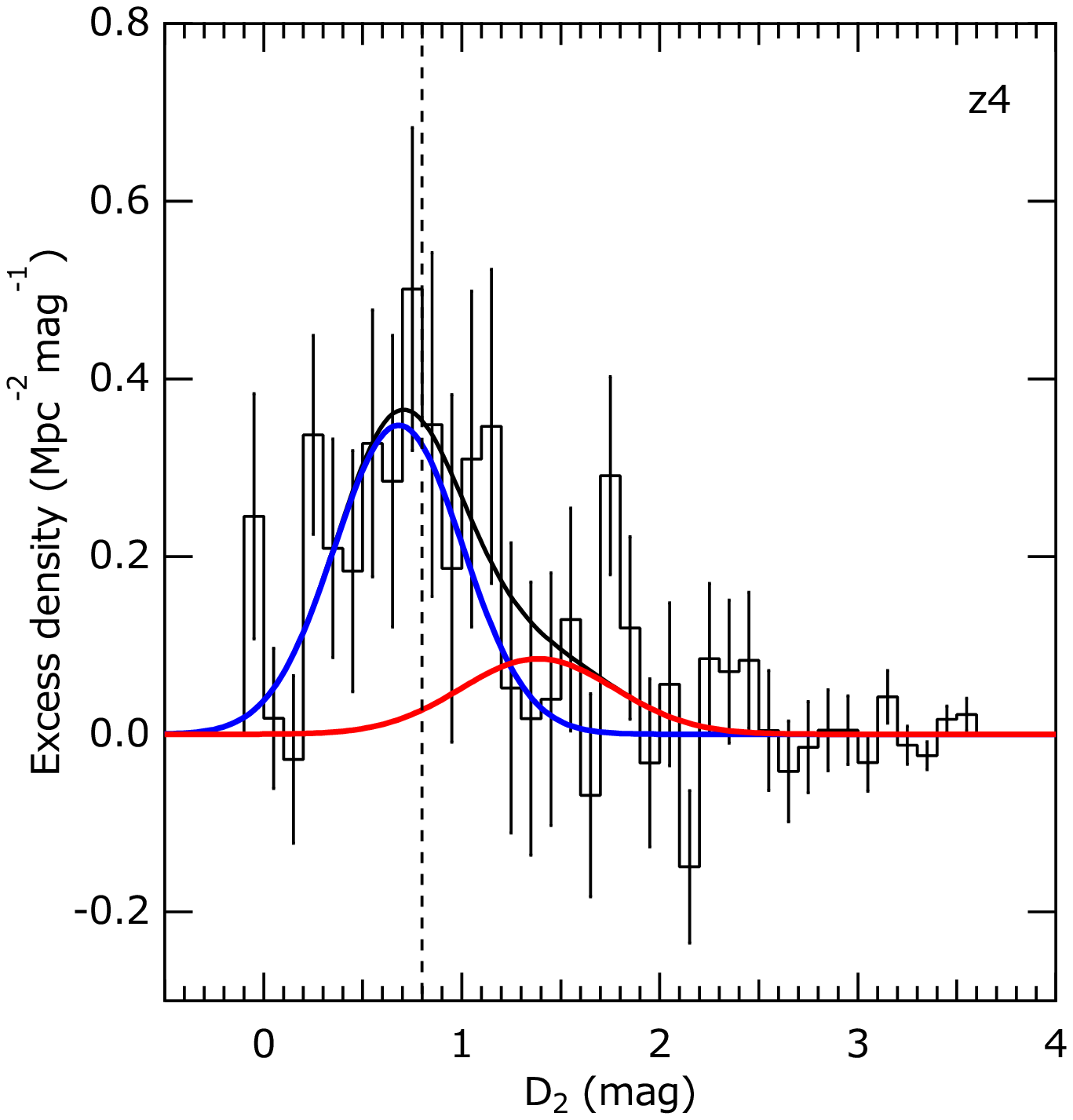} 
  \caption{Color ($D_{2}$) distributions derived for redshift-matched 
     AGNs and $< M_{50\%}$ galaxy samples
     calculated as
     shown in equation~(\ref{eq:n_D}). 
     The panels in the left column are the distributions for
     combined mass groups (M8+M9), and the two panels on the right
     of each row are the distributions for individual
     mass groups of M8 and M9.
     The top three panels are those for redshift z2, and the middle and 
     bottom panels are for z3 and z4, respectively.
     The solid lines represent the results of double Gaussian fitting.
     The dashed vertical lines represent the boundary defining blue and
     red galaxies in this paper.}
  \label{fig:hist_D2}
\end{figure}

Color distributions of galaxies around AGNs were derived by the 
subtraction method described in section~\ref{sec:ana_color},
and the results are shown in Figure~\ref{fig:hist_D1} 
for color parameter $D_{1}$ at z0 to z2 and 
in Figure~\ref{fig:hist_D2} for $D_{2}$ at z2 to z4.
Redshift-matched AGN samples were used for this analysis.
The galaxy samples used are the four-band 
detected
samples, and the threshold magnitude for the galaxy 
was chosen to be $M_{50\%}$, which is an absolute magnitude where
average detection efficiency is 50\%.
The measured values of $M_{50\%}$ are summarized in 
the fourth and sixth columns of Table~\ref{tab:M50_M90}.

The panels at the left end column in the figures show the excess color
density  of $D_{1}$ or $D_{2}$ distributions for combined M8+M9 groups, 
and they are placed in increasing order of their redshifts from top to 
bottom.
The two panels on the right hand side of each row are those for M8 and M9 
groups at the corresponding redshift, respectively.
In the case of the z4 group, only the distribution of the 
whole mass
group is shown, as the number of samples is too small to split the 
sample into two mass groups for the analysis of mass dependence.
$D_{1}$ is defined as a difference between magnitudes at rest frame
wavelengths of 270~nm and 380~nm for redshift z0, z1 and z2 groups,
and $D_{2}$ is a color between magnitudes at 165~nm and 270~nm for
redshift groups of z2, z3 and z4.
The magnitudes were calculated by performing SED fitting using the 
EAZY software \citet{Brammer+08} as described in section~\ref{sec:ana_color}.
The color distributions were fitted with a double Gaussian model in 
which the two Gaussian functions correspond to red (smaller $D_{1}$) 
and blue (larger $D_{1}$) galaxy types.
The fitted model functions are plotted in the same figures.
As can be seen from the figure, blue galaxies dominate the clustered
component in our galaxy samples.
In the fit to the $D_{1}$ distribution for z2-(M8+M9) group shown in
Figure~\ref{fig:hist_D1}, the Gaussian 
parameters for each component were not well constrained.
This is partly because there is a tendency for the distribution of the 
blue component to shift to the redder side as the redshift increases, and 
the components overlap in higher fractions for that group.
Thus we fixed the Gaussian parameters of the red component to those
obtained for z1-(M8+M9) group.
In the case of z3 and z4 shown in Figure~\ref{fig:hist_D2}, the red 
component is barely visible, thus the Gaussian parameters for the red component 
were fixed to those obtained for $D_{2}$ distribution of z2-(M8+M9).

The fitting for the M8 and M9 samples were performed by fixing the 
Gaussian parameters to those obtained for the combined M8+M9 samples
except for the z0 group.
In the case of z0, the peak positions of the blue component are slightly
different between the M8 and M9 groups.
Thus, the Gaussian parameters
for the blue component were derived independently.
Those fitting parameters are summarized in Table~\ref{tab:fit_D}.

The fractions of the blue galaxy for the combined mass groups are
around 0.7--1.0, and there is a marginal indication that the blue 
fraction increases at higher redshifts.
Care needs to be taken in comparison of the different
redshifts by noting the difference in the 
galaxy samples.
The galaxy samples at higher redshift are biased to more luminous
galaxies, and it is also expected that the fraction of red galaxies
increases along with the increase in luminosity as observed in the 
color-magnitude diagrams of the local galaxies.
Thus the observed redshift dependence of the blue fraction is 
biased to the red component at higher redshift, and it is expected
that the increase in the blue fraction is more or less enhanced
if the bias is corrected.

\begin{figure}
  \begin{center}
    \includegraphics[width=0.6\textwidth]{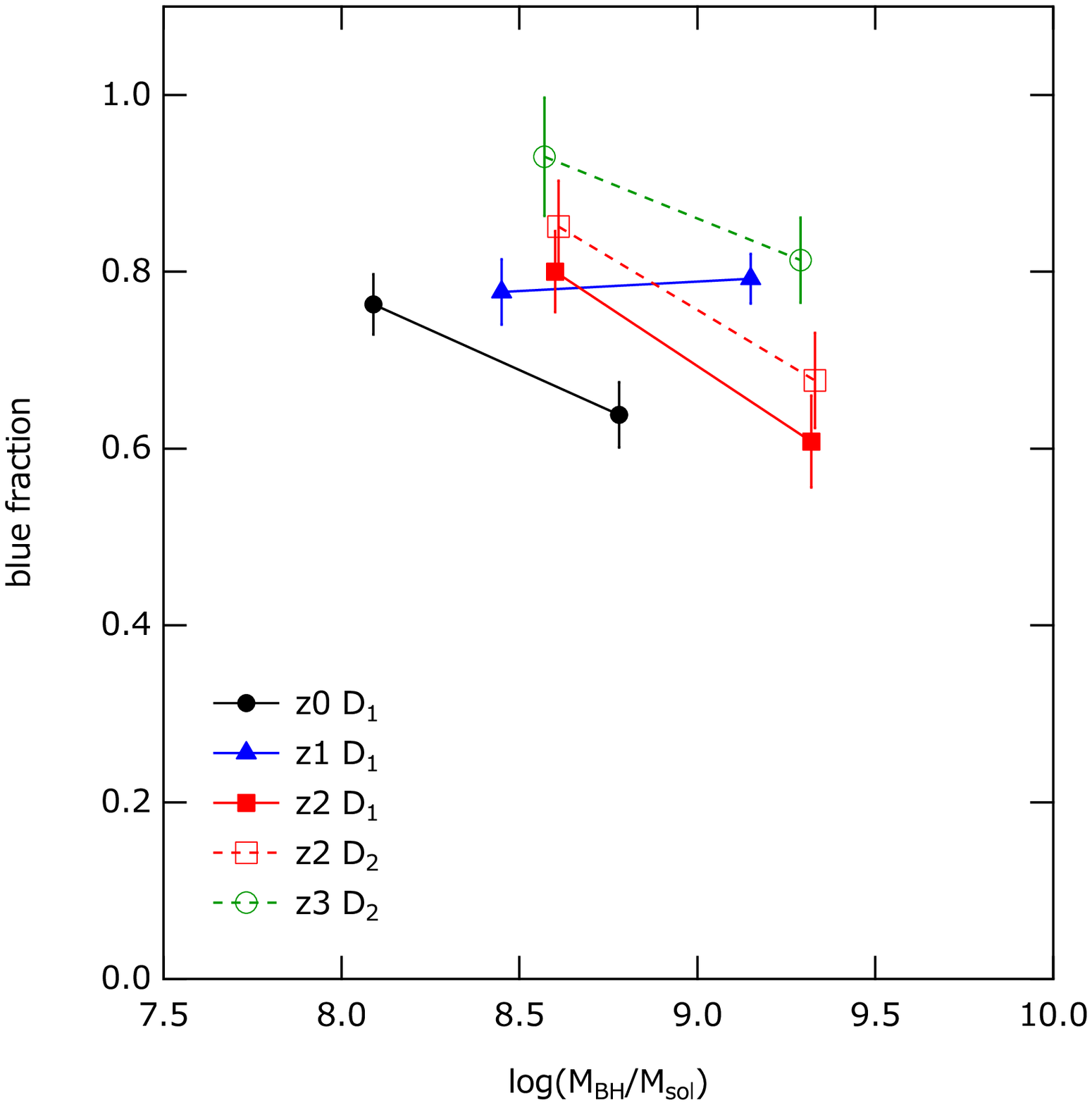} 
  \end{center}
  \caption{Blue galaxy fraction 
at the central region ($<$ 2~Mpc from AGNs)
as a function of BH mass.
These were derived from the fittings shown in Figure~\ref{fig:hist_D1} and 
\ref{fig:hist_D2}.
Galaxy samples 
brighter than $M_{50\%}$ were used.}
  \label{fig:fb_z}
\end{figure}

As already mentioned above, the peak position of the blue 
galaxy is found to shift to the redder side as redshift increases.
We also found that the peak position shifts to the redder side for
brighter galaxies of the same redshift, so the shift of the peak 
position to the redder side at higher redshift is partly due 
to the dependence of the color on the absolute magnitude of 
the galaxy.
The obtained blue galaxy fractions are plotted in Figure~\ref{fig:fb_z}
as a function of BH mass for each redshift group.
A decreasing trend of blue fraction as BH mass is found for all redshifts
except for z1.
The result of z1 might have suffered from the sample variance.
This can be tested by adding more samples from the upcoming 
HSC-SSP survey.

In the previous work of \citet{Shirasaki+16} that used SDSS and UKIDSS 
for galaxy samples, 
the blue galaxy fraction was less than $\sim$~0.2 at redshift group 
z0 of this work, as the galaxy sample was strongly biased to red-type 
galaxies.
In that work they also obtained the result that 
red galaxies more luminous than $M_{\rm IR} = -20$ were strongly clustered
around AGNs of $M_{\rm BH} > 10^{8.2}$~M$_{\solar}$.
To test the consistency with the previous result on the cross-correlation 
with red galaxies, we calculated the cross-correlation lengths for red
galaxies in the z0-M8 and z0-M9 groups separately.

We extracted red galaxies defined as $D_{1} \ge 1.4$,
and restricted their brightness to $M_{\lambda 310} < $ $M_{90\%} (= -18.8)$.
To calculate the cross-correlation function, we should know the average
number density of these galaxy types at the corresponding redshift,
$\bar{\rho}_{\rm 0,red}$.
It was estimated as $\bar{\rho}_{0} (1-f_{\rm blue})$ using 
the blue fraction determined for z0-M8 group.
Then we obtained the cross-correlation lengths for red galaxy samples as
$r_{0,\rm red} = 6.35 \pm 1.22$ for z0-M8, and
$r_{0,\rm red} = 8.73 \pm 1.02$ for z0-M9.
The values of $r_{0,\rm red}$ are plotted in Figure~\ref{fig:r0_MBH} and compared
with the previous work.
They are consistent with each other.

\begin{table}
  \tbl{Fitting result for color distribution.}{%
  \begin{tabular}{ccrccccc}
      \hline
redshift 
 & BH mass
 & $n_{\rm AGN}$
 & $\mu_{\rm blue}$$^{\rm a}$
 & $\sigma_{\rm blue}$$^{\rm b}$
 & $\mu_{\rm red}$$^{\rm c}$ 
 & $\sigma_{\rm red}$$^{\rm d}$
 & $f_{\rm blue}$$^{\rm e}$ \\
 & $\log{(M_{\rm BH}/M_{\solar})}$ & & & & & \\
      \hline
\multicolumn{8}{c}{$D_{1}$, combined mass group}\\
0.6--1.0 (z0) & 7.0--11.0 (M8+M9) &  982 & 0.55$\pm$0.05 & 0.38$\pm$0.05 & 1.59$\pm$0.19 & 0.56$\pm$0.08 & 0.71$\pm$0.09 \\
1.0--1.5 (z1) & 7.0--11.0 (M8+M9) & 1000 & 0.73$\pm$0.02 & 0.21$\pm$0.02 & 1.86$\pm$0.06 & 0.35$\pm$0.04 & 0.78$\pm$0.03 \\
1.5--2.0 (z2) & 7.0--11.0 (M8+M9) & 1016 & 0.88$\pm$0.04 & 0.36$\pm$0.04 & 1.86          & 0.35          & 0.70$\pm$0.04 \\
\multicolumn{8}{c}{$D_{2}$, combined mass group}\\
1.5--2.0 (z2) & 7.0--11.0 (M8+M9) & 1006 & 0.27$\pm$0.04 & 0.30$\pm$0.05 & 1.42$\pm$0.13 & 0.37$\pm$0.08 & 0.76$\pm$0.06 \\
2.0--2.5 (z3) & 7.0--11.0 (M8+M9) & 1162 & 0.43$\pm$0.04 & 0.36$\pm$0.04 & 1.42          & 0.37          & 0.86$\pm$0.05 \\
2.5--3.0 (z4) & 7.0--11.0 &  396 & 0.69$\pm$0.10 & 0.33$\pm$0.09 & 1.42          & 0.37          & 0.79$\pm$0.17 \\
\multicolumn{8}{c}{$D_{1}$, individual mass group}\\
0.6--1.0 (z0) & 7.0-- 8.4 (M8)    &  491 & 0.45$\pm$0.04 & 0.35$\pm$0.04 & 1.59          & 0.56          & 0.76$\pm$0.03 \\
              & 8.4--11.0 (M9)    &  491 & 0.71$\pm$0.04 & 0.32$\pm$0.04 & 1.59          & 0.56          & 0.64$\pm$0.04 \\
1.0--1.5 (z1) & 7.0-- 8.8 (M8)    &  500 & 0.73          & 0.21          & 1.86          & 0.35          & 0.78$\pm$0.04 \\
              & 8.8--11.0 (M9)    &  500 & 0.73          & 0.21          & 1.86          & 0.35          & 0.79$\pm$0.03 \\
1.5--2.0 (z2) & 7.0-- 9.0 (M8)    &  508 & 0.88          & 0.36          & 1.86          & 0.35          & 0.80$\pm$0.05 \\
              & 9.0--11.0 (M9)    &  508 & 0.88          & 0.36          & 1.86          & 0.35          & 0.61$\pm$0.05 \\
\multicolumn{8}{c}{$D_{2}$, individual mass group}\\
1.5--2.0 (z2) & 7.0-- 9.0 (M8)    &  503 & 0.27          & 0.30           & 1.42          & 0.37         & 0.85$\pm$0.05 \\
              & 9.0--11.0 (M9)    &  503 & 0.27          & 0.30           & 1.42          & 0.37         & 0.68$\pm$0.05 \\
2.0--2.5 (z3) & 7.0-- 8.9 (M8)    &  581 & 0.43          & 0.36           & 1.42          & 0.37         & 0.93$\pm$ 0.07 \\
              & 8.9--11.0 (M9)    &  581 & 0.43          & 0.36           & 1.42          & 0.37         & 0.81$\pm$ 0.05 \\
      \hline
  \end{tabular}}\label{tab:fit_D}
  \begin{tabnote}
Redshift-matched AGN datasets and four-band ($griz$) detected galaxy brighter than $M_{50\%}$ 
were used.
$^{\rm a}$mean of the $D_{1}$ or $D_{2}$ distribution for blue component,
$^{\rm b}$standard deviation of the $D_{1}$ or $D_{2}$ distribution for blue component,
$^{\rm c}$mean of the $D_{1}$ or $D_{2}$ distribution for red component,
$^{\rm d}$standard deviation of the $D_{1}$ or $D_{2}$ distribution for red component,
$^{\rm e}$fraction of blue component.
  \end{tabnote}
\end{table}

\subsection{Luminosity functions of galaxies around AGNs}
\label{sec:result_LF}

\begin{figure}
  \begin{center}
    \includegraphics[width=0.48\textwidth]{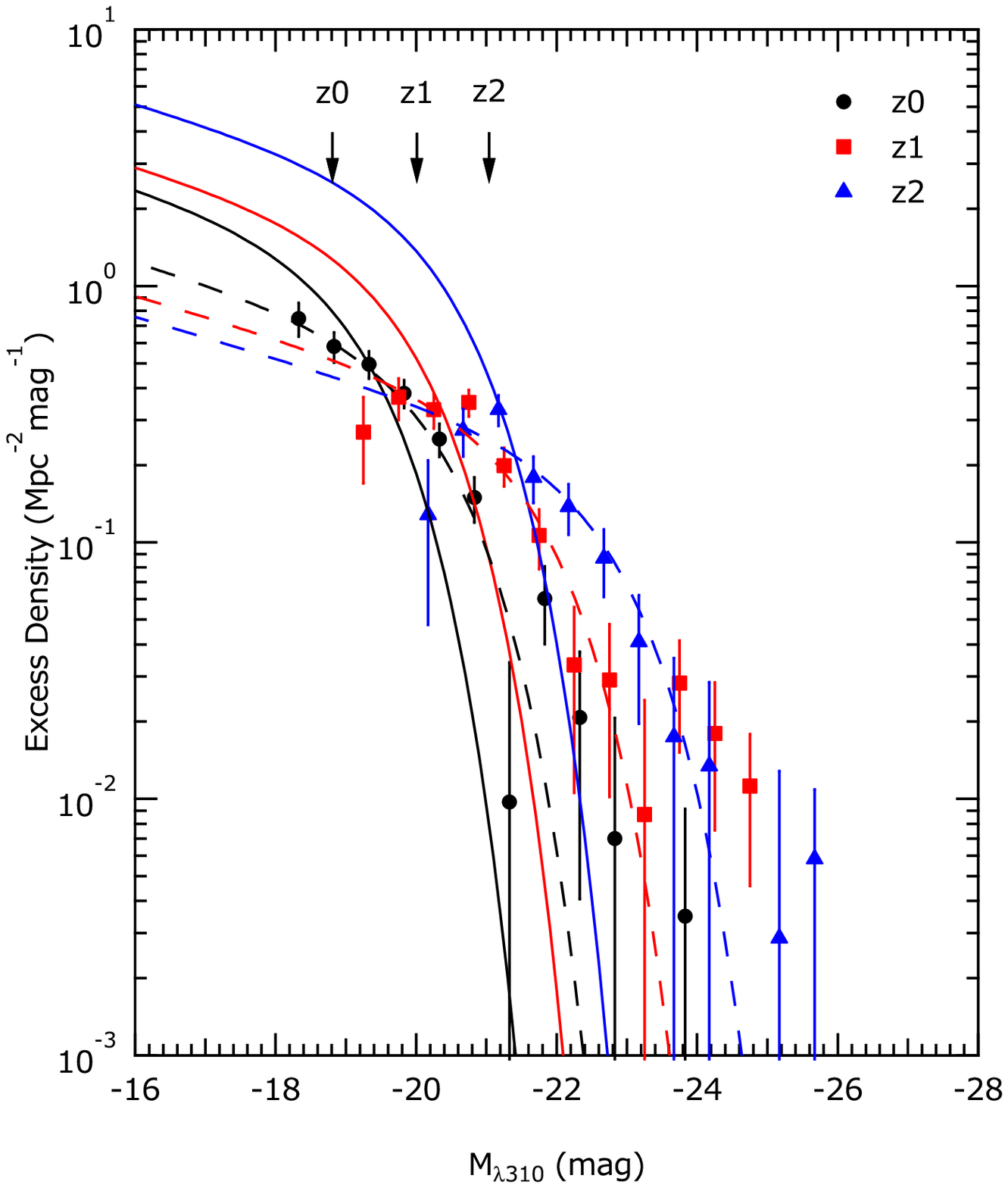} 
    \includegraphics[width=0.48\textwidth]{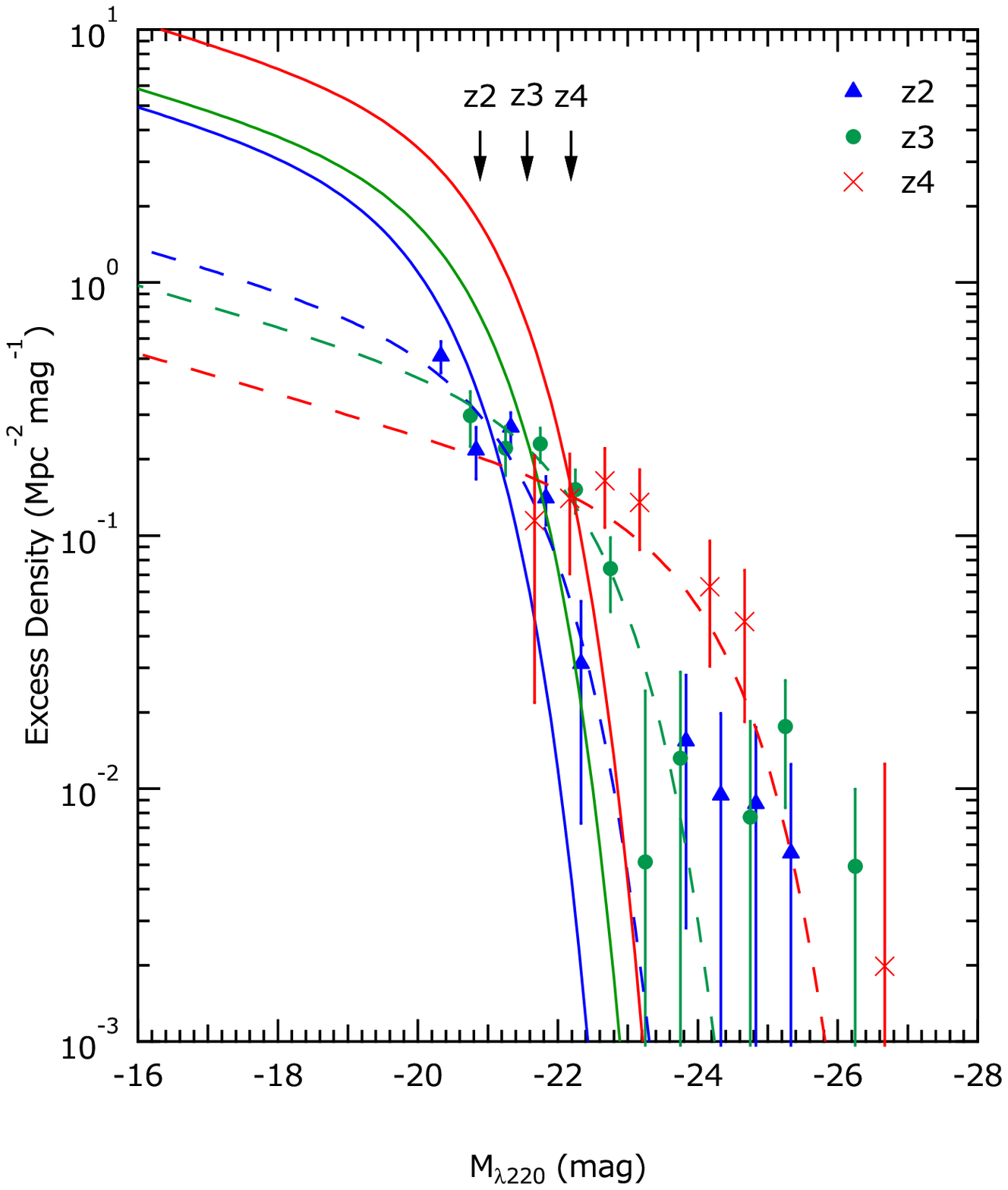} 
  \end{center}
  \caption{Absolute magnitude distributions of galaxies around AGNs
   calculated as in equation~(\ref{eq:n_M}). Detection efficiencies
   were corrected and plotted for $M<M_{50\%}$.
   The left panel shows the distributions 
measured in $M_{\lambda 310}$
for z0, z1 and z2 groups.
   The right panel shows the distributions 
measured in $M_{\lambda 220}$
for z2, z3 and z4 groups.
   The solid lines represent luminosity function calculated by the
   parametrization used in this work and scaled by a factor 
   calculated from the cross-correlation length obtained for $M_{50\%}$
   galaxy samples.
   The scaling factor was calculated as in equation~(\ref{eq:excess_density}).
   The dashed lines represent Schechter functions fitted to the data points.
   The arrows at the top of the panels indicate the 90\% detection limit.
  }
\label{fig:hist_M}
\end{figure}

Next we examined the absolute magnitude distributions of the 
galaxy clustered around AGNs.
The left (right) panel of Figure~\ref{fig:hist_M} shows the 
distributions of $M_{\lambda 310}$ ($M_{\lambda 220}$) for z0, z1 and z2 (z2, z3 and z4) 
redshift groups.
The excess densities were calculated as in equation~(\ref{eq:n_M}) and
corrected for their detection efficiencies calculated as described in
section~\ref{sec:cc_length}.
They are plotted for $M_{\lambda 310}, M_{\lambda 220} < M_{50\%}$.

In the same figure 
luminosity functions scaled by a factor corresponding to the estimated 
cross-correlation length are also plotted with solid lines.
The luminosity function was calculated by using the parameterization 
derived in section~\ref{sec:cc_length}.
The scaling factor was calculated as:
\begin{equation}
\frac{n_{\rm IN} - n_{\rm OUT}}{\overline{\rho}_{0}} = 
\frac{2 C(\gamma) r_{0}^{\gamma}}{3-\gamma}
\left(
   \frac{r_{\rm max,IN}^{3-\gamma} - r_{\rm min,IN}^{3-\gamma}}
        {r_{\rm max,IN}^{2} - r_{\rm min,IN}^{2}}
-  \frac{r_{\rm max,OUT}^{3-\gamma} - r_{\rm min,OUT}^{3-\gamma} }
        {r_{\rm max,OUT}^{2} - r_{\rm min,OUT}^{2}},
\right)
\label{eq:excess_density}
\end{equation}
where $n_{\rm IN}$ ($n_{\rm OUT}$) represents the surface number density
at an inner (outer) region of the AGN field defined by annulus from
$r_{\rm min,IN}$ to $r_{\rm max,IN}$ 
(from $r_{\rm min,OUT}$ to $r_{\rm max,OUT}$), and the right-hand side of the equation
is calculated by integrating the equation~(\ref{eq:model}).
The boundary radii are
$r_{\rm min,IN} = 0.2$,
$r_{\rm max,IN} = 2.0$,
$r_{\rm min,OUT} = 7.0$, and
$r_{\rm max,OUT} = 9.8$.
The arrows at the top of the figure indicate 90\% detection limits.
These results show overdensity against expectations from the luminosity
function at magnitudes brighter than $M_{*}$, which results in larger
clustering of bright galaxies as obtained in section~\ref{sec:result_cc_luminosity}.

\begin{figure}
  \begin{center}
    \includegraphics[width=0.8\textwidth]{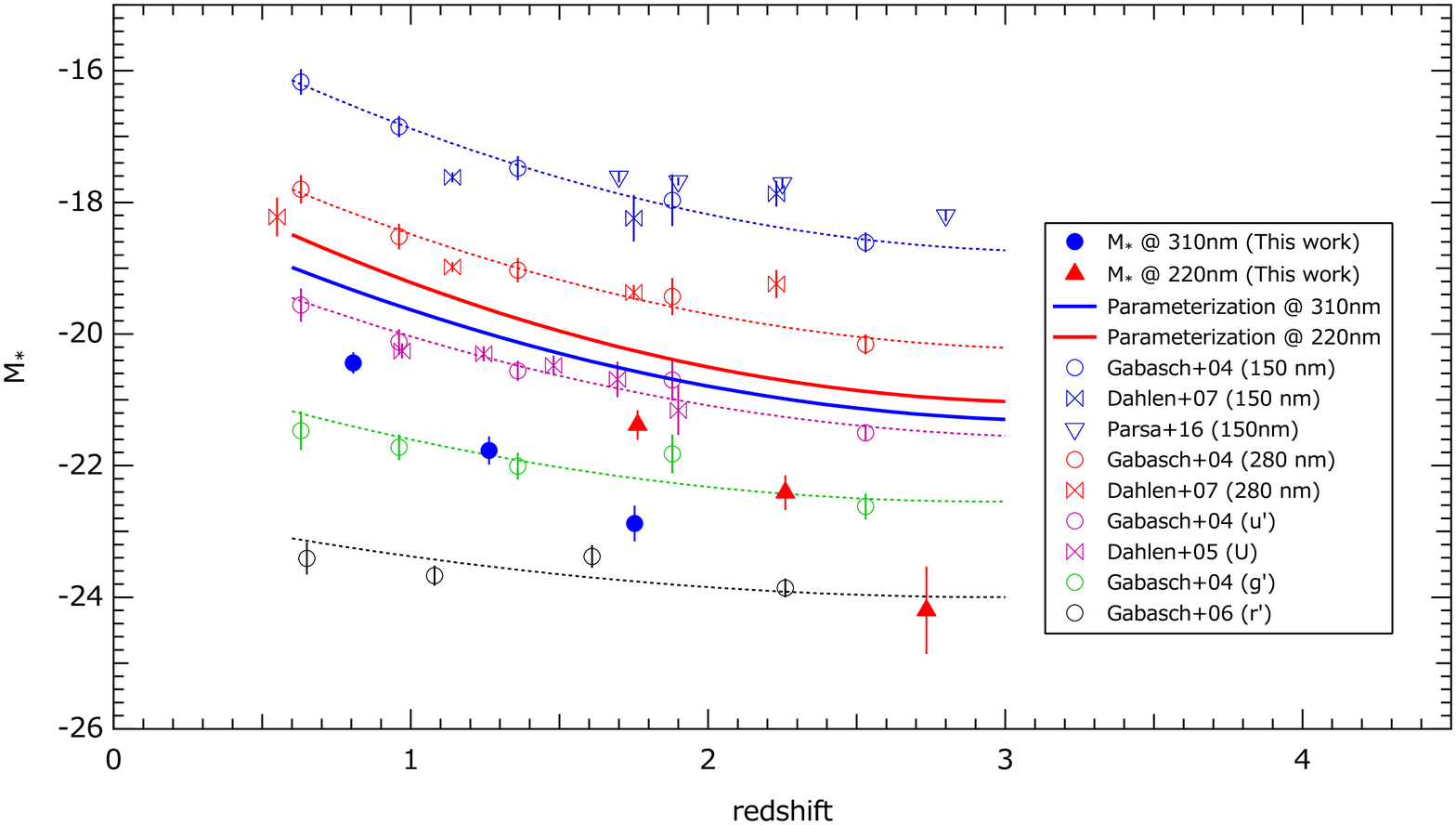} 
  \end{center}
  \caption{Comparison of $M_{*}$ parameters measured for clustering
galaxies around AGNs (filled circles and filled triangles)
and those calculated from the parametrization (thick solid lines).
$M_{*}$ values derived in literature 
\citep{Gabasch+04,Gabasch+06,Dahlen+05,Dahlen+07,Parsa+16}
are also plotted.
 }
  \label{fig:Compare-Mstar}
\end{figure}

To test if the overdensity can be explained by the uncertainty of $M_{*}$
parameter used in the parametrization of luminosity function model,
we measured the $M_{*}$ parameters for the observed magnitude distributions
by fitting the Schechter function to them 
and compared them with the 
values given by the parametrization.
In the fitting $\alpha$ was fixed to $-1.2$.
The results are summarized in Table~\ref{tab:comp_Mstar} and compared
in Figure~\ref{fig:Compare-Mstar} with the parametrization (thick solid lines)
and literature values (open markers).
The obtained 
$M_{*}$
are typically smaller (i.e. brighter) 
than 
$M_{*}$ given by the parameterization
by $>$1~mag, well exceeding the 
uncertainty of 
$M_{*}$, 
0.28~mag, expected for the parametrization.
Thus it is unlikely that the overdensity is attributable only to the 
uncertainty of the assumed luminosity function.

\begin{table}
  \tbl{Comparisons of $M_{*}$ parameters
measured for clustering
galaxies around AGNs and those calculated from the parametrization.
}{%
  \begin{tabular}{cccccc}
\hline
redshift group & average redshift & wavelength 
   & $M_{*}$(clust)$^{a}$ 
   & $M_{*}$(param)$^{b}$
               & $\mathnormal{\Delta}M_{*}^{c}$ \\
 & & nm & mag & mag & mag \\
\hline
z0 & 0.81 & 310 & -20.4$\pm$0.2 & -19.3 & 1.1 \\
z1 & 1.26 & 310 & -21.8$\pm$0.2 & -20.0 & 1.8 \\
z2 & 1.75 & 310 & -22.9$\pm$0.3 & -20.6 & 2.3 \\
z2 & 1.76 & 220 & -21.4$\pm$0.2 & -20.3 & 1.1 \\
z3 & 2.26 & 220 & -22.4$\pm$0.3 & -20.7 & 1.7 \\
z4 & 2.74 & 220 & -24.2$\pm$0.7 & -21.0 & 3.2 \\
\hline
  \end{tabular}}
  \label{tab:comp_Mstar}
  \begin{tabnote}
$^{a}$ $M_{*}$ parameters measured for clustering galaxies around AGN,
$^{b}$ $M_{*}$ parameters calculated from the parametrization as described
in section~\ref{sec:cc_length},
$^{c}$ $M_{*}$(param) $-$ $M_{*}$(clust).
  \end{tabnote}
\end{table}

\begin{figure}
  \begin{center}
    \includegraphics[width=0.32\textwidth]{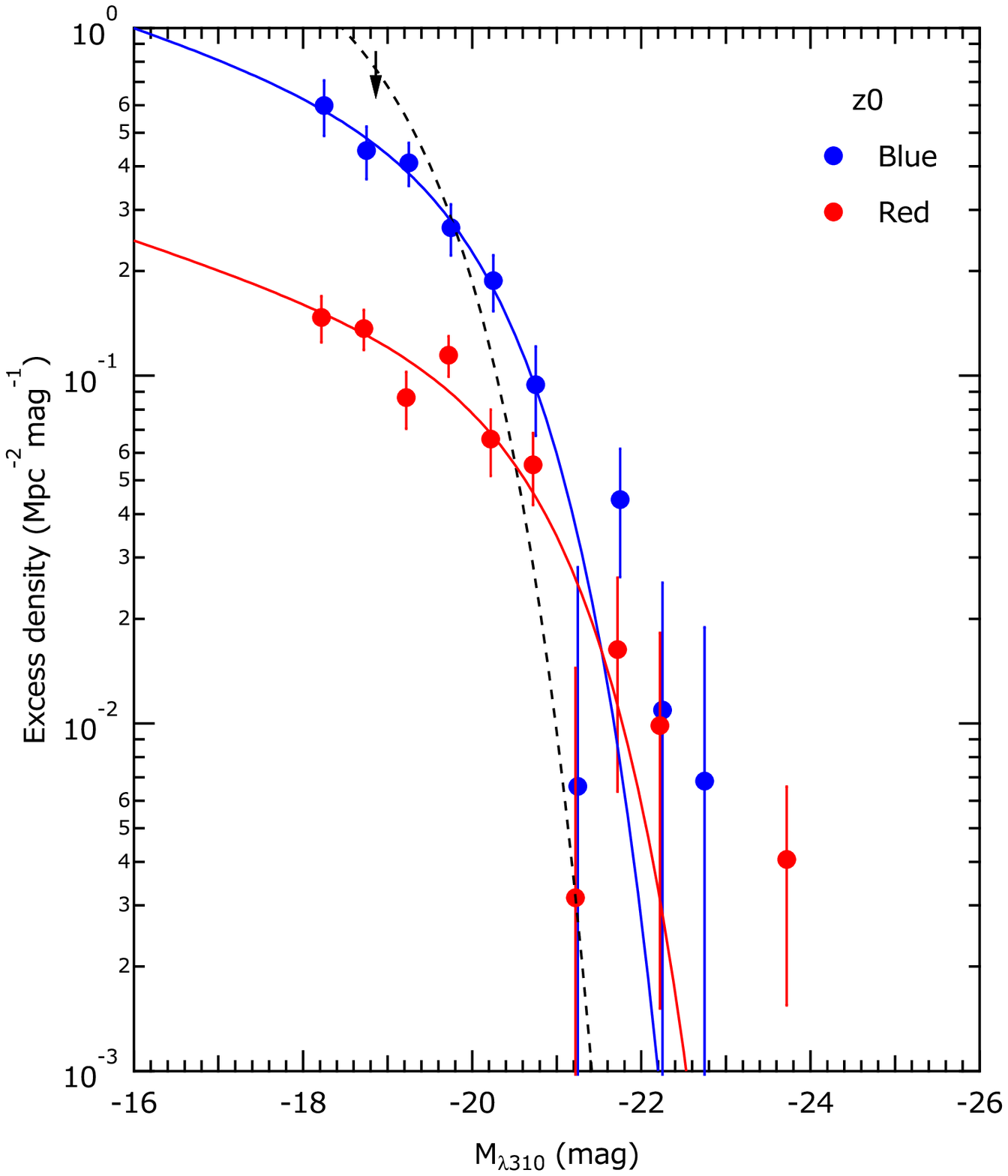} 
    \includegraphics[width=0.32\textwidth]{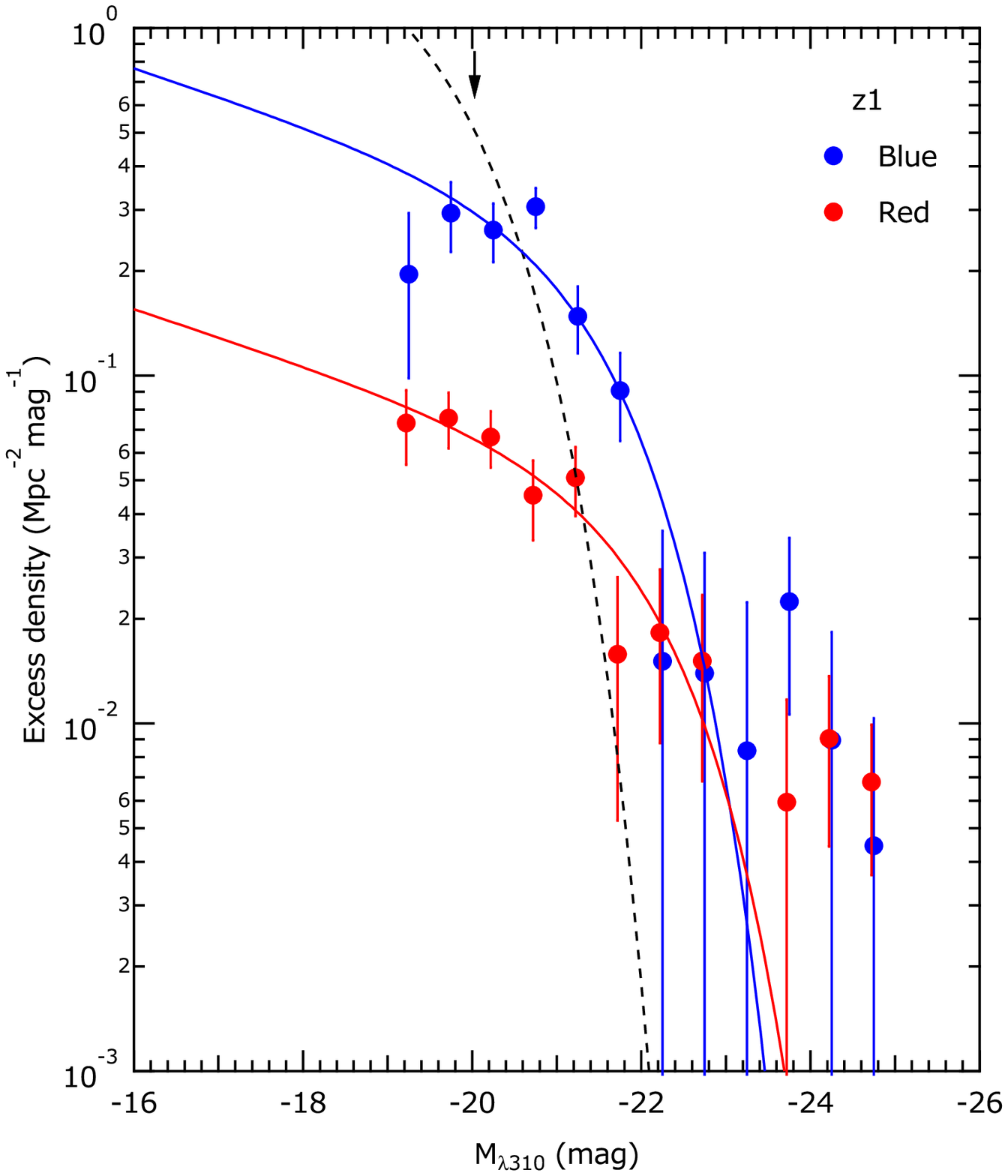} 
    \includegraphics[width=0.32\textwidth]{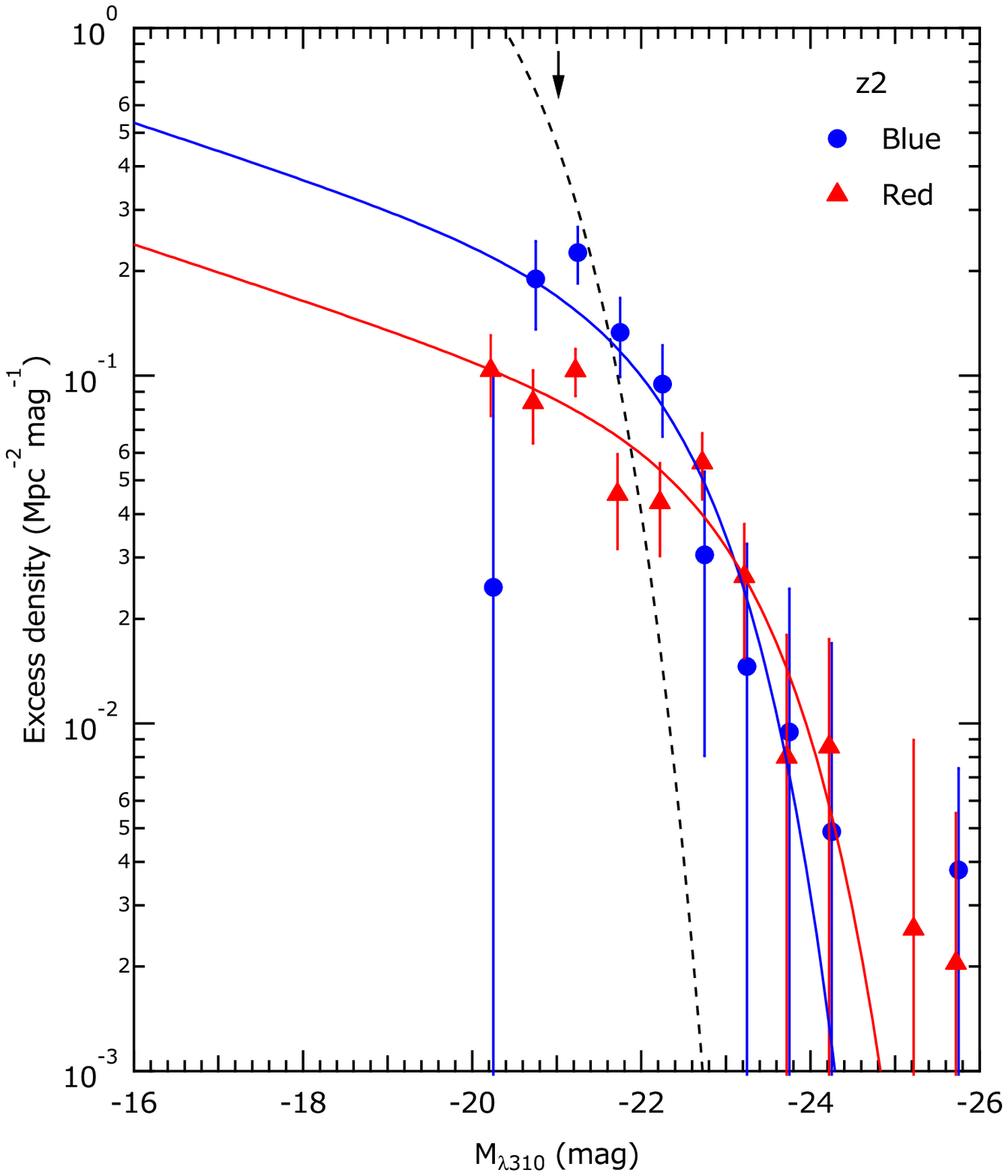} 
    \includegraphics[width=0.32\textwidth]{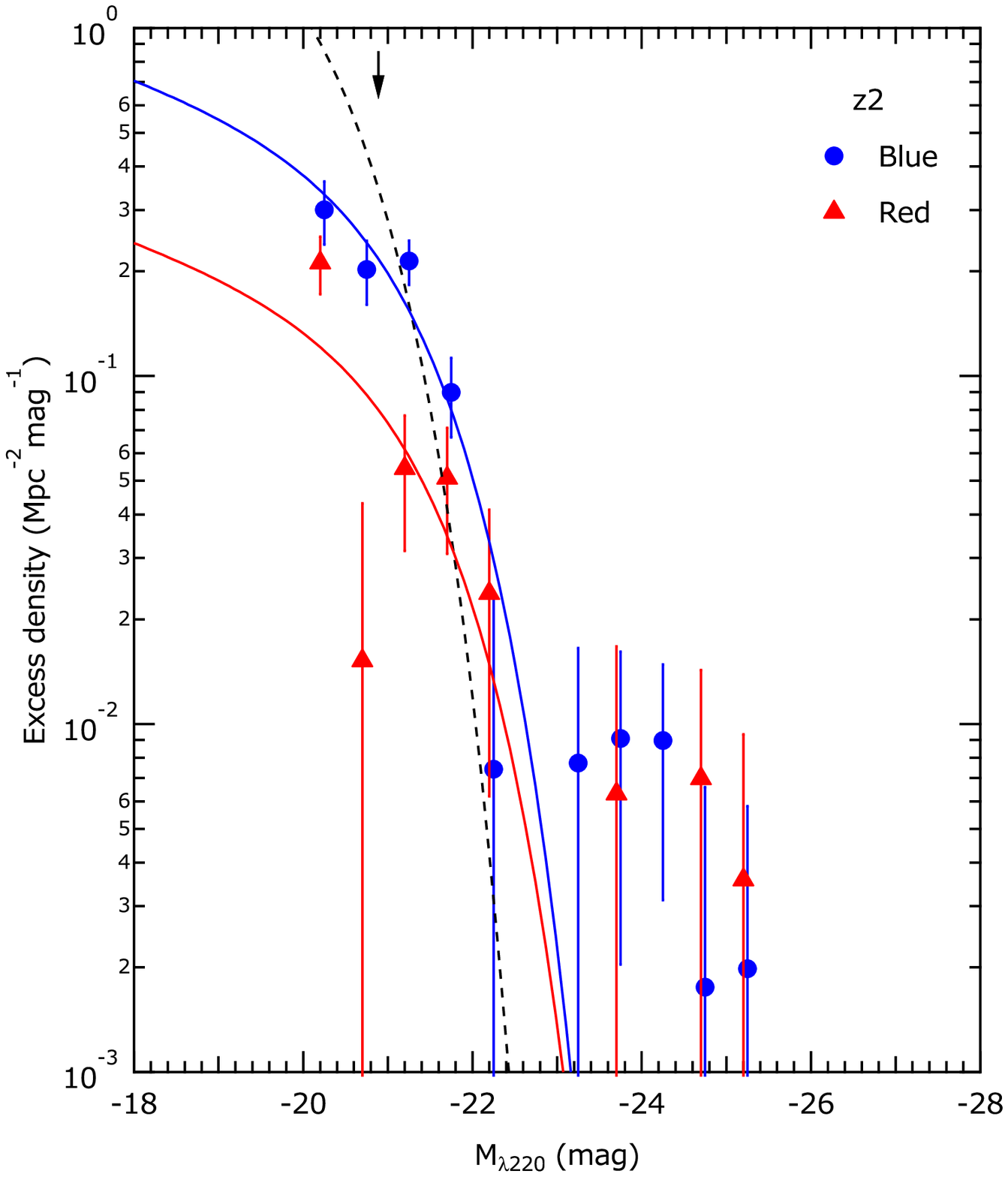} 
    \includegraphics[width=0.32\textwidth]{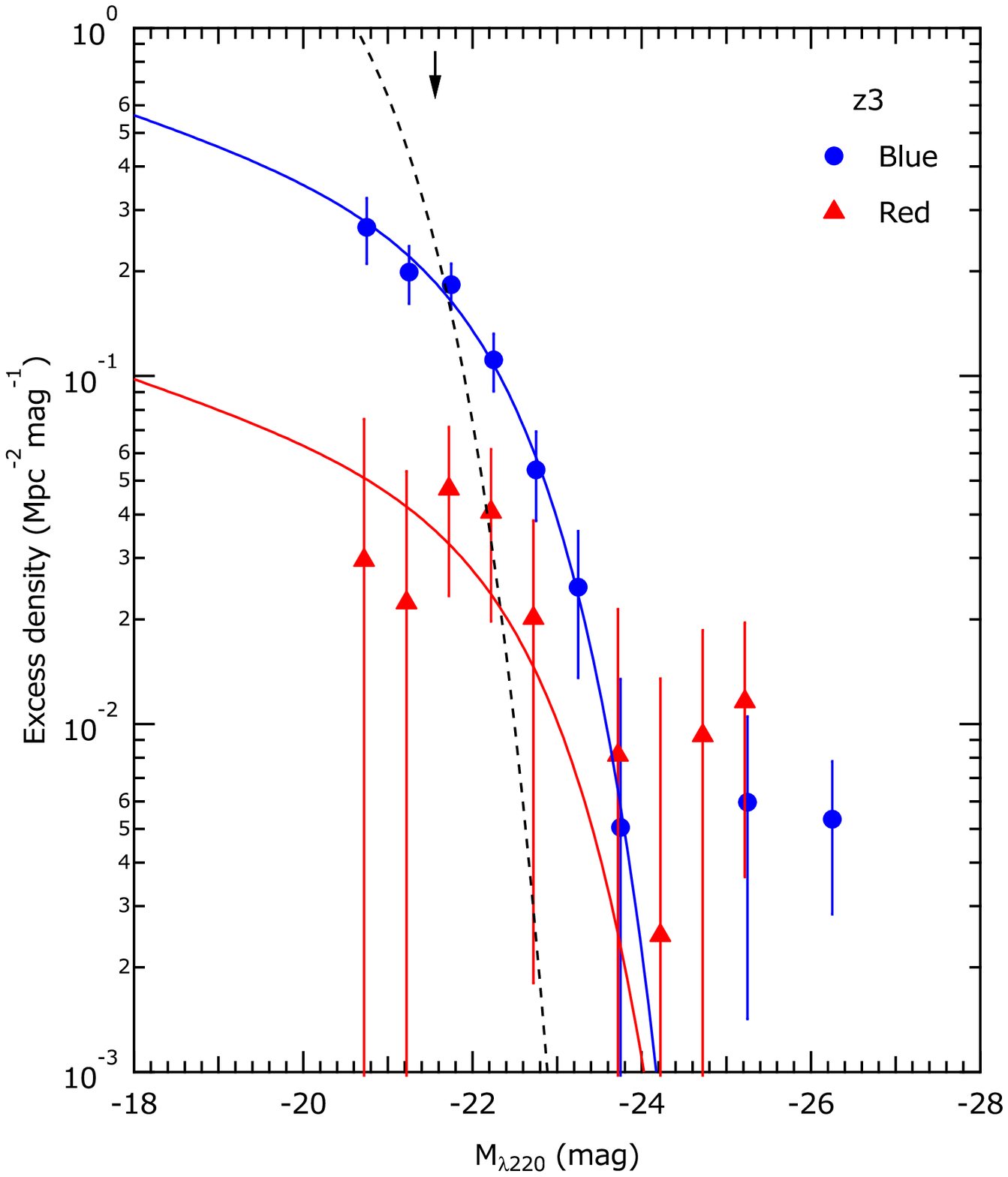} 
    \includegraphics[width=0.32\textwidth]{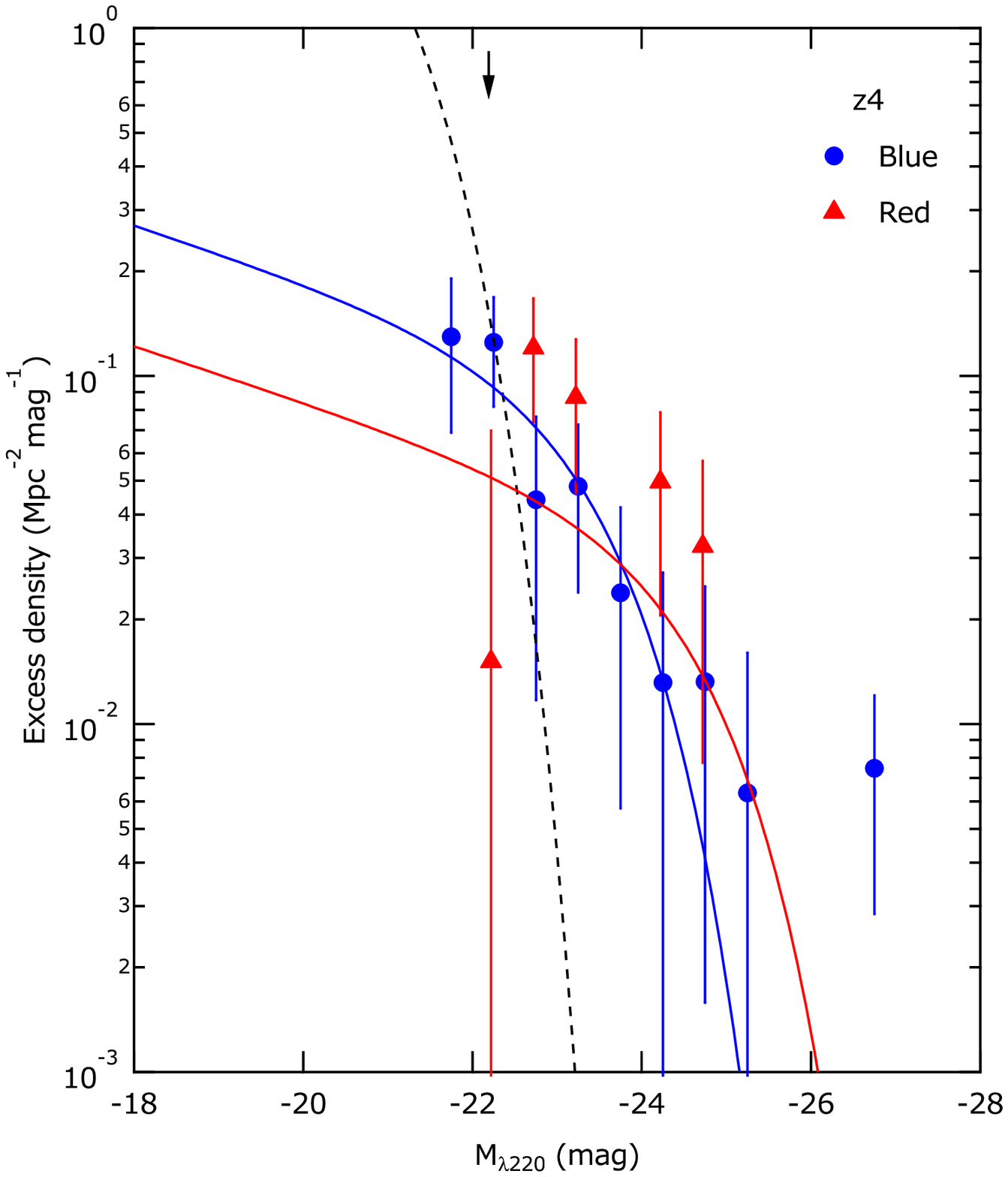} 
  \end{center}
  \caption{Absolute magnitude distributions for blue and red galaxies
  calculated as in equation~(\ref{eq:n_M}).
   The dashed lines represent luminosity function calculated by the
   parametrization used in this work and scaled by a factor 
   calculated from the cross-correlation length.
 The scaling factor was calculated as in equation~(\ref{eq:excess_density}).
   The solid lines represent Schechter functions fitted to the data points.
   The arrows at the top of the panels indicate the 90\% detection limit.}
   \label{fig:hist_M_BR}
\end{figure}

To investigate the galaxy type that contributes to the overdensity
at the bright end, we derived the magnitude distributions for blue and
red galaxies separately.
They were classified at $D_{1} = 1.4$ or $D_{2} = 0.8$.
Figure~\ref{fig:hist_M_BR} shows the comparisons between the magnitude 
distributions for blue and red galaxies.
Both galaxy types contribute to the overdensity.

\begin{figure}
  \begin{center}
    \includegraphics[width=0.6\textwidth]{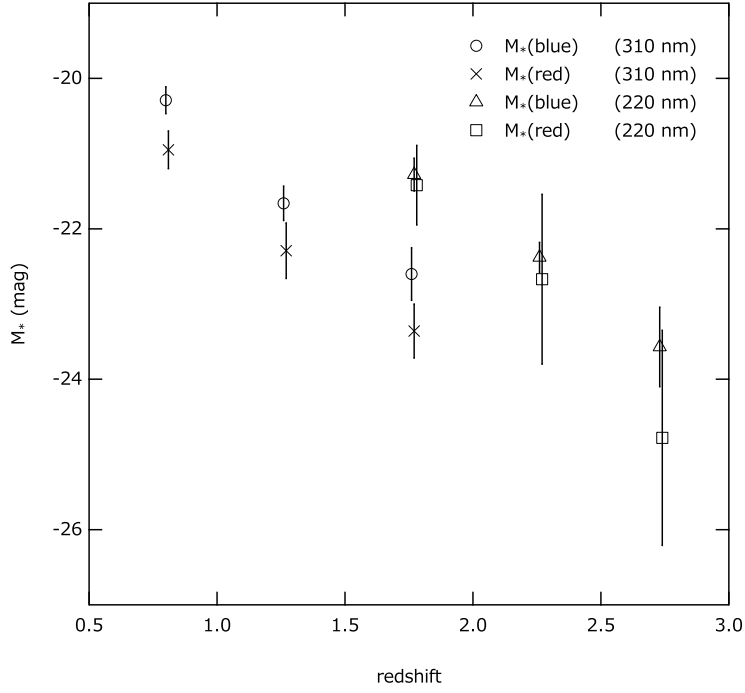} 
  \end{center}
  \caption{Comparison of $M_{*}$ parameters measured for blue and red galaxies.
  These are obtained by fitting Schechter functions to the data points
  in Figure~\ref{fig:hist_M_BR}.
}
  \label{fig:Mstar_BR}
\end{figure}

$M_{*}$ parameters 
measured for the distributions are summarized
in Table~\ref{tab:comp_Mstar_BR} and plotted in Figure~\ref{fig:Mstar_BR}.
The obtained 
$M_{*}$
parameters for red galaxies are systematically
brighter than those for blue galaxies by more than two sigma at redshift z0,
z1, and z2 for measurements at 310~nm.
No statistically significant difference is found for the values for z2, z3, 
and z4 measured at 220~nm, which is mostly due to poor statistics and lower 
resolution of $D_{2}$ parameters on the deconvolution of the two components.

\begin{figure}
  \begin{center}
    \includegraphics[width=0.32\textwidth]{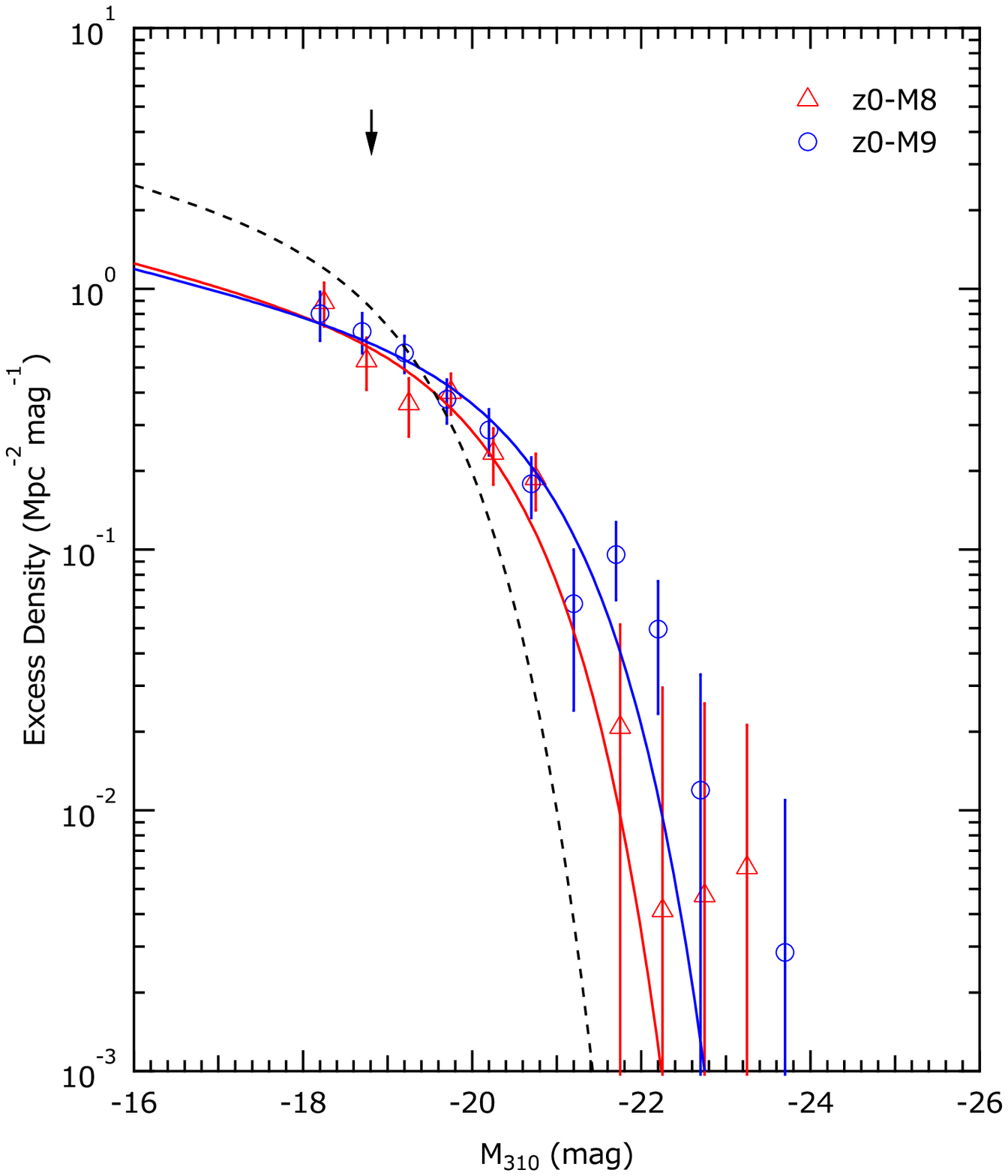} 
    \includegraphics[width=0.32\textwidth]{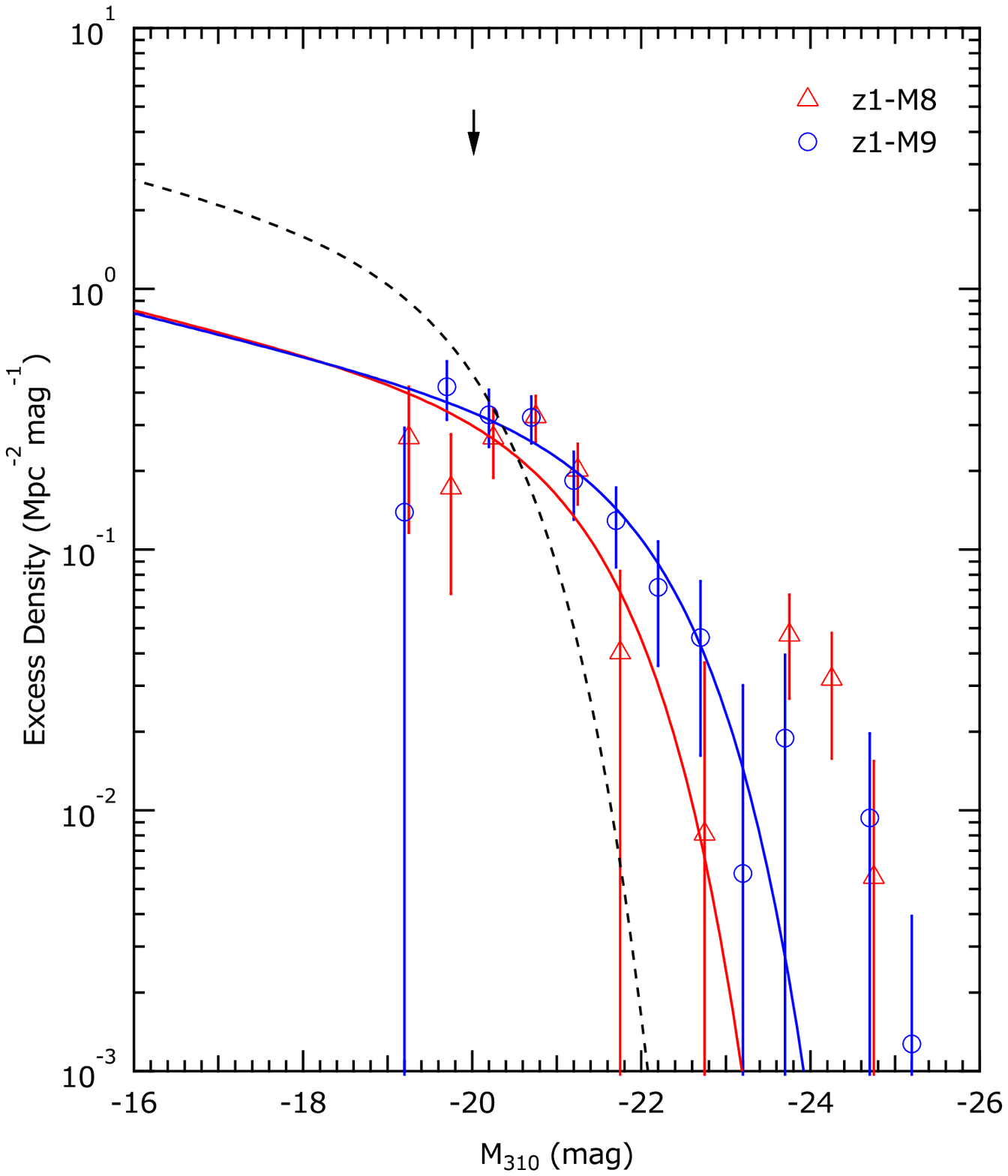} 
    \includegraphics[width=0.32\textwidth]{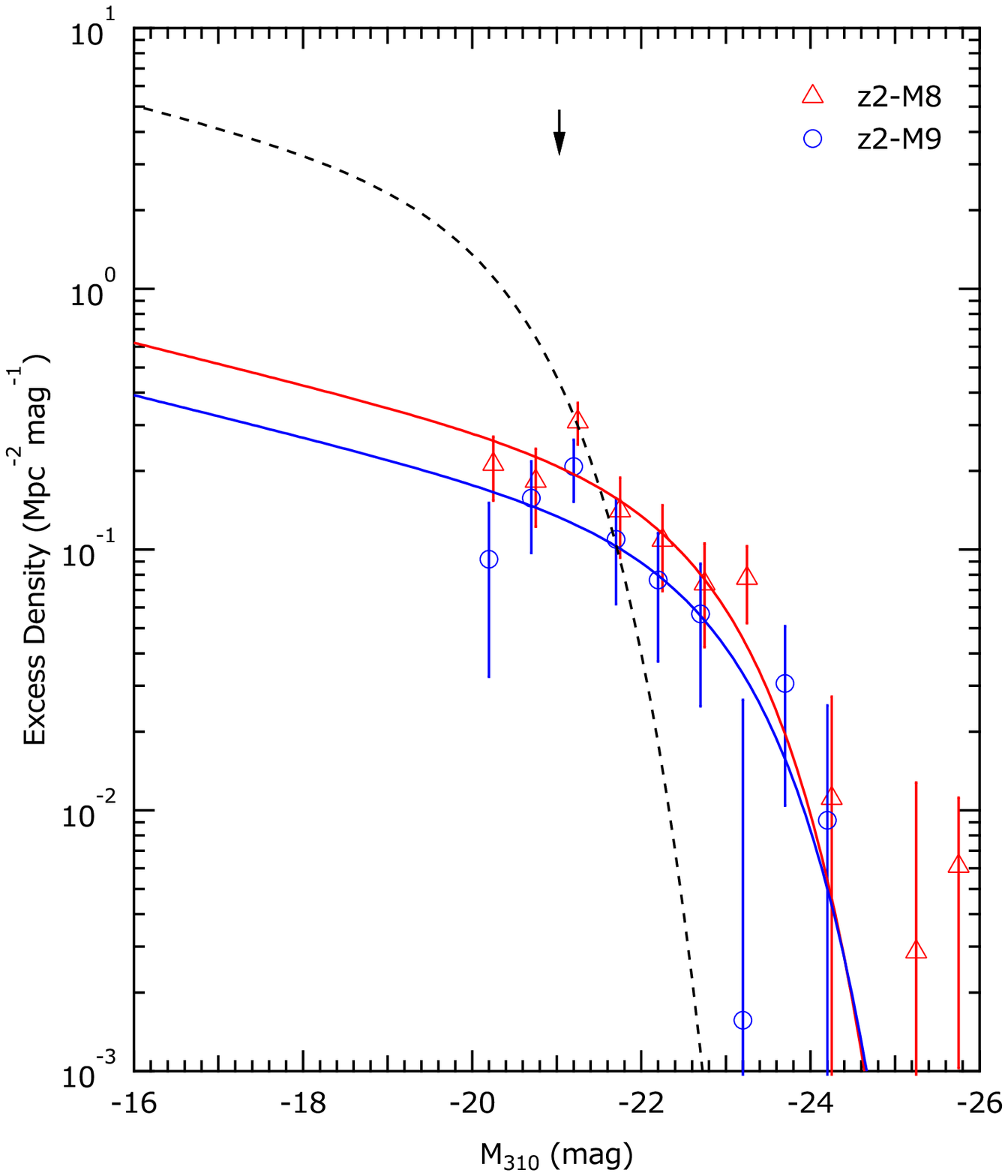} 
    \includegraphics[width=0.32\textwidth]{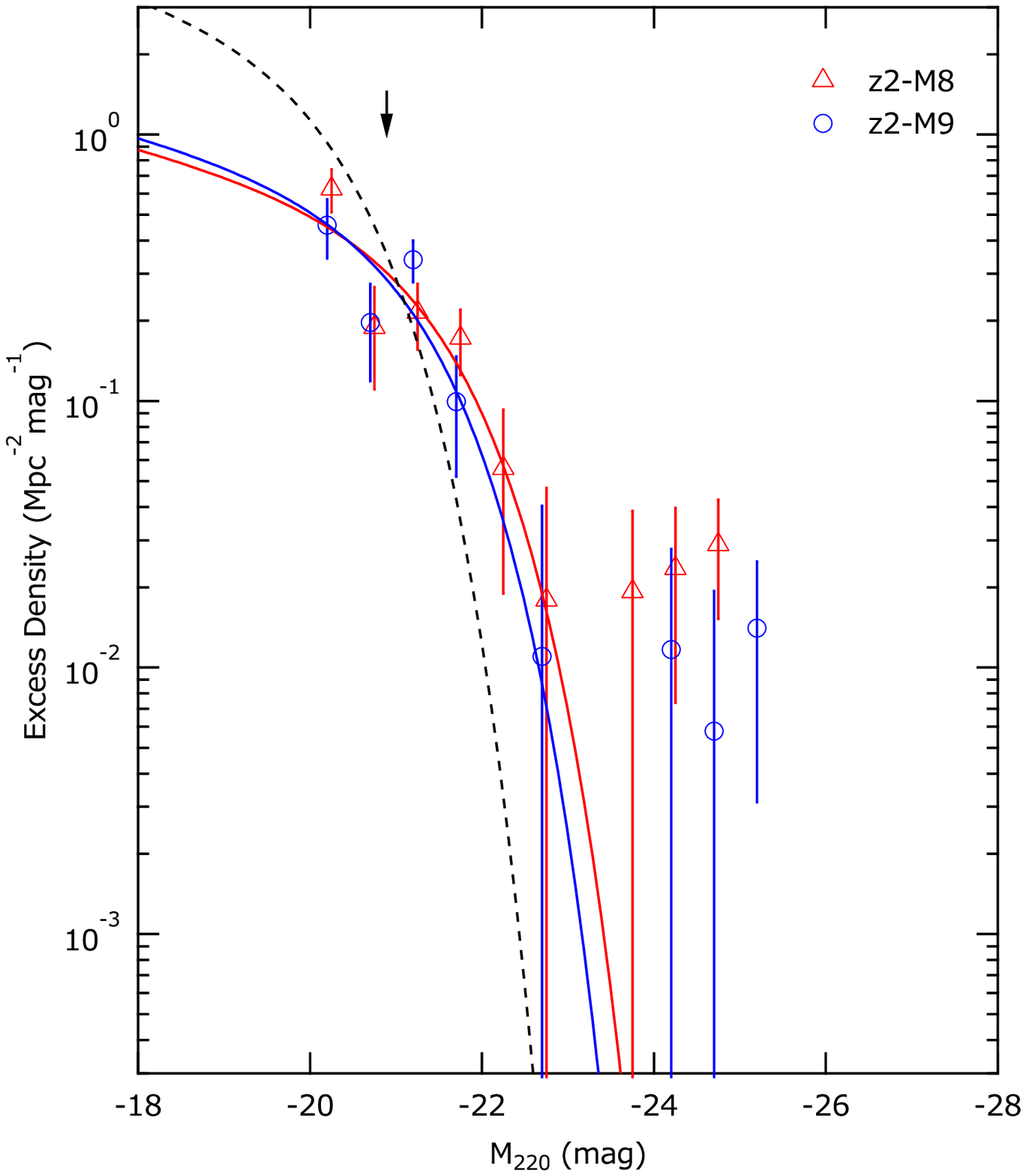} 
    \includegraphics[width=0.32\textwidth]{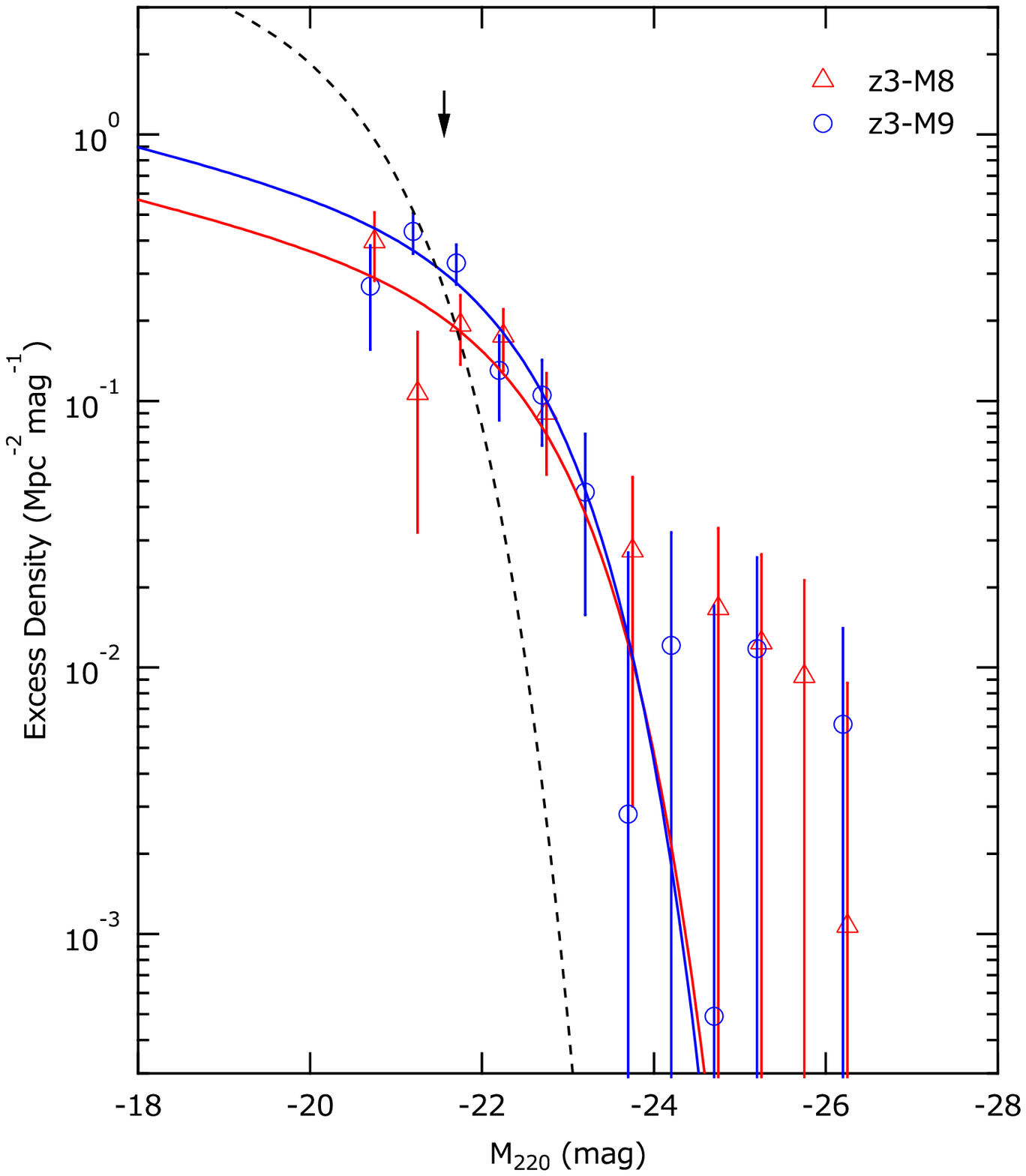} 
  \end{center}
  \caption{Absolute magnitude distributions for M8 and M9 mass groups.
   The dashed lines represent luminosity 
   function calculated by the
   parametrization used in this work and scaled by a factor 
   calculated from the cross-correlation length.
   The scaling factor was calculated as in equation~(\ref{eq:excess_density}).
   The solid lines represent Schechter 
   functions fitted to the data points.
   The arrows at the top of the panels indicates the 90\% detection limit.
   }
   \label{fig:hist_M_M8M9}
\end{figure}

\begin{figure}
  \begin{center}
    \includegraphics[width=0.6\textwidth]{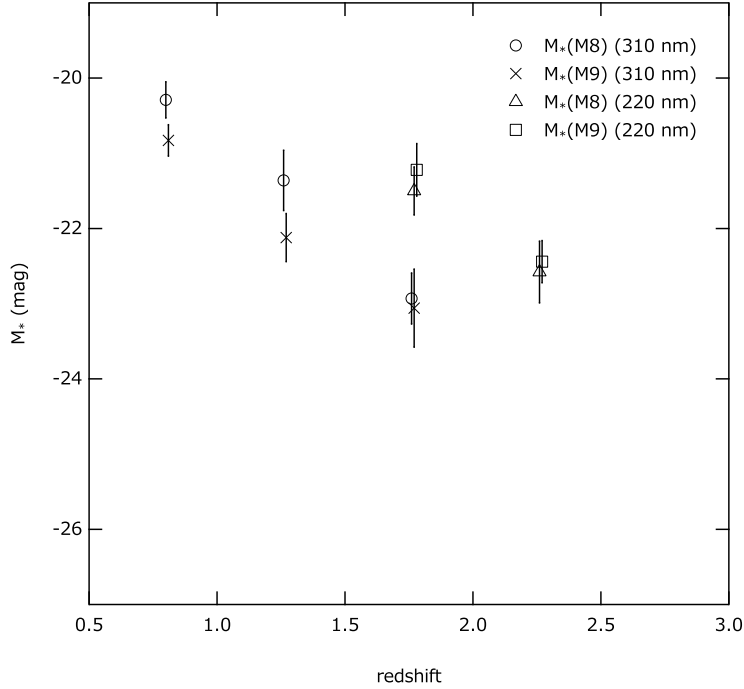} 
  \end{center}
  \caption{Comparison of $M_{*}$ parameters measured for M8 and M9 mass groups.
  These are obtained by fitting Schechter functions to the data points
  in Figure~\ref{fig:hist_M_M8M9}.
}
  \label{fig:Mstar_M89}
\end{figure}

We also investigated whether there is a difference in the luminosity 
functions of clustering galaxies for lower and higher BH mass groups.
Figure~\ref{fig:hist_M_M8M9} shows the comparisons between the magnitude 
distributions for M8 and M9 mass groups, and 
$M_{*}$
parameters measured for them
are summarized in Table~\ref{tab:comp_Mstar_M8M9} and plotted in 
Figure~\ref{fig:Mstar_M89}.

The obtained 
$M_{*}$
parameters for M9 group are systematically
brighter than those for M8 group by more than two sigma at redshifts z0
and z1.
No statistically significant difference is found for the values at 
redshift z2 (both for 310~nm and 220~nm) and z3.
The results may indicate the enhancement of bright galaxies around AGNs
with higher BH mass ($>10^{8.5}M_{\solar}$) at redshift $<$ 1.5, while
at redshift $\ge$ 1.5 the luminosity function of galaxy is similar 
around AGNs with BH mass of $\ge 10^{8}M_{\solar}$.

\begin{table}
  \tbl{Comparisons of $M_{*}$ parameters measured for clustering galaxies 
around AGNs for blue and red galaxy types.}{%
  \begin{tabular}{cccc}
\hline
redshift & wavelength &
  $M_{*}$(blue) &  
  $M_{*}$(red) \\
   & nm         & mag               &  mag               \\
\hline
z0 & 310        & -20.29$\pm$0.18   &  -20.95$\pm$0.25   \\
z1 & 310        & -21.66$\pm$0.23   &  -22.29$\pm$0.37   \\
z2 & 310        & -22.60$\pm$0.35   &  -23.36$\pm$0.36   \\
z2 & 220        & -21.28$\pm$0.22   &  -21.42$\pm$0.53   \\
z3 & 220        & -22.38$\pm$0.20   &  -22.67$\pm$1.13   \\
z4 & 220        & -23.57$\pm$0.53   &  -24.78$\pm$1.43   \\
\hline
  \end{tabular}}
  \label{tab:comp_Mstar_BR}
\begin{tabnote}
\end{tabnote}
\end{table}

\begin{table}
  \tbl{Comparisons of $M_{*}$ parameters measured for clustering galaxies 
around AGNs of M8 and M9 groups.}{%
  \begin{tabular}{cccc}
\hline
redshift
   & wavelength 
   & $M_{*}$(M8) 
   & $M_{*}$(M9) \\
   & nm         & mag               &  mag               \\
\hline
z0 & 310        & -20.29$\pm$0.24   &  -20.83$\pm$0.21   \\
z1 & 310        & -21.36$\pm$0.40   &  -22.12$\pm$0.32   \\
z2 & 310        & -22.93$\pm$0.34   &  -23.06$\pm$0.52   \\
z2 & 220        & -21.50$\pm$0.32   &  -21.22$\pm$0.35   \\
z3 & 220        & -22.58$\pm$0.41   &  -22.44$\pm$0.28   \\
\hline
  \end{tabular}}
  \label{tab:comp_Mstar_M8M9}
\end{table}
\subsection{AGN linear bias and dark matter halo mass hosting AGNs}
\label{sec:bias}

The AGN-galaxy cross-correlation functions can be used to measure 
the linear bias of AGN distribution then to estimate the average
mass of dark matter haloes hosting the AGNs.
Using the relation between the halo mass and its bias obtained by
numerical simulation, we can estimate the halo mass by 
assuming that the AGNs are located in haloes with a similar bias.

To derive the linear bias of AGNs from the AGN-galaxy cross-correlation 
function, we should know also about the auto-correlation function
of galaxies.
It is not, however, determined only from the dataset used in this 
work.
Thus, we assume that auto-correlation functions measured 
by \citet{Zehavi+11}, which were derived by using the SDSS galaxies 
at redshifts $<$ 0.25, doesn't change up to redshift 3.
Since the galaxy auto-correlation functions measured at higher redshifts
\citep[e.g.,][]{Ishikawa+15}
are almost the same as those measured by \citet{Zehavi+11}, it can
be a good approximation.

The auto-correlation function of galaxies was assumed to be expressed as:
\begin{equation}
\xi_{\rm GG} = (1 - f_{\rm blue}) \xi_{\rm GG,red} + f_{\rm blue} \xi_{\rm GG,blue}
\end{equation}
where $\xi_{\rm GG,red}$ and $\xi_{\rm GG,blue}$ are the auto-correlation function
of red and blue galaxies.
We adapted 
$\xi_{\rm GG,red} = (r/6.6 h^{-1}{\rm Mpc})^{-1.9}$  and
$\xi_{\rm GG,blue} = (r/3.6 h^{-1}{\rm Mpc})^{-1.7}$ from \citet{Zehavi+11}.
$f_{\rm blue}$ represents the fraction of blue galaxies and the values
described in Table~\ref{tab:fit_D} were adapted.
Then auto-correlation function of AGNs is calculated as:
\begin{equation}
\xi_{\rm AA} = \xi^{2}_{\rm AG} / \xi_{\rm GG}.
\label{eq:xi_AG}
\end{equation}

In the previous subsections we showed that the galaxies around
AGNs show larger clustering at higher luminosities 
(i.e. $<M_{*}$) than ordinary galaxies (i.e. $>M_{*}$).
As the galaxy auto-correlation function used here is the one 
for $>M_{*}$, we also need to derive the AGN-galaxy cross-correlation
function for $>M_{*}$ galaxies.
The primary cause of the larger clustering of our galaxy samples 
is due to the evolution of, i.e. decrease in, the $M_{*}$ parameter 
in the luminosity function at neighbors of AGNs as shown in the
section~\ref{sec:result_LF}.
To reduce the effect of the $M_{*}$ evolution and derive 
$\xi_{\rm AG}$ for $<M_{*}$ galaxies, 
we calculated the cross-correlation lengths from 
the magnitude distributions of the clustering galaxies
$n_{\rm fit}(M)$ which were obtained by fitting the Schecher function 
to the observation in estimating 
$M_{*}$ for them.
The parameter $\alpha$ of the Schecher function was fixed to $-$1.2.
%%and equation~(\ref{eq:excess_density}).
%%
The magnitude distributions $n_{\rm fit}(M)$ were shown in 
Figure~\ref{fig:hist_M} as dashed lines.

$n_{\rm fit}(M)$ is expressed as:
\begin{equation}
n_{\rm fit}(M) = f(r'_{0}) \phi(M; M_{*}=
M_{*}({\rm clust})  ),
\label{eq:sigma_clust}
\end{equation}
where $f(r'_{0})$ is the multiplying factor calculated as in
equation (\ref{eq:excess_density}) for a given correlation length $r'_{0}$, 
$\phi(M;M_{*}=$ $M_{*}({\rm clust})$)
is the luminosity function as given by $M_{*}({\rm clust})$, which is
$M_{*}$ measured for galaxies clustring around AGNs.
The values of $\phi_{*}$ and $\alpha$ are given by the parametrization
used in this work.
Thus $r'_{0}$, which is the cross-correlation length corrected for the $M_{*}$ 
evolution and thus the one for $>M_{*}$ galaxies, is determined by solving the 
equation~(\ref{eq:sigma_clust}) 
for $r'_{0}$.
The obtained values of $r'_{0} (= r'_{\rm AG})$ are summarized in 
Table~\ref{tab:bias_halomass}.
Using $r'_{\rm AG}$, $\xi_{\rm AG}$ can be expressed as:
\begin{equation}
\xi_{\rm AG} = (r / r'_{\rm AG})^{-\gamma}.
\end{equation}
$\xi_{\rm AA}$ calculated by equation (\ref{eq:xi_AG}) was fitted to the power 
law function of the form  $\xi_{\rm AA} = (r/r_{\rm AA})^{-\gamma}$.
The obtained $r_{\rm AA}$ and $\gamma$ values are summarized in 
Table~\ref{tab:bias_halomass}.

The linear bias of AGNs $b_{\rm AGN}$ is calculated as:
\begin{equation}
b_{\rm AGN} = \frac{\sigma_{\rm 8,AGN}}{\sigma_{\rm 8,DM}},
\end{equation}
where $\sigma_{\rm 8,AGN}$ and $\sigma_{\rm 8,DM}$ are the
rms fluctuations of the AGN and the dark matter density 
within spheres of comoving radius of 8 $h^{-1}$Mpc, respectively.
\begin{equation}
\sigma_{\rm 8,AGN} = J_{2}(\gamma)^{1/2}\left(\frac{r_{\rm AA}}{8}\right)^{\gamma/2},
\end{equation}
\begin{equation}
J_{2}(\gamma) = \frac{72}{(3-\gamma) (4-\gamma) (6-\gamma) 2^{\gamma}},
\end{equation}
and $\sigma_{\rm 8, DM}$ is
\begin{equation}
\sigma_{\rm 8,DM} = \sigma_{8} \frac{D(z)}{D(0)},
\end{equation}
where $D(z)$ is the linear growth factor given as
\begin{equation}
D(z) = \frac{5 \Omega_{\rm m} E(z)}{2} 
    \int_{z}^{\infty} \frac{1+y}{E^{3}(y)} dy,
\end{equation}
\begin{equation}
E(z)^{2} = \Omega_{\rm m} (1 + z)^{3} + \Omega_{\Lambda}.
\end{equation}
Then $b_{\rm AGN}$ is calculated from $r_{\rm AA}$ as:
\begin{equation}
b_{\rm AGN} = \left( \frac{r_{AA}}{8} \right)^{\gamma/2}
   J_{2}(\gamma)^{1/2} \left( \sigma_{8} \frac{D(z)}{D(0)} \right)^{-1}.
\end{equation}
The values of $b_{\rm AGN}$ are summarized in Table~\ref{tab:bias_halomass}.

The relation between dark matter halo mass and bias of dark matter
halo is derived by \citet{Sheth+01} as:
\begin{equation}
b = 1 + \frac{1}{\sqrt{a} \delta_{c}} \left[
    \sqrt{a} (a \nu^{2}) + \sqrt{a} b (a\nu^{2})^{1-c} 
    - \frac{(a\nu^{2})^{c}}{(a\nu^{2})^{c} + b(1-c)(1-c/2)} 
    \right]
\label{eq:b_Mh}
\end{equation}
where, $a$ = 0.707, $b$ = 0.5, $c$ = 0.6,
%% $\nu = \delta_{sc}(z)/\sigma(M_{\rm h})$,
$\delta_{c}$ = 1.686, which is a critical over density required for 
collapse in the spherical model.
$\nu$ is defined as:
\begin{equation}
\nu = \frac{\delta_{c}}{\sigma(M_{\rm h})} \frac{D(0)}{D(z)}
\end{equation}
$\sigma(M_{\rm h})$ is rms density fluctuation of dark matter halo with 
mass $M_{\rm h}$, and is parametrized by \citep{Bosch+02}:
\begin{equation}
\sigma(M_{\rm h}) = \sigma_{8} \frac{f(u)}{f(u_{8})}
\end{equation}
where $u_{8} = 32\Gamma$ with $\Gamma = 0.173$, 
\begin{equation}
u = 3.804 \times 10^{-4} \Gamma \left( \frac{M_{\rm h}}{\Omega_{\rm m}} \right)^{1/3},
\end{equation}
\begin{equation}
f(u) = 64.087 (1 + 1.074 u^{0.3} - 1.581 u^{0.4} + 0.954 u^{0.5} - 0.185 u^{0.6})^{-10}
\end{equation}
Thus the mass of the dark matter halo can be estimated using 
Equation~(\ref{eq:b_Mh}) from the AGN bias parameter.
The mass estimates $M_{\rm h}$ are summarized in Table~\ref{tab:bias_halomass}.

\begin{figure}
  \begin{center}
  \includegraphics[width=0.8\textwidth]{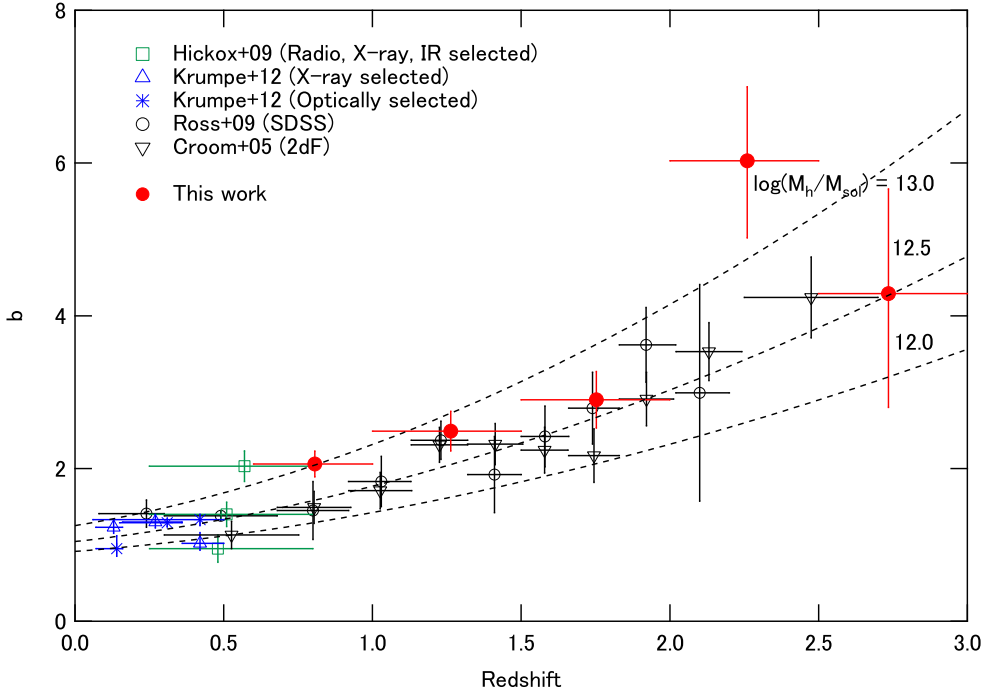} 
  \end{center}
  \caption{
Evolution of the AGN bias. Solid circles are the result
obtained in this work by assuming no evolution in galaxy 
auto-correlation function for $z =$ 0--3 and correcting the 
AGN-galaxy cross-correlation for the $M_{*}$ evolution around 
AGNs. The dashed line represents the bias of the dark matter haloes
calculated using the relation of equation~(\ref{eq:b_Mh})\citep{Sheth+01}
for $\log{(M_{h}/M_{\solar})} =$ 13.0, 12.5, and 12.0 from top to
bottom. As a comparison results obtained by literature are also 
plotted as labeled in the annotation.
The results from \citet{Hickox+09,Krumpe+12} are derived from the
AGN-galaxy cross-correlation, and those from \citet{Ross+09,Croom+05}
are from AGN auto-correlation.
}
  \label{fig:bias_vs_z}
\end{figure}

The obtained AGN biases $b_{\rm AGN}$ are plotted as a function of redshift
with solid circles in Figure~\ref{fig:bias_vs_z}.
The AGN biases obtained in the literature~\citep{Hickox+09,Krumpe+12,Ross+09,Croom+05}
are also shown for the comparison.
In the figure, the bias expected for three halo masses,
$\log{(M_{\rm h}/M_{\solar})}$ = 12.0, 12.5, and 13.0, is plotted
with dashed lines.
The obtained bias at redshift $<$ 2.0 is consistent with those 
obtained in other studies, and hosted by dark matter haloes with
average mass in the range of $10^{12.5}$--$10^{13.0}$M$_{\solar}$.
The large scatters of our results at $>$ 2.0 is mostly due to the
dominance of $<M_{*}$ galaxies in our samples, which introduce
large error in the correction of $r_{\rm AG}$ to $r'_{\rm AG}$.

\begin{table}
  \tbl{AGN linear bias and the dark matter halo mass}{%
  \begin{tabular}{cccccc}
\hline
redshift & $r'_{\rm AG}$$^{a}$ & $r_{\rm AA}$$^{b}$  & $\gamma_{\rm AA}$$^{c}$ 
         & $b_{\rm AGN}$$^{d}$ & $\log(M_{\rm h}/M_{\solar})$$^{e}$ \\
         & $h^{-1}$Mpc & $h^{-1}$Mpc & & & \\
\hline
z0       & 5.5$\pm$0.3 & 6.4$\pm$0.6       & 1.77   
         &  2.06$\pm$0.17 & 13.0$^{+0.1}_{-0.2}$ \\
z1       & 5.3$\pm$0.3 & 6.3$\pm$0.7       & 1.79   
         &  2.49$\pm$0.26 & 12.9$\pm$0.2 \\
z2       & 5.4$\pm$0.4 & 6.1$\pm$0.9       & 1.76   
         &  2.90$\pm$0.37 & 12.6$\pm$0.2 \\
z3       & 6.9$^{+0.6}_{-0.7}$ & 11.2$\pm$2.0 & 1.83   
         &  6.03$^{+0.97}_{-1.01}$ & 13.3$^{+0.2}_{-0.3}$ \\
z4       & 5.5$^{+1.0}_{-1.1}$ & 6.8$^{+2.6}_{-2.4}$ & 1.80   
         &  4.29$^{+1.37}_{-1.49}$ & 12.5$^{+0.4}_{-0.8}$ \\
\hline
  \end{tabular}}\label{tab:bias_halomass}
\begin{tabnote}
$^{a}$ AGN-galaxy cross-correlation length for $>M_{*}$ galaxies
       around AGNs.
$^{b}$ AGN auto-correlation length.
$^{c}$ power law index of AGN auto-correlation function.
$^{d}$ AGN bias.
$^{e}$ Average mass of the dark matter haloes hosting the AGNs.
\end{tabnote}
\end{table}

\section{Discussion and Conclusion}

%%
%% r0 : M_BH dependence
%%

The dependence of AGN-galaxy cross-correlation on redshift,
galaxy luminosity, and BH mass was examined.
The galaxy samples used in this analysis are dominated with 
blue-type galaxies, and the fraction of blue-type galaxies 
is more than 70\%.
No significant dependence on the BH mass was found for the 
cross-correlation length derived using whole galaxy samples.
Considering that the error of BH mass estimation can be as large
as 0.4 dex as explained in section~\ref{sec:agn}, the BH mass 
dependence can be diluted in the comparison with the small mass
range, which is $\sim$1 dex in our sample for each redshift group.
To investigate the relation between BH mass and the host halo mass,
it is required to test the mass dependence for wider BH mass
range using a more accurate mass estimator.

However, we obtained an indication that cross-correlation with 
red galaxies is larger for larger BH mass at the lowest redshift 
group as shown in Figure~\ref{fig:r0_MBH}, which is consistent 
with the previous studies \citep{Komiya+13,Shirasaki+16} in which
galaxy samples are strongly biased to a red type.
The BH mass dependence for the cross-correlation of red 
galaxies and AGN obtained in this work is marginal, but 
it is encouraging to see that we obtained the positive 
correlation for different datasets. 

What does this different BH mass dependence for red and 
blue galaxies mean ?
It is unrealistic to think about direct and physical 
connection between BHs and galaxies separated by more than 
1 Mpc. 
One possible idea to explain such a dependence is
the interaction at a scale of haloes (i.e. groups/clusters
interaction), which causes concurrent
encounters of constituent galaxies and induces ram pressure 
stripping of interstellar medium (ISM) and/or hot gas in 
galaxy haloes, and also brings about galaxy mergers and 
subsequent star formation~\citep{Mihos+04}.
This may result in the increase in the number density of red 
galaxies, and also the $M_{*}$ evolution.
Since those encounters happen to only a part of galaxies,
the number density of the dominant component of blue galaxies 
is almost unchanged.
%%
%%
%%
%%
%%

%%
%% luminosity dependence of cross-correlation length
%%
The redshift dependence of the cross-correlation length 
was found as shown in Figure~\ref{fig:r0_z}, and it turned 
out to be most primarily due to the luminosity dependence 
of galaxy clustering and its redshift evolution.
The luminosity dependence of galaxy clustering has been reported
in literature~\citep[e.g.][]{Zehavi+11,Ishikawa+15}.
They measured auto-correlation lengths for galaxy samples 
divided by their luminosity, and obtained $r_{0}$= 4--5~$h^{-1}$Mpc for 
the lower luminosity groups and $\sim$10~$h^{-1}$Mpc for the 
most luminous group.
Our measurements indicate larger clustering than those 
obtained by the literature, which will be discussed in a 
later part of this section.
The observed luminosity dependence can be explained as a result of 
a shift in the 
$M_{*}$
parameter by $>$1 mag to the 
luminous side as shown in section~\ref{sec:result_color}.
This indicates that the star formation activity is relatively larger 
around AGNs in our sample compared to those in a general field at 
the same redshift.
%%

%%
%% fit
%%
%%
%%
%%

%%
%%
%%

%%
%%
%% blue fraction
%%
The fraction of red galaxies in the environment can be a measure
of the degree of progress on galaxy formation and evolution in 
that system.
The fraction may depend on the frequency of events that 
trigger the star-formation activity, such as major/minor 
merger and/or interaction with a nearby galaxy, and also the time 
scale of the star-formation quenching. 
If the frequency of the triggering event is higher and the time
scale of the quenching is shorter, the galaxy formation and
evolution rapidly proceed and the red galaxy fraction
is increased in such an environment.
In such an environment, it is also expected that the SMBHs
rapidly increase their mass through the mass accretion caused 
by a similar kind of event which triggers the star formation 
in the galaxies around it.

As blue galaxies are dominated, the transformation of only a
small fraction of blue galaxies to red galaxies can significantly 
affect the clustering strength of the red galaxies.
This may be an explanation for the clustering of whole 
galaxies (both blue and red galaxies) almost being independent 
of the BH mass, while the clustering of red galaxies depend on 
the BH mass as reported 
in \citep{Komiya+13,Shirasaki+16}.
The peak position of the color distributions 
for blue galaxies was measured as shifting
to the redder side as the redshift increased.
This is at least partly due to the the luminosity dependence of 
the peak position.
This result may indicate that the luminous star-forming galaxies 
at these redshifts are reddened by dust or by some other mechanism.

%%
%% bias & halo mass
%%
AGN bias was calculated from the AGN-galaxy cross-correlation 
length.
As described above we found that the galaxies around AGNs of our 
samples have significantly larger clustering due to the evolution 
of $M_{*}$ of the luminosity function.
The galaxy auto-correlation function, which was required to derive the
AGN auto-correlation and then to calculate AGN bias, was assumed 
to be expressed as a linear combination of auto-correlation functions
for blue and red galaxies obtained for redshift $<$ 0.25~\citep{Zehavi+11}
with a mixing ratio determined from the color distribution analysis.
Thus we needed to correct the cross-correlation length obtained
from the galaxies clustered more strongly than ordinary galaxies
to match with the clustering properties of galaxies used in 
\citet{Zehavi+11}.

The correction was made 
by finding the cross-correlation length
which gives the galaxy excess density as calculated by extrapolating 
the luminosity function at magnitude $M>M_{*}$.
The luminosity function was obtained by fitting to the data as 
demonstrated in Figure~\ref{fig:hist_M}.
Using the corrected cross-correlation length, we obtained evolution
of AGN bias as shown in Figure~\ref{fig:bias_vs_z}, which is consistent 
with the existing results.
The dark matter halo mass which has the same bias as AGNs
in our samples is in the range of $10^{12.5}$--$10^{13.0}$M$_{\solar}$.
%% which is a typical mass that was measured in other 
%% studies~\citep{Ross+09,Croom+05}.
%%

It should be noted, however, that the result obtained here was
derived under the assumption that galaxy clustering at the investigated
redshifts is expressed with the same correlation function as measured 
at $z<0.25$.
To overcome this situation, we need to measure the galaxy auto-correlation
function for the same galaxy samples which were cross-correlated with
the AGN samples.

If we estimate the galaxy auto-correlation from the AGN-galaxy 
cross-correlation lengths ($\sim$16~$h^{-1}$Mpc) obtained for 
$M<M_{90\%}=-20.3$ galaxies in the redshift group z2 ($z$ = 1.5--2.0) 
and the AGN auto-correlation lengths ($\sim$6~$h^{-1}$Mpc) obtained 
in this work, we obtain $r_{\rm GG} \sim 40\pm6~h^{-1}$Mpc using
the relation of equation (\ref{eq:xi_AG}).
This auto-correlation length is extremely large compared
to the value $\sim$10~$h^{-1}$Mpc for the brightest galaxies obtained
by \citet{Zehavi+11,Ishikawa+15}.
This is the indication that the spatial distribution of galaxies with
absolute magnitude $M<M_{*}$ is not independent of that of AGNs.
The interaction between cluster and/or group galaxies could produce
both the AGN phenomena and starburst galaxies in the same region of 
a few Mpc, which results in the association of the overdense region of 
$<M_{*}$ galaxies with AGNs.

If we assume that the overdensity of $<M_{*}$ galaxies is restricted
to the region surrounding the AGNs, e.g. within the sphere of 10~$h^{-1}$Mpc 
radius, the effective volume occupied by the overdensity region is 0.4\% 
(for 1235 AGNs) of the total volume at redshift 1.5--2.0.
Thus contribution of the overdensity to the auto-correlation function 
measured in the total volume is reduced by 1/250.
The relative overdensity of the clustering with $r_{0} = 16~h^{-1}$Mpc to
that with $r_{0} = 10~h^{-1}$Mpc is calculated as 
$\sim \left( \frac{16}{10} \right)^{1.8} = 2.3$ assuming the power law density 
profile with power index of 1.8.
Thus the overdense region can occupy up to $\sim$1/2 of the total volume 
without conflicting with the clustering measurements in literature.
The relative overdensity can be larger if the effective volume occupied
by the overdensity is small enough to be consistent with the clustering
measurement.
It should be noted that the galaxy bias can be larger at locations 
where the rapid progress of galaxy evolution is occurring and so 
is not necessarily the same as the bias of the dark matter halo.
The galaxy evolution is governed by a baryonic physics such as
gas cooling, star formation, and supernova/AGN feedback, which are
mostly (not entirely) unrelated with the underling dark matter,
thus the bias of galaxies selected with specific properties such as
color and luminosity can deviate from the halo bias~\citep{Baugh-13}.
Although much progress has been made to relate the galaxy clustering 
to the distribution of dark matter, we still lack the robust relation
which is applicable to various types of galaxies.
As we discussed above, the galaxy bias can overwhelm the bias of the halo 
hosting the galaxies.
Thus care should be taken when estimating the average mass of the 
dark matter haloes from the clustering properties of luminous galaxies,
which usually assume the equality between the galaxy and halo biases.
We demonstrated that HSC-SSP imaging data alone can 
be a powerful tool to investigate the environment of AGNs at 
intermediate redshifts up to 3.0.
The current S15b dataset covers 
240~deg$^{2}$ 
of sky which is only
1/5 of the planned survey area.
Using the final dataset, we can significantly improve the measurement
on the characteristics of environmental galaxies around AGNs, their 
dependence on types of AGN, and also their redshift evolution.
It is also important to combine datasets taken at other wavelengths,
especially infrared bands, which improves the selection of galaxies 
that are located at the AGN redshift and also the distinction of red-
and blue-type galaxies at higher redshifts, and enable measurement of
the properties of dust-obscured populations that are not detected 
in the HSC.
The analysis method used in this work can be adapted to other 
types of extragalactic objects with known redshift.

In this work we focused on AGNs that have BHs with relatively 
larger mass of $M_{\rm BH} \ge 10^{7} M_{\solar}$,
It is also required to investigate the AGNs with lower BH mass and
compare the difference in their environments to understand 
their effect on the evolution of SMBH.
Those AGNs will be identified by the follow-up observations of the
candidate AGNs discovered in the HSC-SSP dataset.
The future Subaru Prime Focus Spectrograph (PFS) survey project will
also provide dataset of those AGNs, and it will provide an essential 
dataset to directly measure the environment of individual AGN.

\begin{ack}

We would like to thank the anonymous referee for the constructive 
feedback, which helped us in improving the paper.
We would like to thank Michael Strauss for the valuable comments.
TN is financially supported by JSPS KAKENHI (16H01101 and 16H03958).
The Hyper Suprime-Cam (HSC) collaboration includes the astronomical 
communities of Japan and Taiwan, and Princeton University.  The HSC 
instrumentation and software were developed by the National 
Astronomical Observatory of Japan (NAOJ), the Kavli Institute for 
the Physics and Mathematics of the Universe (Kavli IPMU), the 
University of Tokyo, the High Energy Accelerator Research 
Organization (KEK), the Academia Sinica Institute for Astronomy and 
Astrophysics in Taiwan (ASIAA), and Princeton University.  Funding
was contributed by the FIRST program from Japanese Cabinet Office, 
the Ministry of Education, Culture, Sports, Science and Technology 
(MEXT), the Japan Society for the Promotion of Science (JSPS),  
Japan Science and Technology Agency  (JST),  the Toray Science  
Foundation, NAOJ, Kavli IPMU, KEK, ASIAA,  and Princeton University.
This paper makes use of software developed for the Large Synoptic Survey
Telescope. We thank the LSST Project for making their code available as 
free software at 
http://dm.lsstorp.org.
Funding for the Sloan Digital Sky Survey IV has been provided by
the Alfred P. Sloan Foundation, the U.S. Department of Energy Office of
Science, and the Participating Institutions. SDSS-IV acknowledges
support and resources from the Center for High-Performance Computing at
the University of Utah. The SDSS web site is www.sdss.org.
The Pan-STARRS1 Surveys (PS1) have been made possible through contributions 
of the Institute for Astronomy, the University of Hawaii, the Pan-STARRS 
Project Office, the Max-Planck Society and its participating institutes, 
the Max Planck Institute for Astronomy, Heidelberg and the Max Planck 
Institute for Extraterrestrial Physics, Garching, The Johns Hopkins 
University, Durham University, the University of Edinburgh, Queen's 
University Belfast, the Harvard-Smithsonian Center for Astrophysics,
the Las Cumbres Observatory Global Telescope Network Incorporated, 
the National Central University of Taiwan, the Space Telescope Science
Institute, the National Aeronautics and Space Administration under Grant
No. NNX08AR22G issued through the Planetary Science Division of the NASA
Science Mission Directorate, the National Science Foundation under 
Grant No. AST-1238877, the University of Maryland, and Eotvos Lorand 
University (ELTE)
SDSS-IV is managed by the Astrophysical Research Consortium for the 
Participating Institutions of the SDSS Collaboration including the 
Brazilian Participation Group, the Carnegie Institution for Science, 
Carnegie Mellon University, the Chilean Participation Group, 
the French Participation Group, Harvard-Smithsonian Center for Astrophysics, 
Instituto de Astrof\'isica de Canarias, The Johns Hopkins University, 
Kavli Institute for the Physics and Mathematics of the Universe (IPMU) / 
University of Tokyo, Lawrence Berkeley National Laboratory, 
Leibniz Institut f\"ur Astrophysik Potsdam (AIP),  
Max-Planck-Institut f\"ur Astronomie (MPIA Heidelberg), 
Max-Planck-Institut f\"ur Astrophysik (MPA Garching), 
Max-Planck-Institut f\"ur Extraterrestrische Physik (MPE), 
National Astronomical Observatories of China, New Mexico State University, 
New York University, University of Notre Dame, 
Observat\'ario Nacional / MCTI, The Ohio State University, 
Pennsylvania State University, Shanghai Astronomical Observatory, 
United Kingdom Participation Group,
Universidad Nacional Aut\'onoma de M\'exico, University of Arizona, 
University of Colorado Boulder, University of Oxford, University of Portsmouth, 
University of Utah, University of Virginia, University of Washington, University of Wisconsin, 
Vanderbilt University, and Yale University.

\end{ack}

\begin{appendix}
\section*{}
In this appendix, the group and parameter names that are frequently refereed 
to in the text are summarized in Table~\ref{tab:param_desc}.

\begin{table}[h]
  \tbl{Summary of the names of the AGN dataset groups and of the parameters frequently refereed
to in the text.}{
  \begin{tabular}{cp{0.7\textwidth}}
\hline
group or parameter name & description \\
\hline
z0, z1, z2, z3       & names of the redshift group defined in Figure~\ref{fig:AGN_M-z}\\
M8, M9               & names of the black hole mass group defined in Figure~\ref{fig:AGN_M-z}\\
$M_{*}$$^{a}$        & characteristic absolute magnitude of the Schechter function as defined in equation~(\ref{eq:schechter})\\
$M_{50\%}$, $M_{90\%}$         & absolute magnitudes where the detection efficiency is 50\% and 90\% for a given redshift group\\
$M_{\lambda 310}$, $M_{\lambda 220}$ & absolute magnitudes at wavelengths of 310~nm and 220~nm in the source frame\\
$D_{1}$, $D_{2}$     & color parameters defined for redshift 0.6--2.0 and 1.5--3.0, see section~\ref{sec:ana_color} for their definitions \\
$f_{\rm blue}$       & blue galaxy fraction\\
$\rho_{0}$$^{b}$     & average number density of galaxies which are brighter than the threshold magnitude and can be detected at the redshift of AGN\\
$\xi$($r$), $\omega$($r_{p}$)& correlation function and projected correlation function\\
$r_{0} (=r_{\rm AG}$), $\gamma$    & cross-correlation length and power law index of the AGN-galaxy cross-correlation function as defined in equation~(\ref{eq:cc_func})\\
$r_{\rm AA}$, $r_{\rm GG}$   & auto-correlation lengths of AGNs and galaxies\\
$M_{\rm BH}$         & black hole mass\\
$M_{\rm h}$          & dark matter halo mass\\
$b_{\rm AGN}$            & AGN bias \\
\hline
  \end{tabular}
}\label{tab:param_desc}
\begin{tabnote}
$^{a}$ Where necessary this parameter is distinguished as 
$M_{*}({\rm param})$ or $M_{*}({\rm clust})$ for describing
the $M_{*}$ given by the parametrization discussed in section~\ref{sec:cc_length}
and that measured from this observational data.
$^{b}$ This parameter is calculated integrating the product of luminosity function
and detection efficiency function as given in equation~(\ref{eq:rho0}). 
The luminosity function is given by the parametrization as described in 
section~\ref{sec:cc_length}, and the detection efficiency function is measured
for each AGN dataset as described in section~\ref{sec:cc_length}.
\end{tabnote}
\end{table}

\end{appendix}

\clearpage
%%%
% See the manual for the detail.
%%%

\end{document}